\documentclass[final,5p,times,twocolumn,authoryear]{elsarticle}

\usepackage{stfloats}
\usepackage{amssymb}
\usepackage{amsmath}
\usepackage{lineno}
\usepackage{xcolor,colortbl}
\usepackage{soul}
\usepackage{gensymb} 
\usepackage{url}
\usepackage{subcaption}
\usepackage{hyperref}
\setcitestyle{authoryear,round,semicolon,aysep={,},yysep={\&}}
\graphicspath{ {./figures/} } 
\journal{International Journal of Heat and Mass Transfer}

\begin{document}

\begin{frontmatter}

\title{Quantifying Injection-Driven, Interphase Mass Transfer within Porous Media via Time-Elapsed X-ray micro-Computed Tomography} 

\author[label_UTK_CEE] {\textbf{Christopher A. Allison}} 
\author[label_Sydney]     {Ruotong Huang}
\author[label_UCL]     {Anindityo Patmonoaji}
\author[label_ANU]     {Lydia Knuefing}
\author[label_UTK_CEE] {\textbf{Anna L. Herring}} 

\affiliation[label_UTK_CEE]{organization={Department of Civil and Environmental Engineering, University of Tennessee - Knoxville},
            city={Knoxville},
            state={TN},
            country={USA}}      
\affiliation[label_Sydney]{organization={School of Civil Engineering, 
            University of Sydney},
            city = {Sydney},
            state = {NSW},
            country = {Australia}
            }
\affiliation[label_UCL]{organization={Department of Chemical Engineering, University College London},
            city={London},
            country={United Kingdom}
            } 
\affiliation[label_ANU]{organization={Department of Materials Physics, Research School of Physics, Australian National University},
            city={Canberra},
            country={Australia}
            }       
\begin{abstract} 
Understanding interphase mass transfer is essential for a variety of applications in porous media, from groundwater remediation to geologic energy storage. While X-ray micro-Computed Tomography ($\mu$CT) is a measurement technique that can provide critical \textit{in situ} observations, quantifying transient mass transfer requires models and workflows capable of compensating for the spatial and temporal resolution constraints. Current literature presents three analytical frameworks for evaluating interphase mass transfer using time-lapsed sequences of X-ray $\mu$CT scans: the Slice-Averaged Concentration (SAC) approach, the Non-Classified per-Cluster (NPC) approach, and the Classified per-Cluster (CPC) approach. We systematically evaluate the frameworks and results for each of the three approaches as applied to several sequences of time-lapsed  $\mu$CT scans, each observing hydrogen dissolution experiments at varying injection rates. Estimates of system-scale properties – e.g., average mass transfer coefficients – vary within an order of magnitude, increasing as injection rate decreases. However, approach estimates diverged when approximating more complex, pore-scale phenomena, such as aqueous solute concentration. Per-Cluster approaches preserve more information, enabling quantification of important localized physicochemical processes such as dissolution fingering and cluster remobilization. Given the lack of ground-truth benchmarks for mass-transfer kinetic data in the conditions of interest to fields such as geologic storage, additional research is still needed to understand how these approaches vary under different state and flow conditions. The utility of one approach over another is ultimately determined by the desired level of system detail, at the cost of the computational resources required to achieve it. Our results provide a framework for researchers to select analytical approaches based on the desired level of physical detail.

\end{abstract}
\begin{graphicalabstract}

\includegraphics[width = \textwidth]{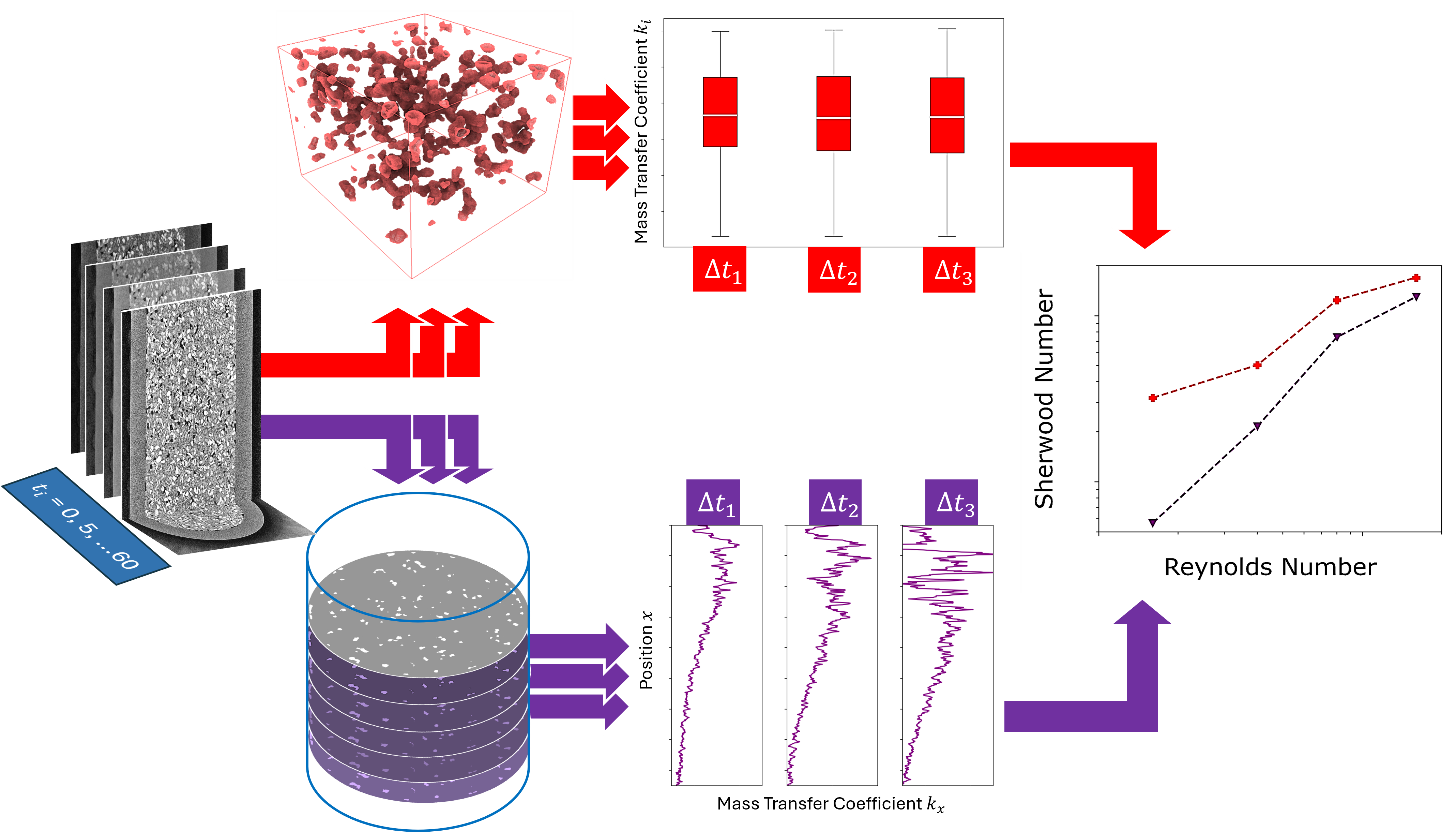}
\end{graphicalabstract}
\begin{highlights} 
\item Literature establishes three approaches for estimating mass transfer from µCT data.

\item We systematically investigate each approach as applied to time-lapsed sequences of µCT images.

\item All approaches consistently estimated mass transfer coefficients within an order of magnitude.

\item The scale of the information preserved depends on the approach selected.

\end{highlights}

\end{frontmatter}


\section{Introduction}
\label{sec:intro}
\subsection{Background}
\label{subsec:back_intro}
Understanding the dynamics of interphase mass transfer as a function of bulk transport conditions is a key element in the management and control of a wide variety of engineered and natural processes, from resource extraction and component manufacturing to water treatment and pollution control \citep{ENVE_tranport, sep_process_king}. Characterizing interphase mass transfer and transport in porous media is particularly important for groundwater contamination and remediation \citep{Donaldson_1997, Miller_1990}, geologic energy storage \citep{Crotogino_2017, Higgs_2023}, and geologic carbon sequestration \citep{IPCC2022, Zahasky_Krevor_2020}. Extensive work has gone into developing empirical and analytical expressions to characterize mass transfer as a function of complex variables that describe system-scale and pore-scale phenomena. For multiphase fluid systems in 3-D geologic media, direct observation of \textit{in situ} variables and their development is often obscured by the media and vessels in which they occur. X-ray micro-Computed Tomography (X-ray $\mu$CT) is a non-destructive imaging technique utilized to visualize optically obscured internal structures and dynamic processes in porous media \citep{Wildenschild2013_betti, Bultreys_2016}. Much research has gone into developing X-ray $\mu$CT into a technology that provides accurate, spatially resolved data necessary to estimate mass transfer properties, while accounting for the inherent limitations of imaging. Properties such as phase boundaries and cluster volumes, as well as interactions among media, fluids, and gases, can be quantified \citep{Huang_2021, Schlüter_2014}. However, using $\mu$CT to evaluate the propagation of time-dependent phenomena, analogous to measuring a ``pulse'' in a fixed observation window, entails a trade-off between sampling fast enough in time to characterize the process and imaging the space well enough to accurately measure geometries. Quick scans of a large system will have a lower voxel resolution and potentially a higher incidence of artifacts and noise than slow scans of a small region of interest \citep{WILDENSCHILD_2002, Wildenschild2013_betti, Schlüter_2014}.  Both the resolution in time and space will affect the accuracy of the measured rates, and any analytical basis for characterizing the pulse through the imaged system must also account for these limitations through models and/or assumptions, at the cost of what processes related to mass transfer are preserved in the final results.\par
\subsection{Mass Transfer Measurement} 
\label{sec:mass_transfer_measurement}
Dissolution of a substance into a solvent is often described in terms of a driving force (\ref{eq:mass_drive}). Here, $\frac{\partial m_{\text{g}}}{\partial t}$, the rate at which the gaseous solute mass can dissolve into the solvent, is related to how quickly dissolved solute concentration ($C_{\text{g}}$) can be diffused into the bulk solvent at position ($x$), through the interfacial area ($A$) between the gas and the solvent. The diffusion coefficient ($D_{g,w}$) quantifies the ease with which dissolved solute molecules move through the solvent under the influence of molecular forces. The dissolved solute lingering at the interface is treated as a boundary layer, or film, and allows Eq. (\ref{eq:mass_drive}) to be simplified to Eq. (\ref{eq:film_theory}), where $C_{\text{sol}}$ is the maximum concentration of the solute, the solubility limit, at the interface, and $\delta$ is the distance from the interface to the bulk concentration of the solvent, $C_{\text{g}}$. \citep{sep_process_king, Miller_1990}
\begin{align}
    \left(\frac{\partial m_{\text{g}}}{\partial t}\right)^{\text{cluster}} &=  -D_{g,w}A\frac{\partial C_{\text{g}}}{\partial x}  \label{eq:mass_drive}\\
    \left(\frac{d m_{\text{g}}}{d t}\right)^{\text{cluster}} &= -D_{g,w}A\frac{C_{\text{sol}} - C_{\text{g}}}{\delta} \label{eq:film_theory}
\end{align}
The boundary layer is a thin, relatively stagnant layer of saturated solvent that separates the bulk transporting solvent from the solute source. The thickness of this layer is determined by the net transport properties (advection and diffusion) of the bulk phase and the solute's diffusion through the boundary layer. Thus, the rate of mass transfer between the solute source and the bulk transport phase is limited by diffusion and the thickness of the boundary layer $\delta$.  Due to the complexity of characterizing $\delta$, $D$ and $\delta$ are lumped into a single term, the mass transfer coefficient, $k$ (Eq. \ref{eq:mass_trans_def}) \citep{sep_process_king, Miller_1990}.
\begin{equation}
    \label{eq:mass_trans_def}
    \frac{dm_{\text{g}}}{d t}=  \rho^{gas}\frac{d V_{\text{g}}}{d t} = -kA(C_{\text{sol}} - C_{\text{g}})
\end{equation}
Eq. (\ref{eq:mass_trans_def}) assumes that the density of the solute gas ($\rho^{gas}$) is constant for small perturbations to temperature and pressure, and that the change in cluster mass is directly proportional to the cluster's change in volume ($V_{g}$).\par
Eq. (\ref{eq:mass_trans_def}) describes mass transfer in a wide variety of complex systems. However, one limitation of \textit{in situ} mass transfer measurements is the difficulty or inability to measure the concentration of dissolved species in the bulk fluid. $\mu$CT can only resolve different phases by the extent to which each phase attenuates the incident X-ray beam, meaning that $\mu$CT can only resolve the spatial distribution of dissolved species that significantly alter the attenuation as a function of concentration, applicable for solutes such as salts \citep{Liyanage_2019}, but this is not the case for most gases. Currently, in existing literature, there are two categories of methodologies that can be used to quantify mass transfer events captured in the X-ray tomograms to solve Eq. (\ref{eq:mass_trans_def}) for gases: the Slice-Averaged approach \citep{Patmonoaji_2023, Lv_2024} and per-Cluster approaches \citep{Ruotong_H_2023, Lv_2024}.\par
\subsubsection{Existing µCT Mass Transfer Studies}
\newcommand{\quarter}{\dimexpr0.48\linewidth\relax}
\begin{figure*}[htb!]
    
    \begin{subfigure}{\quarter}
        \centering
        (a)\\\includegraphics[width=0.9\linewidth]{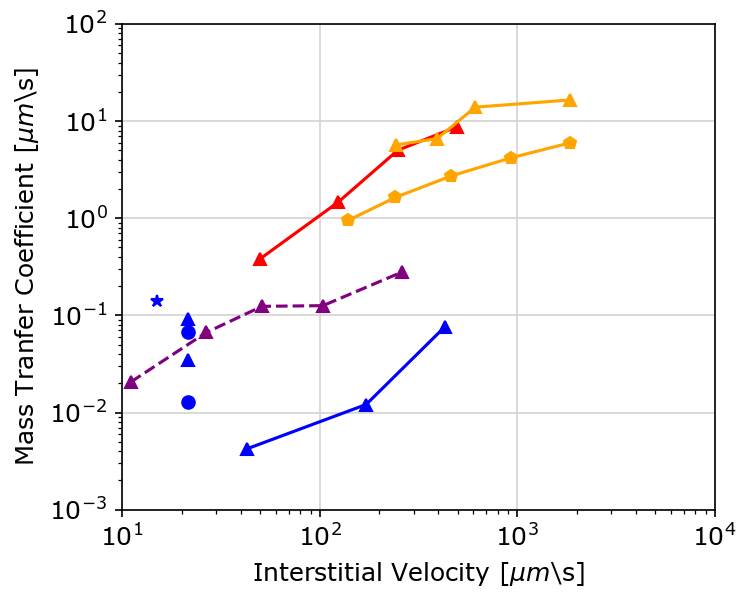}
    \end{subfigure}%
    \begin{subfigure}{\quarter}
        \centering
        (b)\\\includegraphics[width=0.9\linewidth]{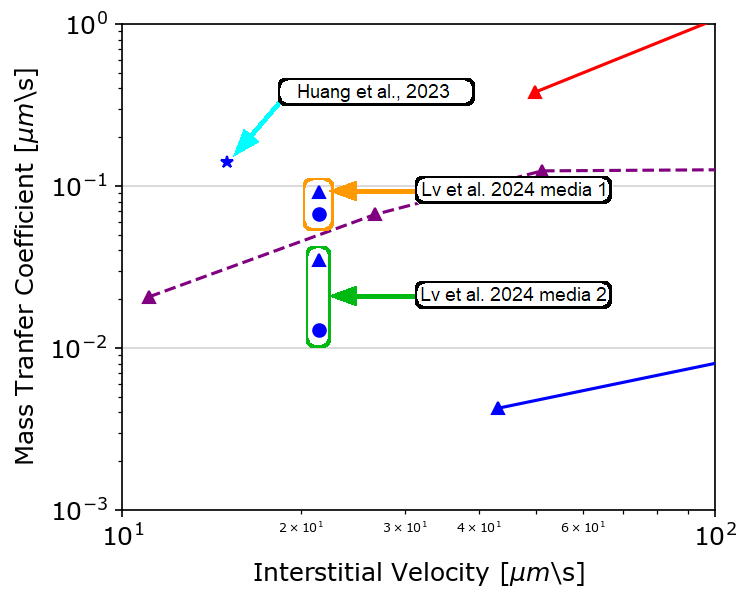}
    \end{subfigure}%
    
    \begin{subfigure}{\quarter}
        \centering
        (c)\\\includegraphics[width=0.9\linewidth]{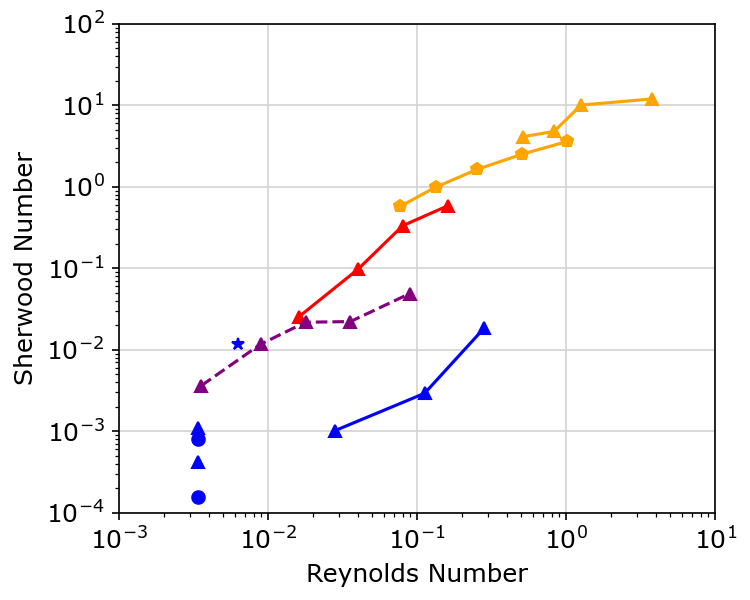}
    \end{subfigure}%
    \begin{subfigure}{\quarter}
        \centering
        \includegraphics[width=\linewidth]{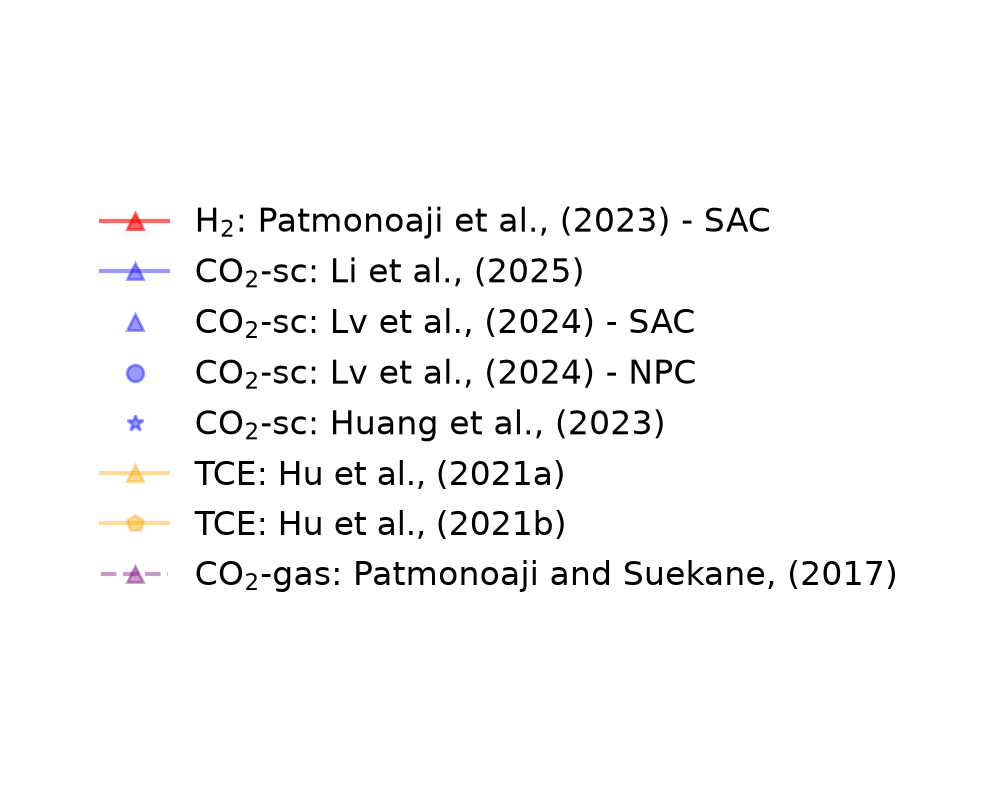}
    \end{subfigure}%
    \caption {A cross-study plot of the resulting mass transfer coefficients from various studies, against the interstitial velocity (a), and in dimensionless form (c) as the Sherwood number and the Reynolds number. Figure (b) is the same plot as (a) but zoomed in on the per-Cluster method studies (CPC: \citet{Ruotong_H_2023} and NPC: \citet{Lv_2024}), which are annotated for clarity. All plots share the same legend. The marker shape denotes the approach used for the data; SAC: triangles, NPC: filled circles, pentagons: \citet{Hu_2021b}, and CPC: stars. Each color indicates the solute evaluated by the study: hydrogen gas (red), trichloroethylene (TCE: orange), gaseous carbon dioxide (purple), and supercritical carbon dioxide (blue). Note: \citet{Pat_Suek_2017}, \citet{Hu_2021a, Hu_2021b}, and \citet{Li_2025} present their Reynolds numbers in terms of Darcy velocity (u); we have converted these to an interstitial velocity basis ($\nu$) for this comparison, using the porosities given in each study.}
    \label{fig:cross_study_plots}
\end{figure*}
\newcommand{\short}{3.0cm}
\begin{table*}[htb!]
    \caption{Experimental setup and conditions for the studies plotted in Figure \ref{fig:cross_study_plots}}
    \resizebox{\textwidth}{!}{%
    \begin{tabular}{ l c c c c c} \hline \hline
        Study & Solute & Media & Approach \#1 & Approach \#2 & Notes\\\hline
        \citet{Pat_Suek_2017} & CO$_2$ & plastic packing & SAC & --- & (25 °C, 1.0 bar) \\
        \citet{Hu_2021a}& TCE & glass bead pack & SAC & --- & (25 °C, 1.0 bar)\\
        \citet{Hu_2021b} & TCE & sand pack & Mixed  & --- & (25 °C, 1.0 bar)\\
        \citet{Ruotong_H_2023} & CO$_2$ & Bentheimer sandstone &CPC & --- &  (45 °C, 86.2 bar)\\ 
        \citet{Patmonoaji_2023} & H$_2$ & plastic packing & SAC & --- & (25 °C, 1.0 bar)\\
        \citet{Lv_2024} & CO$_2$ & Berea and unnamed sandstones & SAC & NPC &  (40 °C, 80.0 bar) \\
        \citet{Li_2025} & CO$_2$ & sand pack & SAC & --- & (40 °C, 80.0 bar) \\
    \end{tabular}}
    
    \label{tab:cross_study}
\end{table*}
In general, any methodology to quantify mass transfer via X-ray $\mu$CT must also account for $\mu$CT limitations (limited resolution in time and space) through models and/or assumptions. Many studies lack ground-truth measurements to assess the accuracy and validity of any part of the workflow pipeline, from image acquisition to data evaluation. Often, replicate data is also lacking. Both missing aspects stem, in part, from the time and resources $\mu$CT imaging requires, the sensitivity of mass transfer in porous media to the specifics of the experimental system, and the capability of $\mu$CT to measure elements of mass transfer.\par
To date, no standardized workflows exist for acquiring, processing, and evaluating $\mu$CT data for mass transfer for specific properties or data qualities, and all are situationally dependent. Specific image acquisition and processing pipelines depend on the user's available resources to capture the phenomena and process the resulting large volumes of data. The goal of the current study is to evaluate established analytical approaches for quantifying mass transfer via $\mu$CT, understand how they handle the limited information available from $\mu$CT imaging, and identify when their underlying assumptions fail.\par
Each analytical approach is a framework of calculations and assumptions; these frameworks allow the phenomena captured by the $\mu$CT scans (changing gas-phase volumes with time) to be applied to idealized equations, such as Eq. (\ref{eq:mass_trans_def}), to estimate mass transfer properties. The predominant distinction between the Slice-Average Concentration (SAC) and per-Cluster (Non-Classified (NPC) and Classified (CPC)) approaches lies in the assumptions that either simplify the system's dimensionality or alter how the system is sampled to approximate its values. While these assumptions are mathematically simple, each influences how the estimates should be interpreted in the broader context of the system and may begin to fail under a variety of conditions. Given that there are often no ground-truth measurements of interphase mass transfer in porous media, the estimates of each approach need to be evaluated in aggregate to understand where these frameworks may begin to fail. Similar or converging estimates would indicate that the estimated values are independent of the approach selected, and thus independent of the underlying assumptions, while dissimilar/diverging estimates would indicate otherwise and would need to be evaluated. Figure \ref{fig:cross_study_plots} plots the mass transfer estimates from several studies evaluating porous media with $\mu$CT, against the injection rate of the advecting solvent. Table \ref{tab:cross_study} lists the experimental conditions and approach used by each study in Figure \ref{fig:cross_study_plots}. To enable generalization of each study's results, we have presented the values in dimensional form (mass transfer coefficient:$ k$ and interstitial advective velocity $\nu$) and in dimensionless form (Sherwood number: Sh [Eq. \ref{eq:Sh}] and Reynolds number: Re [Eq. \ref{eq:Re}]).
\begin{align}
    Sh &= \frac{\text{total mass transfer}}{\text{diffusive mass transport capability}} = \frac{k d_{\text{50}}}{D_{g,w}} \label{eq:Sh}\\
    Re &= \frac{\text{Inertial fluid forces}} {\text{Viscous fluid forces}} = \frac{\rho_{\text{w}} \nu d_{\text{50}}}{\mu _{\text{w}}} \label{eq:Re}
\end{align}
where $\rho_{\text{w}}$, $\mu _{\text{w}}$, and $\nu$ are the density, viscosity, and interstitial velocity of the injected water, respectively; $d_{50}$ is the mean particle size (or characteristic length) of the porous medium.\par
Each of the studies presented in Table \ref{tab:cross_study} and Figure \ref{fig:cross_study_plots} was chosen to form a complementary pair (i.e., evaluating the same analyte but using a different analytical approach), except for the hydrogen (H$_2$) dataset from \citet{Patmonoaji_2023}, which the current study will complement. Note that some datasets and studies were not included either because they did not have a complement pair \citep[$N_2$, $CH_4$, and $C_2H_6$ data sets]{Patmonoaji_2023} or because the study evaluated parameters not properly encapsulated by Figure \ref{fig:cross_study_plots} \citep{jiang_2017, Patmonoaji_2021}. Additionally, because of how we classified the approaches, studies like \citet{Hu_2021b} are considered a mixture of the SAC and per-Cluster approaches. Multiple aspects of each approach, e.g., the image processing workflow or the resulting calculations, are not mutually exclusive.\par
As shown in Figure \ref{fig:cross_study_plots}, most studies (including those not plotted: \citet{jiang_2017} and \citet{Patmonoaji_2021}) employ the SAC approach. At the time of the current study, two studies use a per-Cluster approach (CPC: \citet{Ruotong_H_2023}, NPC: \citet{Lv_2024}), which differ in how they sample the resulting measurement pools of isolated clusters; both evaluate only one advective condition (highlighted in Figure \ref{fig:cross_study_plots}.b). Notably, \citet{Lv_2024} is unique as, at the time of the current study, it is the only study to evaluate multiple analytical approaches, comparing the NPC and SAC approaches for supercritical CO$_2$ trapped in several geologic media. Though \citet{Lv_2024} did not evaluate the approaches over multiple advective injection rates, nor analyze the implications behind the frameworks, their findings provide several insights relevant to the current study:
\begin{itemize}
    \item For the same media, the estimated mass transfer coefficient for the SAC is greater than that for the NPC.
    \item For the same media, the different estimates for each approach fall within the same order of magnitude.
    \item The difference in approach estimates (i.e., $\Delta = k_{SAC} - k_{NPC}$) varies across porous media, under the same advective conditions.
\end{itemize}
\citet{Patmonoaji_2021} (not in Figure \ref{fig:cross_study_plots}) similarly found that altering the size fractions in granular packing yielded different mass transfer coefficients for each of the tested compositions, using the same approach and advective conditions. The results from \citet{Patmonoaji_2021} and \citet{Lv_2024} indicate that studies under similar conditions (e.g., \citet{Ruotong_H_2023} and \citet{Li_2025}; Table \ref{tab:cross_study}) cannot be meaningfully compared unless they are effectively in the same media, due to the dependence of mass transfer on pore-scale processes (e.g., interface geometries and advective flow paths). Thus, the goal of the current study is to compare the estimates from each approach at various advective injection rates, under more constrained conditions. Meanwhile, we will systematically analyze the assumptions and calculation frameworks to understand what is being imposed on the system, and where those implications may fail. To constrain the number of variables to advection-driven mass transfer, each approach will be applied to the same sequences of $\mu$CT data evaluating the same analyte gas, in the same porous media, over several advective injection rates. This will be accomplished by reevaluating the experimental H$_2$ dissolution tomogram sequences from \citet{Patmonoaji_2023}, now using the CPC approach developed by \citet{Ruotong_H_2023} and the NPC approach detailed in \citet{Lv_2024}, in comparison with the original estimates from the SAC approach.\par

\subsubsection{ Slice-Averaged Concentration Approach} 
The \textit{Slice-Averaged Concentration} (SAC) approach has been used in numerous prior studies \citep{Pat_Suek_2017, jiang_2017, Hu_2021a, Hu_2021b, Patmonoaji_2021, Patmonoaji_2023, Lv_2024, Li_2025}. The approach begins by using the advection-diffusion equation (Eq. \ref{eq:pat_advet_disp}) to model the movement of the injected solvent and the propagation of a mass transfer front through the media, informed by the observed change in gas saturation in time, at a discrete $x$ position. The change in axial saturation is then used to solve for the evolving concentration profile along the principal axis ($x$), resulting in the aqueous solute concentration averaged over each cross-section.
\begin{equation}
       \phi (1-S^{gas})\frac{\partial C}{\partial t} = \frac{\partial}{\partial x}\left[ \phi (1-S^{gas})D_{h}\frac{\partial C}{\partial x} \right] - q\frac{\partial C}{\partial x} - \rho^{gas} \phi \frac{\partial S^{gas}}{\partial t} \label{eq:pat_advet_disp} \\
\end{equation}
Here $S^{gas}$ is the gas phase saturation of the pore space, $\phi$ is the porosity, $D_{h}$ is the hydrodynamic dispersion coefficient (which combines diffusion and dispersion), and $\rho ^{gas}$ is the density of the gas phase. Given the complexity of analytically solving (\ref{eq:pat_advet_disp}), the system is assumed to be at pseudo-steady state, as the dissolution-driven mass transfer is considered significantly slower than the advective transport. Considering the system to be advection dominant in time, and assuming radial symmetry, Eq. (\ref{eq:pat_advet_disp}) is simplified to be a one-dimensional function of the fluid advecting through the media and the change in gas-phase saturation with time (Eq. \ref{eq:pat_conc_x}):
\begin{equation}
    C_{x+\Delta x, t+\Delta t} = C_{x,t} - \rho^{gas} \phi \frac{1}{q} \frac{\Delta S^{gas}}{\Delta t}\Delta x \label{eq:pat_conc_x} \\
\end{equation}
Here, $C_{x,t}$ and $C_{x+\Delta x, t+\Delta t}$ are the dissolved gas concentrations: $C_{x,t}$ is the average concentration of a cross-section at axial position $x$ and current time interval $t$; $C_{x+\Delta x, t+\Delta t}$ is similarly the cross-section average, but defined one voxel length forward and at the next time-interval. The remaining terms $\phi$ and $S^{gas}$ are the same, but averaged over the cylindrical cross-section. $\rho^{gas}$ is the density of the gas phase, and $q$ is the Darcy flux.\par
Eq. (\ref{eq:pat_conc_x}) is solved iteratively with initial condition all $C(x,t=0) = 0$, and the boundary condition that $C(x=0,t) = 0$. The $C(x=0,t) = 0$ boundary condition requires that the axial image cannot be cropped arbitrarily in the axial direction ($x^{*}$); as $x^{*}=0$ must be the injection inlet for $C(x^{*}=0,t) = 0$ to be true at all $t$, otherwise, $C_{x+\Delta x, t+\Delta t}$ is missing critical $C_{x,t}$ information from the prior slice and timestamps. This matters because axial cropping is typically performed to remove regions of poor image quality and imposes acquisition and quality requirements to maintain that $x^{*}=0$ is in the usable region of the image (the vertical extremes in many conical beam X-ray $\mu$CT systems are often lower-quality data regions \citep{Wildenschild2013_betti}). This contrasts with per-cluster approaches; they are a-positional, so regions may be cropped to preserve image quality at the cost of the sample window.\par
With an estimate of average concentration at each axial cross-section, the mass transfer of each slice ($k_{\Delta x, \Delta t}$) is then estimated using Eq. (\ref{eq:pat_mass_trans_coeff}). Here, $k_{\Delta x, \Delta t}$ refers to the mass transfer coefficient as a function of axial position and time-interval to produce a linear series of ($k_{\Delta x}$) values for each time-interval, which we will represent with $\mathbb{K}_{\Delta t} (\Delta x)$.\citep{Patmonoaji_2021, Patmonoaji_2023}
\begin{align}
    &k_{\Delta x, \Delta t} = \frac{\rho^{gas} \phi}{a_{\Delta t,\text{avg}}} \frac{\Delta S^{gas}}{\Delta t} \frac{1}{C_{s} - C^\text{avg}_{\Delta x,t}} \label{eq:pat_mass_trans_coeff}\\
    &\text{where}\ C^{\text{avg}}_{\Delta x,t} = \frac{C_{x,t} + C_{x+\Delta x,t} }{2} \ \text{and} \nonumber \\
    &a_{\Delta t,\text{avg}} = \frac{a_{t} + a_{t-\Delta t}}{2} \nonumber
\end{align}

\subsubsection{Per-Cluster Approaches}
\textit{Per-Cluster} approaches rely on tracking and measuring the changes of individual ``clusters'' (also called ``ganglia'' or more colloquially, ``bubbles'', when referring to gaseous phases). The difference between the per-Cluster and SAC approaches lies in how they use system information from the experimental setup versus the phase information in image sequences. Per-Cluster approaches track the change in volume of individual clusters over a time-interval, using assumptions to correlate the information contained in the clusters' volume change (dissolution) to supplement the unknown aqueous concentration. As a result, the measured values are independent of all other volume changes and position, but reliant on the accuracy of the assumption. This is in contrast to the SAC approach, which uses the advective transport model (Eq. \ref{eq:pat_conc_x}) with the known information about the bulk system (the advective Darcy flux, $q$), to estimate the concentration information not obtained in the $\mu$CT images, but produces results tied to linear position.\par
At the time of this study, \citet{Ruotong_H_2023} and \citet{Lv_2024} are the only other studies that use a per-Cluster approach. Both studies apply Eq. (\ref{eq:mass_trans_def}) to a measurement pool of changes in cluster volume ($\Delta V$) within a time-interval ($\Delta t$), resulting in a list of mass transfer coefficients ($k_{i,j}$) that we will represent with the distribution $\mathbb{K}(k_{i,j}|\ j)$. Notation $i \in [1, 2,..., N]$ represents the sampled volume change; $ N$ is the total number of $\Delta V$ sampled; and $j \in \{1,2,...M\} $ represents the sampled $\Delta t$ with M being the total number of sampled $\Delta t$. However, Eq. (\ref{eq:mass_trans_def}) still requires an assumption about the aqueous solute concentration around each cluster ($C^{gas}_i$) at the time of the event, which is not inherent to $\mu$CT image data; this is where the two studies begin to differ. The simplest assumption is that the bulk solvent concentration is infinitely dilute ($C^{gas}_i \approx 0$) around every cluster, implying the maximum possible dissolution gradient to drive interphase mass transfer, Eq. (\ref{eq:dis_eqn}). 
\begin{equation}
    \label{eq:dis_eqn}
    k_{i,j} = \frac{-\Delta V^{\text{gas}}_{i,j}}{\Delta t_j}\frac{\rho^{gas}}{\Delta Sa_{i,j}} \frac{1}{(C_{\text{sol}} - 0)}
\end{equation}
Here $\frac{-\Delta V^{\text{gas}}_{i,j}}{\Delta t_j}$ is the change in the cluster volume over the observed time-interval (final - initial volume), $\rho^{gas}$ is the density of the gas phase, and $\Delta Sa_{i,j}$ is a time-averaged total surface area of the cluster. In \citet{Lv_2024}, Eq. (\ref{eq:dis_eqn}) is applied to all cluster volume changes, and we will refer to this method as a \textit{Non-Classified} per-Cluster (NPC) and the resulting mass transfer coefficient distribution as ($\mathbb{K}_{NPC}$).\par
\begin{table}[htb!]
    \centering
    \caption{Cluster classification criteria for the CPC approach}
    \begin{tabular}{l p{1.7cm} c}\hline
        Volume change &  Cluster  &  Additional \\
        $V(t_1 \rightarrow t_2$)&Classification&criteria\\ \hline\hline
        $V_1 < V_2 $ & Grown  & $\Delta V _{\%}  > +10\%$\\\hline
        $V_1 > V_2 $ & Partially Dissolved  & $\Delta V _{\%}< -10\%$\\ \hline
        $V_1 \neq 0, V_2 = 0 $& Completely Dissolved & --- \\ \hline
        $V_1 = 0, V_2 \neq 0 $& Snapped Off  & --- \\ \hline
        $\Delta V = 0$ & No change & $ -10\% < \Delta V _{\%} < +10\%$\\ \hline
    \end{tabular}
     \\$\Delta V _{\%} = (V_2 - V_1) / V_1$ 
    \label{tab:class}
\end{table}
Alternatively, \citet{Ruotong_H_2023} restricts the $C^{gas}_i \approx 0$ assumption to a specific classification of volume changes: clusters that completely dissolved in a time-interval. We will refer to this method as a \textit{Classified} per-Cluster approach (CPC), with the resulting distribution denoted by $\mathbb{K}_{CPC}$. The CPC approach requires categorization of each cluster for every time-interval following the change in morphology (volume) that occurred within that time-interval. These volume changes fall into one of several event categories: growth, partially dissolved, completely dissolved, snapped off, and no change (see Table \ref{tab:class}).\par 

Specifically, \citet{Ruotong_H_2023} alters Eq. (\ref{eq:dis_eqn}) to calculate individual mass transfer coefficients for each completely dissolved cluster ($k_{i,j}$ via Eq. \ref{eq:indi_mass_trans}); then estimate a weighted, average mass transfer coefficient ($K_{\text{avg}}$, Eq. \ref{eq:weight_ave}) to reduce error introduced by smaller clusters; and uses the partially dissolved cluster class to back-calculate the aqueous solute concentration ($C_{i,j}$, Eq. \ref{eq:indi_con_grad}) via the estimated $K_{\text{avg}}$.  
\begin{eqnarray}
    k_{i,j} &=& \frac{-\Delta V^{\text{gas}}_{i,j}}{\Delta t_j}\frac{\rho^{\text{gas}}}{Sa^{\text{tot}}_{i,j}\{*\}} \frac{1}{ C_{\text{sol}} - 0} \label{eq:indi_mass_trans}\\
    Sa^{\text{tot}}_{i,j}\{*\} & \approx & \alpha_{seq}\{\frac{1}{2}V^{\text{gas}}_{i,j}\} + \beta_{seq}\\
    K^{seq}_{\text{ave}}& =& \frac{\sum^{M}_{j = 1}\sum^{N}_{i = 1}(k_{i,j} \times Sa^{\text{tot}}_{i,j}\{*\})} {\sum^{M}_{j = 1}\sum^{N}_{i = 1} (Sa^{\text{tot}}_{i,j}\{*\})} \label{eq:weight_ave}\\
    \frac{C_{i,j}}{ C_{\text{sol}}} &=&  1  + \frac{\Delta V^{\text{gas}}_{i,j}}{\Delta t_j } \frac{\rho^{\text{gas}}}{C_{\text{sol}}\ \Delta Sa^{\text{ff}}_{i,j}} \frac{1}{K_{\text{ave}}^{\text{seq}}} \label{eq:indi_con_grad}
\end{eqnarray}
Here, \citet{Ruotong_H_2023} makes several distinctions in Eqs. (\ref{eq:indi_mass_trans}) and \ref{eq:indi_con_grad}: first, $Sa^{\text{tot}}_{i,j}\{*\}$ is the total surface area of a cluster's time-averaged volume change in that interval, while $\Delta Sa^{\text{ff}}_{i,j}$ is the time-averaged fluid-fluid interface. $\alpha_{seq}$ and $\beta_{seq}$ are regression constants obtained by fitting a curve between the total surface area and volume of all completely dissolved clusters in the observed image sequence (see \textit{supplementary materials: CPC Surface Area Calculation}). The justification behind this ``volume-averaged'' surface area is that the evolution of the cluster surface area is not linear in time; rather, it is related to the pore space occupied by the volume, and information about how the surface areas relate to the occupied volumes can be estimated from the existing cluster pool (of clusters that dissolve in an interval). Note that the change in volume over the unobserved time interval is treated as linear (i.e., $\frac{1}{2}V^{\text{gas}}_{i,j} = \text{average}(V^{\text{gas}}_{i,j},0)$ ) as we lack any other information about how the clusters evolve in time. A sensitivity analysis of the CPC approach to the surface areas (total versus fluid-fluid) used is given in the supplementary material of \citet{Ruotong_H_2023}; but in short, this methodology was chosen as it underestimated the system-averaged mass transfer coefficient the least (this will be discussed further in the concentration estimates section). Lastly, it is important to note that the fitting is recalculated for each image sequence after data post-processing, as the evolution of the surface area with the cluster volume will depend on the mass transfer conditions in the system.\par
\citet{Ruotong_H_2023} restrict $C^{gas}_i \approx 0$ to completely dissolved clusters because, for an unsteady-state system, there should exist a region in which the saturated water previously occupying that space has been fully displaced by the fresh injected water. This region effectively lags behind the initial pulse of the injected solvent, ``proceeding'' forward as the pore spaces reach an equilibrium with the fresh advecting solvent governed by unsteady-state multiphase fluid dynamics and not measured from $\mu$CT image data. We will refer to the forward-most edge of the advancing fresh water region (in the direction of flow) as the ``dissolution front''. \citet{Ruotong_H_2023} assumes that complete dissolution occurs only in this region behind the dissolution front, as it would be the only region where local concentration is high enough to drive complete dissolution of a fairly insoluble gas (see Table \ref{tab:gas_table}). The SAC approach utilizes a concept similar to this dissolution front in the advective model (Eq. \ref{eq:pat_conc_x}) to simulate the translation of the injected solvent through the system, but the reduction in dimensionality treats the displacement as a stable, uniform front \textit{a priori}. Rather than approximating where the dissolution front may be to classify clusters, the CPC approach uses individual and aggregate information within cluster sub-populations (i.e., how clusters classified as completely dissolved are distributed spatially) to determine its location.\par

\subsubsection{The difference between NPC and CPC approaches}
The distinction between NPC and CPC approaches depends on when in the workflow the two are compared, at the initial image-processing stage or during the subsequent calculation stage. As discussed, each approach has a specific minimum image-processing workflow; i.e., the NPC approach can be applied to the same clusters identified by the CPC image-processing workflow. Due to issues with the image data (see \nameref{sec:methods}: \nameref{subsec:im_pro_curr}), full, native implementations of the NPC and CPC approach image-processing workflows would yield different measurement pools (the cluster volume changes measured in a time interval). In order to compare how the underlying assumptions differentiating each approach operate on the measurement pool, rather than the effects of the image processing workflow on the estimates, both approaches must be applied to the same measurement pool. The CPC approach's stricter minimum requirements (3-phase segmentation) require using CPC-isolated clusters as the common measurement pool for both approaches. However, in doing so, the current study loses the benefit of the NPC's lower minimum image-processing workflow.\par   
Having both approaches sample the same measurement pool then relates the results in sample distributions, as  $\mathbb{K}_{CPC} \subseteq \mathbb{K}_{NPC}$; and the core difference between the approaches lies in the implications behind how each approach samples the measurement pool. Given the understanding that some of the measured morphology changes are not actually representative of \textit{interphase} mass transfer (i.e., clusters moving and merging): 
\begin{itemize}
    \item The CPC approach, by restricting sampled volume changes to a subset of events that approximate a specific system condition, assumes that using all other classifications introduces too much uncertainty and error to the estimated property (i.e., the mass transfer coefficient and local aqueous concentration). 
    \item The NPC approach, by not restricting the events sampled, assumes that a sufficient number of actual interphase mass transfer events are present in the measurement pool to outweigh the uncertainty introduced by displacement events, via the law of large numbers.
\end{itemize}
Going forward, the current study will expand on the implications of these assumptions in the context of relevant mass-transfer properties and the results of the SAC approach. As the CPC approach requires further post-processing filtration of the time-intervals used (see section \nameref{sec:methods}: \nameref{subsec:mobility}), for the comparison of individual intervals, the approaches will be compared on the filtered common intervals. However, for the final evaluation of the system mass transfer coefficient, the NPC will operate natively on the unfiltered CPC identified data pool.\par

\section{Materials}%
\label{sec:materials}%
\subsection{Original Experiment}
The image dataset analyzed in this study is from a larger mass-transfer estimation study by \citet{Patmonoaji_2023}. The datasets are available from the original authors upon request. In the original study, the authors evaluated the time-lapsed dissolution of various gases (hydrogen, nitrogen, methane, and ethane) via X-ray $\mu$CT. Prior to each experiment, the analyte gases were trapped in a water-wet, granular, plastic packing, and multiple high-flow-rate injections were applied from both the inlet and the outlet. Trapping was considered complete when the gaseous analyte was no longer observed in the aqueous effluent. Trapping before freshwater injection ensures that observed changes or movement in the gas phase result from interphase mass transfer rather than physical displacement of gas clusters by the injected freshwater. Imaging began after trapping when freshwater reached the packing inlet.\par
The goal of each experiment was to characterize the interphase mass transfer of the various analyte gases across a range of advective flow rates of injected freshwater. The data acquired from each experiment consist of an initial ``dry'' scan (packing and gas phase only, before trapping) and a time-lapsed sequence of ``wet'' scans (packing, gas, and water, during freshwater injection after trapping). At the end of each experiment, the packing was purged and dried. Each analyte gas was evaluated over four freshwater injection rates (i.e., four experiments) to make an experiment set. Each analyte gas set was evaluated separately, and there was no interaction or contamination between gases.\par
\subsection{Gaseous Hydrogen Data Set}
For the current study, we elected to evaluate the H$_2$ data set from \citet{Patmonoaji_2023}, spanning the following fresh water injection rates: 0.10, 0.25, 0.50, and 1.0 mL/min; parameters relevant to each experiment are listed in Tables \ref{tab:seq_table}, \ref{tab:gas_table}, and \ref{tab:packing_table}. The timing of each scan in their respective experiment and image sequence is presented in Table \ref{tab:scans_times}. The formulations for the Reynolds number (Re), Capillary number (Ca), and Schmidt Number (Sc) are given in Eqs. (\ref{eq:Re}), (\ref{eq:Ca}), and (\ref{eq:Sc}) respectively. 
\begin{table}[ht]
        \centering
        \caption{Evaluated H$_2$ Experiments}%
        \begin{tabular}[width = \columnwidth]{l  c c c c}\hline \hline
            Sequence  & Flow Rate & Re    & Ca               & Sc     \\
                      & [mL/min]  & Eq. (\ref{eq:Re}) 
                                  & Eq. (\ref{eq:Ca})            
                                  & Eq. (\ref{eq:Sc})     \\\hline
            Seq. 0.10 & 0.10      & 0.016 & 5.16 x $10^{-7}$ &  223   \\
            Seq. 0.25 & 0.25      & 0.040 & 1.29 x $10^{-6}$ &  223   \\
            Seq. 0.50 & 0.50      & 0.080 & 2.58 x $10^{-6}$ &  223   \\ 
            Seq. 1.00 & 1.00      & 0.160 & 5.16 x $10^{-6}$ &  223   \\\hline
        \end{tabular}%
        \label{tab:seq_table}
\end{table}%
\begin{table}[ht]
        \centering
        \caption{H$_2$ gas Properties}%
        \begin{tabular}[width = \columnwidth]{l c}\hline \hline
            Pressure [bar] & 1.01 \\\hline
            Temperature [\degree C] & 20 \\\hline
            Density [g/m$_{\text{gas}}^3$] & 90 \\\hline
            Max Solubility [g/m$_{\text{water}}^3$] & 1.5 \\\hline
            Interfacial Tension [mN/m] & 72.9$^*$ \\\hline
        \end{tabular}\\%
        $^*$ \citet{Chow_2020}
        \label{tab:gas_table}
     \vspace{0.5cm}
            \centering
        \caption{Granular Plastic Packing Properties}%
        \begin{tabular}[width = \columnwidth]{l c}\hline \hline
            Mean Particle Size - $d_{\text{50}}$ [$\mu$m] & 300 \\\hline
            Porosity [\%] & 45 $\pm$ 5 \\\hline
            Permeability [Darcy] & 132 \\\hline
        \end{tabular}%
        
        \label{tab:packing_table}
\end{table}%
\begin{align} 
    Ca &= \frac{\text{Viscous fluid forces}}{\text{Capillary forces}} = \frac{\mu _{\text{w}} \nu}{\sigma} \label{eq:Ca}\\
    Sc &= \frac{\text{Viscous diffusion}}{\text{Molecular diffusion}} = \frac{\mu _w}{\rho_w D_{g,w}} \label{eq:Sc}
\end{align}
Here $\rho_{\text{w}}$, $\mu _{\text{w}}$, and $\nu$ are the density, viscosity, and interstitial velocity of the injected water, respectively and $\sigma$ is the interfacial tension between water and H$_2$ at the temperature and pressure indicated in Table \ref{tab:gas_table}.\par

\subsection{Image Acquisition}
\label{subsec:image_acq}
Each experiment was imaged via $\mu$CT (ScanXmate-RB090SS from Comscantechno Co.) under constant settings (116 $\mu$A, 65 kV, and 7.5 W), to ensure equal brightness and contrast for each image. Each scan took 90 s to complete a full scan rotation, during which the system captured 1000 images. After reconstruction, each 3D image consists of a 992 $\times$ 992 $\times$ 992 voxel cube, with a voxel size of 16.472 $\mu m$. The total run duration for each experiment sequence varies, as does the time between scans; see Table \ref{tab:scans_times}.\par
\newcommand{\wide}{0.5cm}
\begin{table*}[htb!]
    \centering
    \caption{Table of Scan times within a sequence}
    \resizebox{\textwidth}{!}{%
    \begin{tabular}{l|p{\wide} p{\wide} p{\wide} p{\wide} p{\wide} p{\wide} p{\wide} p{\wide} p{\wide} p{\wide} p{\wide} p{\wide} p{\wide} p{\wide} p{\wide} p{\wide} p{\wide} p{\wide} p{\wide} p{\wide} p{\wide} } \hline
         scan \# & 1 & 2 & 3 & 4 & 5 & 6 & 7 & 8 & 9 & 10 & 11 & 12 & 13 & 14 & 15 & 16 & 17 & 18 & 19  \\ \hline \hline
        Sequence & \multicolumn{19}{c}{Time of scan in the sequence }\\
        \text{[ml/min]} & \multicolumn{19}{c}{\text{[minutes]}}\\\hline 
        0.10& 00 & 05 & 10 & 15 & 20 & 25& 30 & 40 & 50 & 60 & 80 & 100 & 120 & 150 & 180 & 210 & 240 & 270 & 300\\
         0.25& 00 & 05 & 10 & 15 & 20 & 25& 30 & 40 & 50 & 60 & 70 & 80 & 90 & 100 & 117 & - & - & - & - \\
         0.50& 00 & 05 & 10 & 15 & 20 & 25& 30 & 35 & 40 & 45 & 50 & 55 & 60 & - & - & - & - & - & - \\
         1.00& 00 & 04 & 08 & 12 & 16 & 20& 24 & 28 & 32 & 36 & 40 & - & - & - & - & - & - & -  & - \\ \hline \hline

    \end{tabular}}
      Note: the scan duration and image settings of each scan are constant and described in the \textit{\nameref{subsec:image_acq}} section.

    \label{tab:scans_times}
\end{table*}
\section{Methods}
\label{sec:methods}
\begin{figure}[htb!]
    \centering
    \includegraphics[width=\linewidth, height = 10cm]{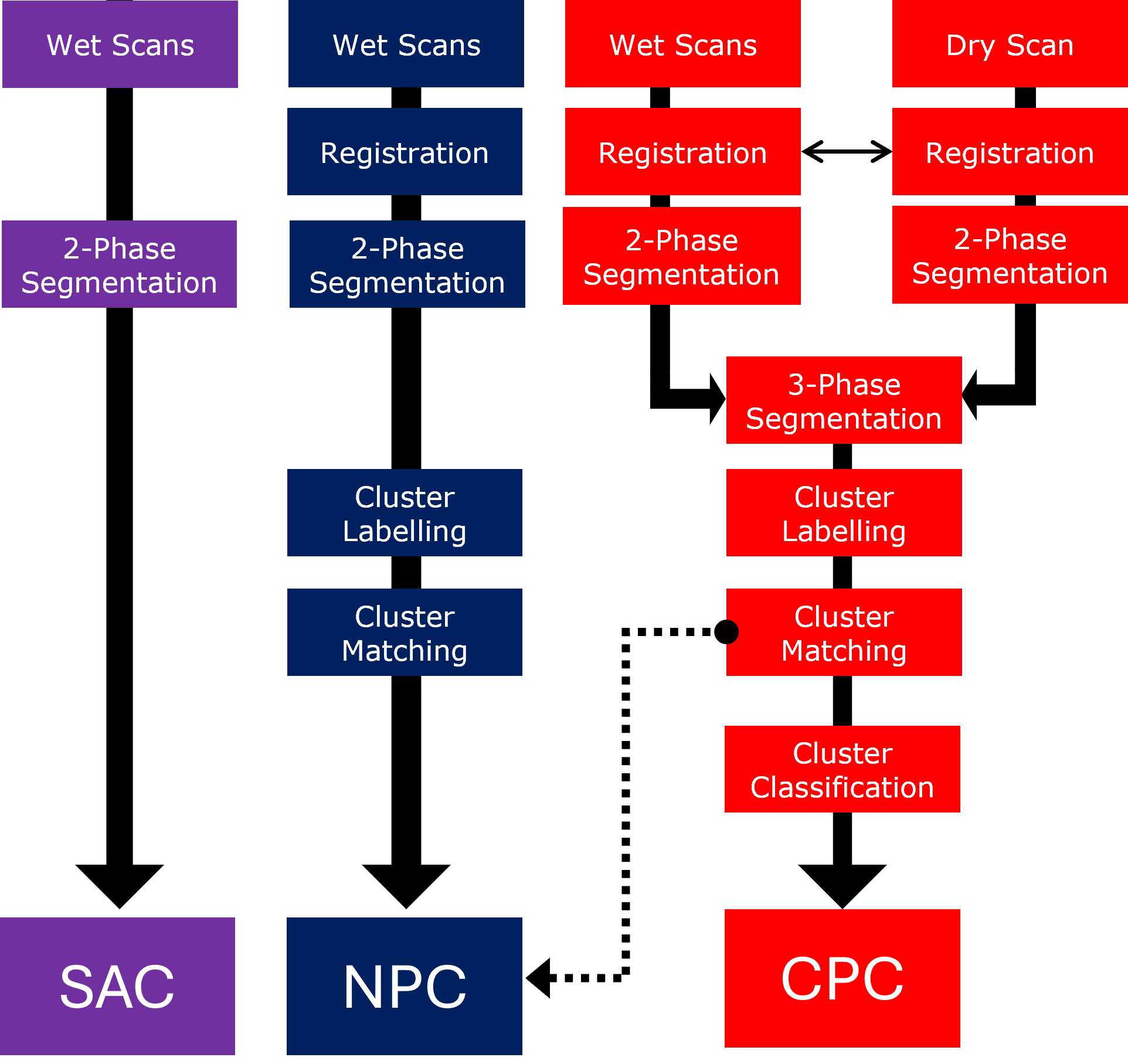}
    \caption{Cartoon of the image-processing workflow outlined for each analytical approach, juxtaposing commonalities and differences. E.g., both per-Cluster approaches share initial (gas-else) 2-phase segmentation of the wet scan, and differ from that point onward. The dashed line from the CPC to the NPC workflow indicates that the NPC approach can be applied to data from the CPC workflow (as we did in the current study).}
    \label{fig:img_pro_pipes}
\end{figure}

The analytical approaches evaluated in the current study have a required minimum level of image processing; see Figure \ref{fig:img_pro_pipes}. Although other important image properties (resolution, contrast, signal-to-noise ratio) and processing steps (smoothing, de-noising, and segmentation) will affect image-derived quantities \citep{Schlüter_2014, Wildenschild2013_betti, Huang_2021}, the comparative analysis herein focuses only on the steps relevant to the mass-transfer estimation approaches themselves.\par
At a minimum, the SAC approach \citep{Pat_Suek_2017} requires image processing up to 2-phase segmentation of the wet scans, as it observes the aggregate change in gas saturation across each cross-section over time. The NPC approach \citep{Lv_2024} requires further processing of the 2-phase wet scan to isolate and ``track'' changes in each cluster over time. The CPC approach \citep{Ruotong_H_2023} requires the most image processing among the three approaches to enable 3-phase segmentation to separate the gas-water interface from the gas-grain interface, in addition to tracking and classifying the clusters.\par
The current study used the cluster pool yielded by the CPC image-processing workflow (the converging, red branch in Figure \ref{fig:img_pro_pipes}) for the NPC approach rather than running two separate workflows to isolate image-processing differences from the final results. Using the same pool of clusters for both per-Cluster approaches enables comparison of the effects of cluster classification on the resulting estimates, while under-utilizing the simpler image-processing workflow of the NPC approach.\par
\subsection{Image Processing from Patmonoaji et al., 2023}
In the original study \citep{Patmonoaji_2023}, all scans were cropped to the central 600 $\times$ 600 voxels in the XY plane (the area just inside the coreholder), and the central 850 voxels along the Z axis (to remove edge effects and align $x^{*}=0$ with the injection inlet). Median filtering and anisotropic diffusion were used to increase the grey-scale contrast between phases. The original study did not rely on a three-phase segmentation, instead prioritizing the isolation of the gas phase. Because the SAC approach evaluates the phase saturation of each cross-section over time rather than the volume of the clusters, the wet scans do not need to be registered (given that all images are aligned in the axial direction).\par
\subsection{Image Processing for the Current Study}
\label{subsec:im_pro_curr}
All image processing for the current study was performed using \textit{Webmango}, a web-based tool for interfacing with \textit{Mango}. \textit{Mango} is 3-D image-processing software developed by the Australian National University and uses the high-performance systems at the NCI (National Computational Infrastructure). The current study follows the stricter image-processing workflow developed by \citet{Ruotong_H_2023}, which isolates and matches the clusters over each time-interval and separates the different phase interfaces of each cluster (i.e., the gas-water and the gas-grain interfaces). As discussed, the current study applied the clusters identified by the CPC workflow to the NPC estimates as well, up to the cluster classification.\par
All images were cropped to the contents just outside the coreholder, a 630 $\times$ 630 pixel cylindrical area, 
to avoid possible data loss from the registration process. The coreholder is later masked out prior to the segmentation process. Importantly, the images needed to be cropped in the axial direction to the central 600 pixels in order to remove the extreme vertical subsections in which grain movement and edge artifacts were observed. Cropping in the axial direction limits the size of the spatial window in which we observe the solvent pulse, though this will only alter characterization if the cropped region contained the entirety of the pulse for the one or more time-intervals. To check if the pulse has been cropped out of the initial time intervals, we estimate the time required for the slowest injection rate (0.10 ml/min) to propagate through the cropped region using the mean interstitial velocity. The calculations in Table \ref{tab:pulse_timing} show that the slowest injection rate should propagate through the cropped section in 1.11 minutes; with a 5-minute time between scans, the front propagates $\approx 4\times$ faster than the system is sampled (imaged). This estimate indicates that we should still be able to observe mass transfer events in the first time-interval of each sequence, despite cropping the window. \par
\begin{table}[ht]
        \centering
        \caption{Solvent Propagation Velocity}%
        \begin{tabular}[width = \columnwidth]{l  c c}\hline \hline
        Injection rate    & mL/min      & 0.10               \\
        Cross section     & $\mu m$m    & $9.9\times10^{+3}$ \\
        Porosity          & \%          & 45                 \\
        Porous Area       & $\mu m^2$   & 34.5               \\
        interstitial flux & $\mu m$/min & $2.8\times10^{+3}$ \\
        cropped inlet     & $\mu m$     & $3.2\times10^{+3}$ \\
        travel time       & min         & 1.11               \\
        shortest scan interval     & min         & 5.0       \\ \hline    
        \end{tabular}%
        \label{tab:pulse_timing}
\end{table}%
However, because clusters still exist in the cropped regions, the potential measurement pool for the per-Cluster approaches is smaller. A full native NPC image-processing workflow would not inherently require the same cropping because it does not need 3-phase segmentation, and thus would produce a different measurement pool from this data set than the image processing used in the current study. Because both per-cluster approaches aim to capture a pulse moving in space and time through its interaction with the trapped clusters, as long as the sampling window remains sufficient to capture the pulse, the reduced measurement pools should not affect the outcome.\par
In order to obtain the fluid-fluid interfaces, as outlined by \citet{Ruotong_H_2023}, all scans within an image sequence (i.e., Seq. 0.10, 0.25, 0.50, and 1.00) must be aligned to the same orientation (registered) to voxel-level precision for successful 3-phase segmentation. Registration was performed by seeding matching high-attenuation grain features that persist across the dry and wet scans, into an optimization algorithm that minimizes the negative correlation ratio between grayscale tomograms, introducing perturbations with downscaled similarity transforms \citep{Latham_2008}. Once aligned, each wet scan was segmented into three phases by logical comparison with the grain phases in the dry scan; this process is described in greater detail in the \textit{Supplementary Materials: Segmentation Criteria} section.\par
\begin{figure}[!ht]
    \centering
    \includegraphics[width=0.75\linewidth]{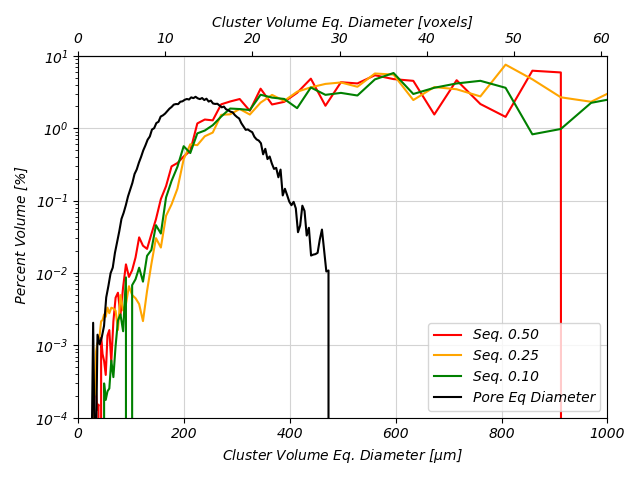}
    \caption{Volume-weighted size distribution using the equivalent spherical diameter derived from the volume of every identified cluster in each Sequence. Seq. 0.50 = red curve, Seq. 0.25 = yellow curve, Seq. 0.10 = green; the pore sizes are the black line.}
    \label{fig:cluster_size_dist}
\end{figure}

Once segmented into three phases, the H$_2$ phase was evaluated and separated into its individual component clusters, each assigned a unique identifier distinct from the original phase labels. Using a pore-network reconstruction of the segmented dry scan, we generate a pore-size distribution to help remove erroneous clusters. Figure \ref{fig:cluster_size_dist} shows the volume-weighted distribution of pore and cluster volumes, simplified into an equivalent spherical diameter. Using the distribution, we removed clusters smaller than the smallest pore size, 50.40 $\mu m$ (15 voxels in volume, $\approx$ 3 voxels in equivalent spherical diameter). Each cluster is evaluated for center of geometry, volume, total surface area, and cluster fluid-fluid interfacial area (or wetted surface area).\par
The cluster-matching workflow operates on pairs of scans in sequential order. The time for each scan is indicated in Table \ref{tab:scans_times}. For example, in Sequence 1.00, the scan at 00 minutes is evaluated pairwise with the following scans: 04 with 08, 08 with 16, and so forth. For ease of discussion, these pairs are referred to by their mid-interval time (Table \ref{tab:interval_times}); e.g., the pairing between scans 12 and 16 is indicated as ``14 minutes''. Cluster classification nomenclature is given in Table \ref{tab:class} for the cluster classification nomenclature; a more detailed description is given in the supplementary material (see section \textit{Cluster Matching And Classification}).\par
\newcommand{\waffle}{0.5cm}
\begin{table*}
    \centering
    \caption{Table of time-intervals for each Sequence} 
    \resizebox{\textwidth}{!}{%
    \begin{tabular}{l|p{\waffle} p{\waffle} p{\waffle} p{\waffle} p{\waffle} p{\waffle} p{\waffle} p{\waffle} p{\waffle} p{\waffle} p{\waffle} p{\waffle} p{\waffle} p{\waffle} p{\waffle} p{\waffle} p{\waffle} p{\waffle} } \hline

        Seq. & 1   & 2   & 3    & 4    & 5    & 6    & 7    & 8    & 9    & 10   & 11   & 12   & 13  & 14    & 15  & 16  & 17  & 18  \\ 
        \text{[ml/min]}  & \multicolumn{18}{c}{Mid-interval time [minutes]}\\ \hline \hline 
        0.10 & 2.5 & 7.5 & 12.5 & 17.5 & 22.5 & 27.5 & 35   & 45   & 55   & 70   & 90   & 110  & 135 & 165   & 195 & 225 & 255 & 285 \\
        0.25 & 2.5 & 7.5 & 12.5 & 17.5 & 22.5 & 27.5 & 35   & 45   & 55   & 65   & 75   & 85   & 95  & 108.5 & -   & -   & -   & -  \\
        0.50 & 2.5 & 7.5 & 12.5 & 17.5 & 22.5 & 27.5 & 32.5 & 37.5 & 42.5 & 47.5 & 52.5 & 57.5 & -   & -     & -   & -   & -   & -  \\
        1.00 & 2.0 & 6.0 & 10.0 & 14.0 & 18.0 & 22.0 & 26.0 & 30.0 & 34.0 & 38.0 & -    & -    & -   & -     & -   & -   & -   & -  \\ \hline \hline     
    \end{tabular}}
    
    \label{tab:interval_times}
\end{table*}
\subsection{Key Calculation Principles}
As stated, the current study follows the procedure outlined in \citet{Ruotong_H_2023}; however, we reiterate several assumptions important to the CPC approach and the discussion moving forward.
\begin{itemize}
    \item It is assumed that clusters change in volume due to dissolution, and said volume change is linear over the observed time-interval.
    \item When a cluster is observed to have completely dissolved in a time-interval, it is assumed that all of the cluster's surface is in contact with the solvent, over a subset of that time-interval. 
    \item The error from approximating continuous geometries from discrete voxels scales inversely with cluster size; thus, smaller clusters introduce more measurement error than larger clusters, and $\mathbb{K}_{CPC}$ will need to be weighted to favor measurements from larger clusters. 
\end{itemize}
The only deviations from the methods of \citet{Ruotong_H_2023} are,
\begin{itemize}
    \item The current study uses a linear best-fit relation function to relate a cluster's total surface area to its volume, rather than a power-law best-fit relation
    \item The current study evaluates $K_\text{ave}$ over each time-interval (\textbf{$K_{\text{ave}}^{\text{int}}$}) in addition to evaluating $K_\text{ave}$ over the entire sequence (\textbf{$K_{\text{ave}}^{\text{seq}}$}).
    \item The current study uses different criteria to threshold time-intervals with significant levels of cluster re-mobilization from analysis; see \textit{Results} section.
\end{itemize}
The NPC approach, as outlined by \citet{Lv_2024}, uses the time-averaged total surface area of the cluster over the entire time step and, correspondingly, the total change in volume over that time step. Additionally, \citet{Lv_2024} originally uses an unweighted arithmetic average to calculate $K_{\text{mean}}^{\text{seq}}$; to minimize further change from the native NPC implementation, we also use the unweighted average when relevant.. Lastly, we back-calculate the concentrations around each cluster using the NPC ($K_{\text{mean}}^{\text{seq}}$) to evaluate the total dilute concentration under the assumption $C^{gas}_{i} \approx 0$, though this calculation was not performed in \citet{Lv_2024}, as it goes against the core assumption.\par
\subsection{Cluster mobilization and interval filtering}
\label{subsec:mobility}
A key finding of \citet{Ruotong_H_2023} was the observation of significant cluster growth during solvent injection and post-trapping, indicative of cluster remobilization; similar observations of cluster remobilization were noted in \citet{Lv_2024}.  The current study also observed considerable cluster growth throughout several time-intervals in each sequence; these cluster growth events manifested as statistical outliers in $\mathbb{K}_{CPC}$. \citet{Ruotong_H_2023} hypothesizes that this re-mobilization is, in part, due to dissolution shrinking the cluster, thereby increasing its capillary pressure enough to overcome the local trapping threshold. Mobilized clusters that merge with a more stable cluster at lower capillary pressure are then what drive the observed growth. On a system scale, sufficient levels of dissolution-induced mobilization should be indicated by a gain in cluster volume proportional to the volume lost by clusters completely dissolving. Not accounting for cluster remobilization and assuming that all volume loss is due to mass transfer will result in an overestimation of the mass transfer present.\par
\begin{figure}[h]
    \begin{subfigure}{0.5\columnwidth}
        \includegraphics[width=\linewidth]{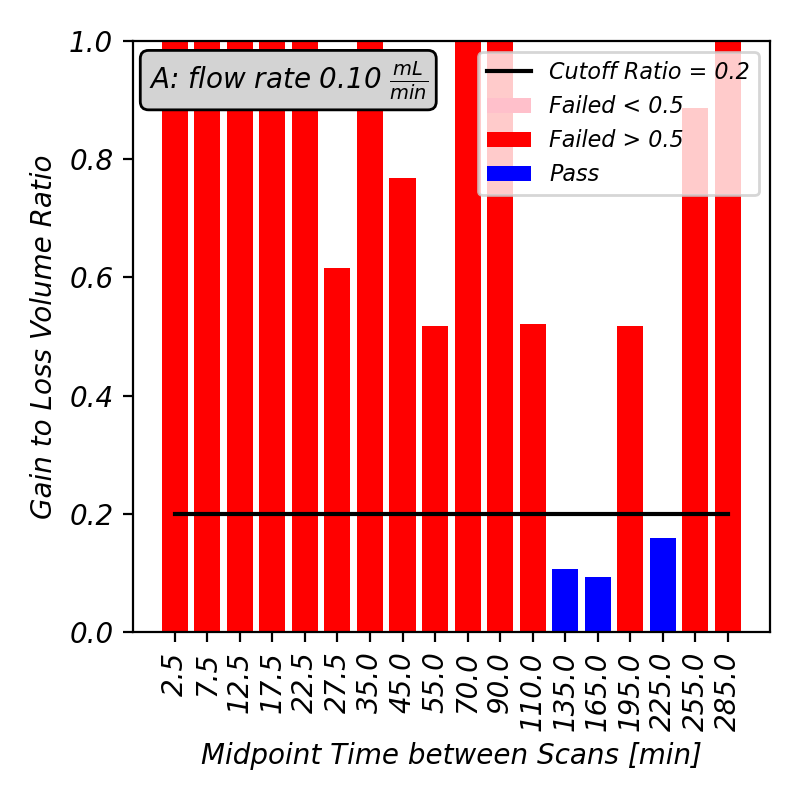}
    \end{subfigure}%
    \begin{subfigure}{0.5\columnwidth}
       \includegraphics[width=\linewidth]{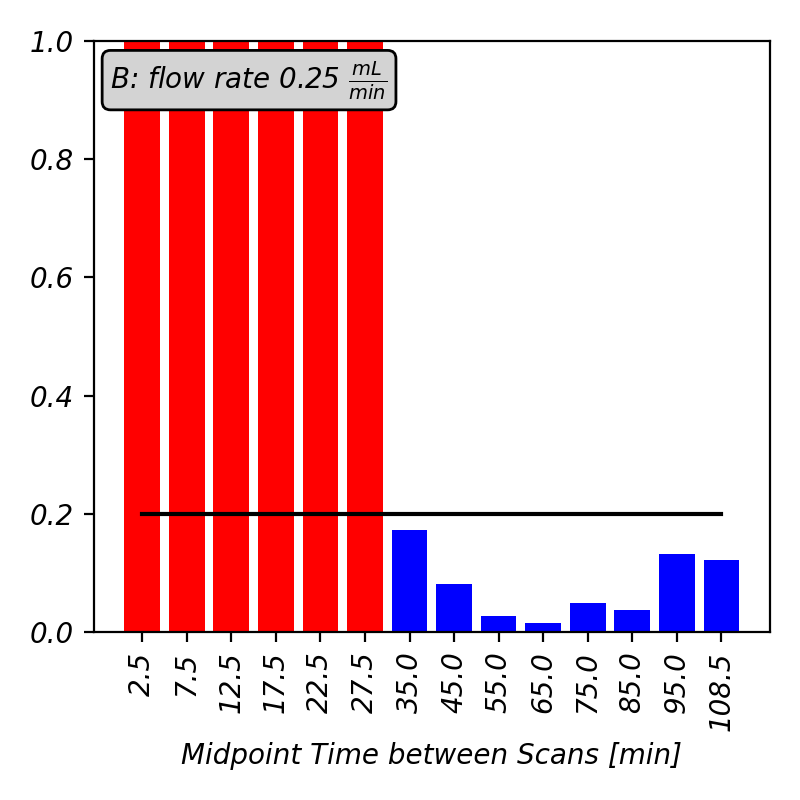}
    \end{subfigure}
    \begin{subfigure}{0.5\columnwidth}
        \includegraphics[width=\linewidth]{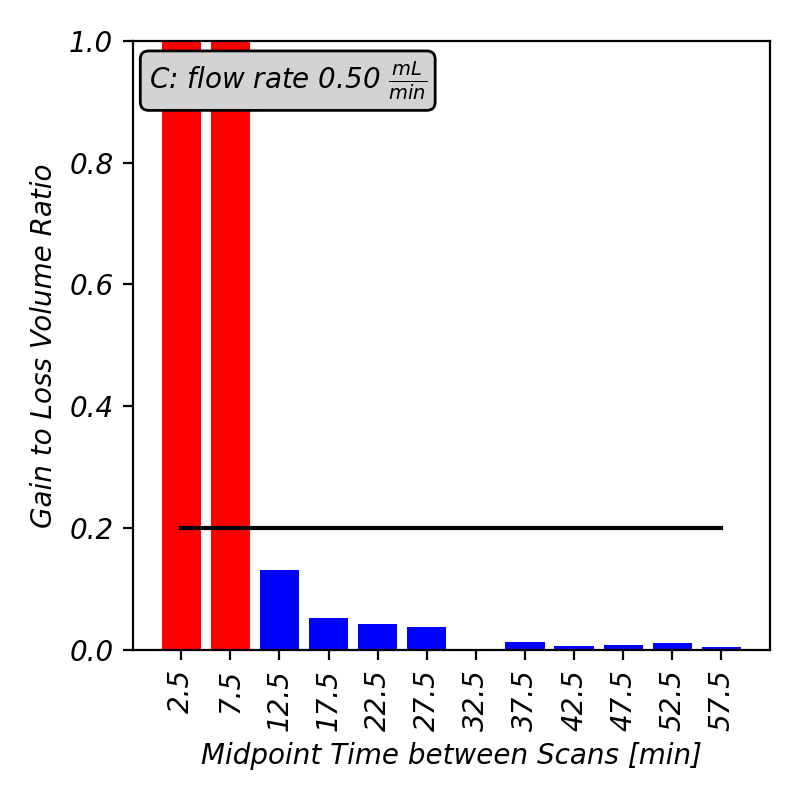}
    \end{subfigure}%
    \begin{subfigure}{0.5\columnwidth}
        \includegraphics[width=\linewidth]{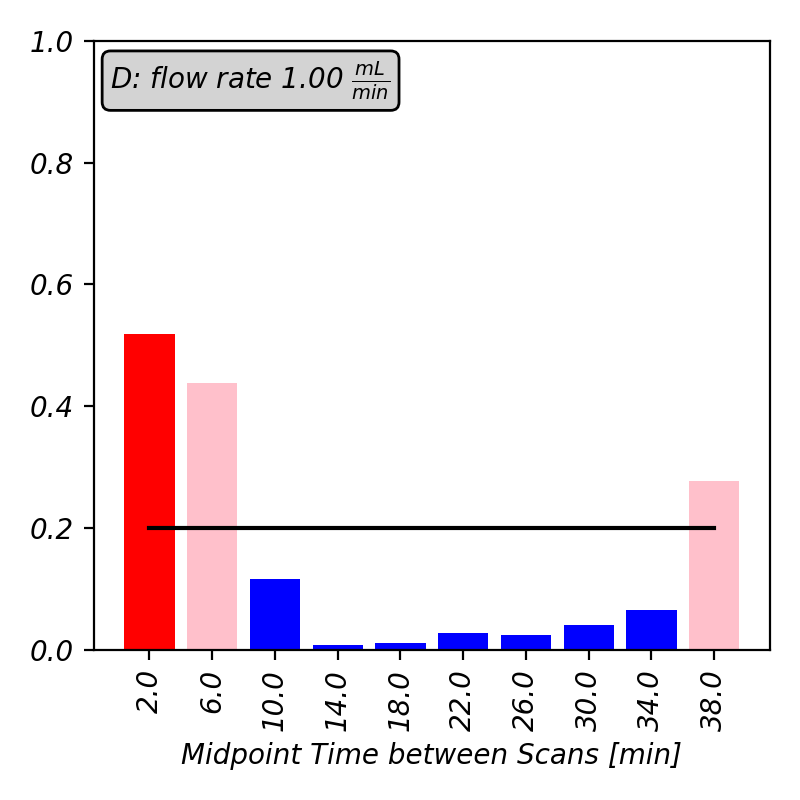}
    \end{subfigure}
     \caption{Bar charts of the $\frac{\Delta \text{volume gained}} {|\Delta \text{volume lost}|}$ in each interval. $\frac{\Delta \text{volume gained}} {|\Delta \text{volume lost}|} \le 0.2$ cutoff are blue, interval values $ 0.2 \le \frac{\Delta \text{volume gained}} {|\Delta \text{volume lost}|} \le 0.5 $ are pink, and interval values $\frac{\Delta \text{volume gained}} {|\Delta \text{volume lost}|} \ge 0.5 $ are in red. The graph is truncated at 1; some values extend to orders of magnitude beyond 1.}
    \label{fig:per_chng_bars}
\end{figure}
To avoid overestimating the mass transferred, \citet{Ruotong_H_2023} removed time-intervals with significant levels of cluster mobilization from the final estimates; however, there is no established method or threshold for classifying acceptable versus unacceptable levels of suspected cluster mobilization. For the current study, we chose a ratio $\frac{\Delta \text{volume gained}} { |\Delta \text{volume lost}|}$ as a threshold to flag time-intervals with flagged high volume gain relative to volume lost.\par
When $\frac{\Delta \text{volume gained}} { |\Delta \text{volume lost}|} = 1$, the volume gained equals the volume lost. Thus, the probability that clusters assumed to have been dissolved were actually remobilized is high. In contrast, low volume-change ratios ($<< 1$) indicate time-intervals with a higher probability of being dissolution-dominant. To increase confidence that the measured clusters truly represent dissolution, we use this ratio to exclude flagged time-intervals from the final estimate, sacrificing the size of the measurement pools to minimize bias introduced into the measurement pool by these displacement events. However, reducing the size of the measurement pool does come with the inherent cost that the final estimate is more sensitive to measurement error and outlier events in the remaining time-intervals.\par
Figure \ref{fig:per_chng_bars} shows the intervals where $\frac{\Delta \text{volume gained}} {|\Delta \text{volume lost}|} \ge 0.2$. Note that of the 26 time-intervals flagged, 17 intervals exceed a ratio of 1.0, which means that the majority of the flagged intervals are not affected by adjusting the threshold to < 1.0. Thresholds greater than 1.0 are likely the result of clusters mobilizing into the field of view (a result of subsetting in the axial direction). In Figure \ref{fig:con_front}(a), we see that the most cluster growth occurs downstream of the direction of flow, close to the top of the packed bed, where a large bubble accumulates and disturbs the packed media (this region was cropped out to avoid complications with 3-phase segmentation). In terms of the elapsed time, the majority of excluded intervals occur early on in each sequence, although Sequences 0.10 and 1.00 do show time-intervals flagged at the end of the experiment. There appears to be a trend in which the mobilization duration in a sequence increases as the injection rate decreases.\par
As we advance with the analysis, we exclude all time-intervals with a volume change ratio ( $\frac{\Delta \text{volume gained}}{|\Delta \text{volume lost}|} \ge 0.2$) to target dissolution-dominant data, rather than remobilization. Using a threshold value of 0.2 allowed for the removal of flagged time-intervals, while ensuring sufficient data remained for subsequent analyses to be representative. The amount of data retained after this mobilization filtering is given in Tables \ref{tab:results_CPC} and \ref{tab:results_NPC} (for the CPC and NPC approaches, respectively), along with the subsequent change in the final system mass transfer coefficient ($K_{\text{ave}}^{\text{seq}}$). Mobilization filtering is only required by the CPC approach; however, to compare the behavior of each approach on each time-interval we compare only the common post-filtration time-intervals. Otherwise, the final mass transfer estimates for the SAC and NPC approaches use the full, unfiltered data set to remain closer to the native implementations. A sensitivity analysis of the per-Cluster approach results as a function of the mobilization threshold is provided in the Supplementary Materials: \textit{Mobilization Filtering Sensitivity} section.\par

\section{Results and Discussion}
\label{sec:results}
\subsection{Mass Transfer Distributions}
\label{subsec:mass_trans_dist}
In an ideal case, under uniform interphase transfer conditions, the population of true mass transfer coefficients $k^{*}_{i,j}$ in a system should approach a singular value. In reality, the population would be expected to resemble a normal distribution, $\mathbb{N}(k^{*}_{i,j})$, due to independent fluctuations in local mass-transfer conditions or small-scale fluctuations inherent to the experimental setup. To approximate $\mathbb{N}(k^{*}_{i,j})$, we would need to be able to measure the interphase mass transfer in the system directly; however, in principle, X-ray $\mu$CT does not directly measure $k^{*}_{i,j}$. Rather, what is measured from the time-lapsed X-ray $\mu$CT sequences are discrete volume changes of connected voxel clusters, identified (segmented) to be the gas phase. These measurements carry inherent errors due to the constrained spatial and temporal resolution of $\mu$CT, which manifest in the calculation of continuous properties from a discrete domain (i.e., cluster morphology and dissolution rate). Through the explicit design of the imaged dissolution experiment, the measured volume changes are highly likely to represent actual interphase mass transfer events, each with a corresponding mass-transfer coefficient ($k_{i,j}$). However, since not all observed volume changes in the measurement pool are interphase mass transfer (e.g., cluster remobilization), any randomly selected measurement event has an unknown probability of sampling a ``false'' population distribution as well as $\mathbb{N}(k^{*}_{i,j})$. The framework of the analytical approaches determines how the measurement pool ($\mathbb{K}(k_{i,j})$) is sampled in order to approximate $\mathbb{N}(k^{*}_{i,j})$.\par
\begin{figure*} [htb!]
    \begin{subfigure}{0.250\linewidth}
        \includegraphics[width=\linewidth, height = 4.0cm]{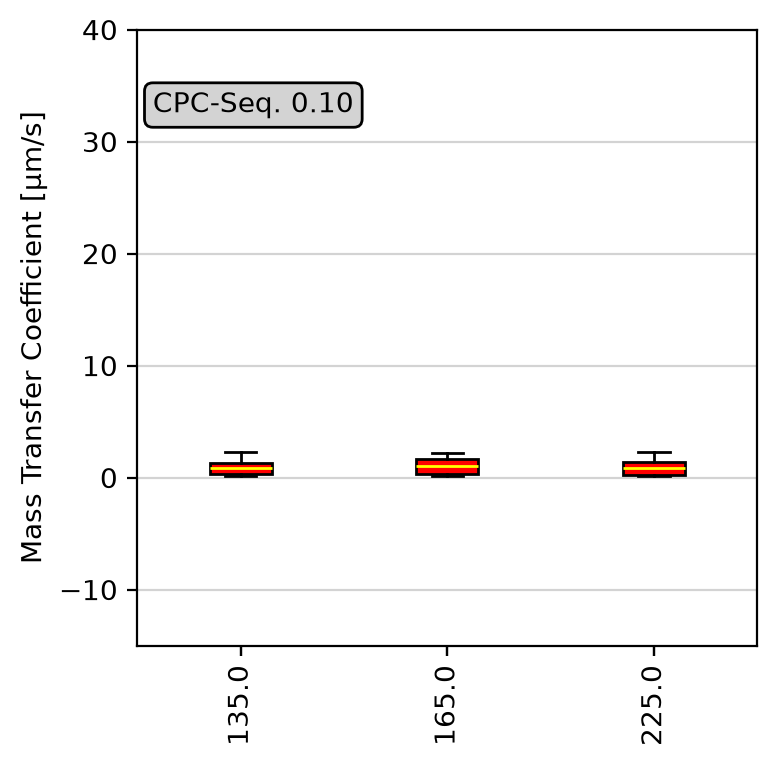}
    \end{subfigure}%
    \begin{subfigure}{0.25\linewidth}
        \includegraphics[width=0.96\linewidth, height = 4.0cm]{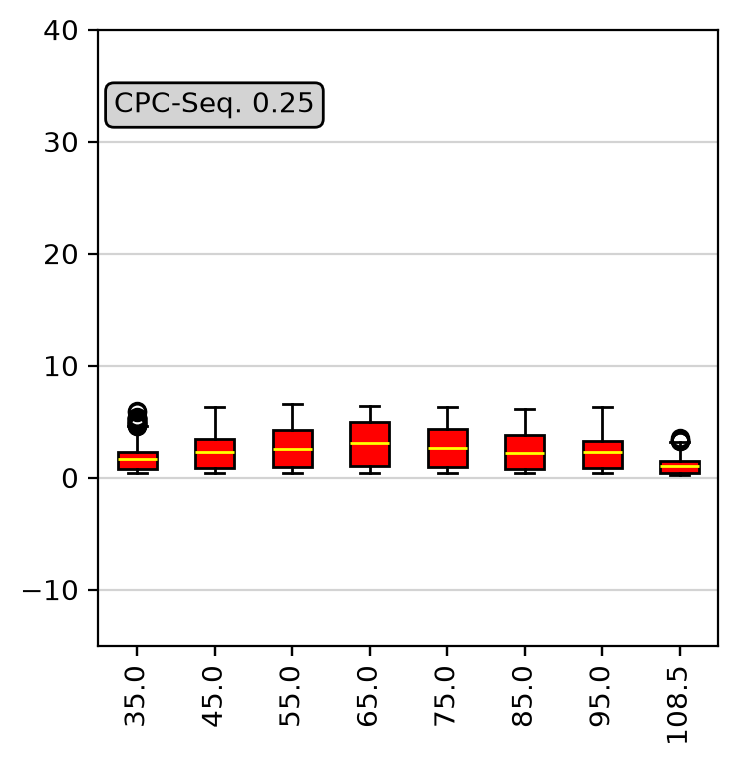}
    \end{subfigure}%
    \begin{subfigure}{0.25\linewidth}
        \includegraphics[width=\linewidth, height = 4.0cm]{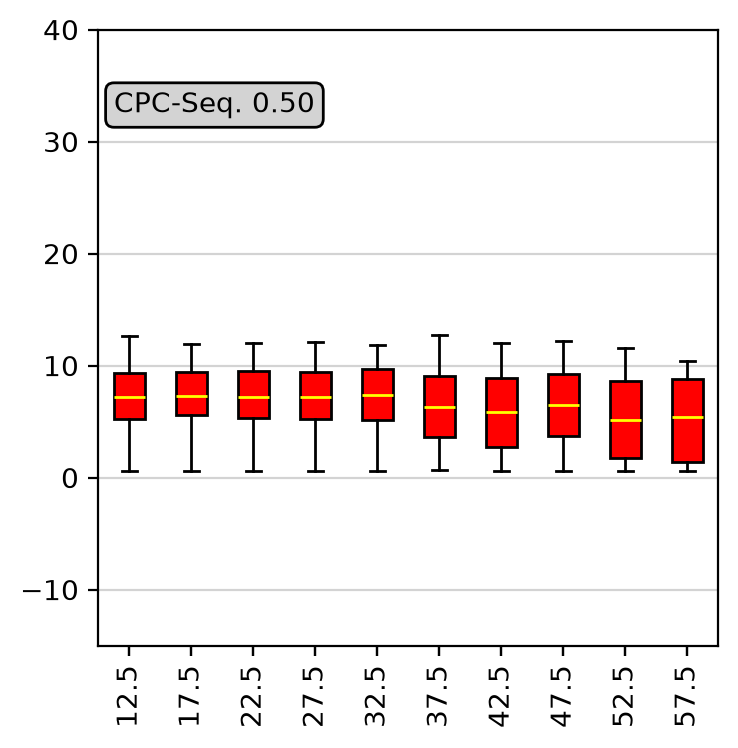}
    \end{subfigure}%
    \begin{subfigure}{0.25\linewidth}
        \includegraphics[width=\linewidth, height = 4.0cm]{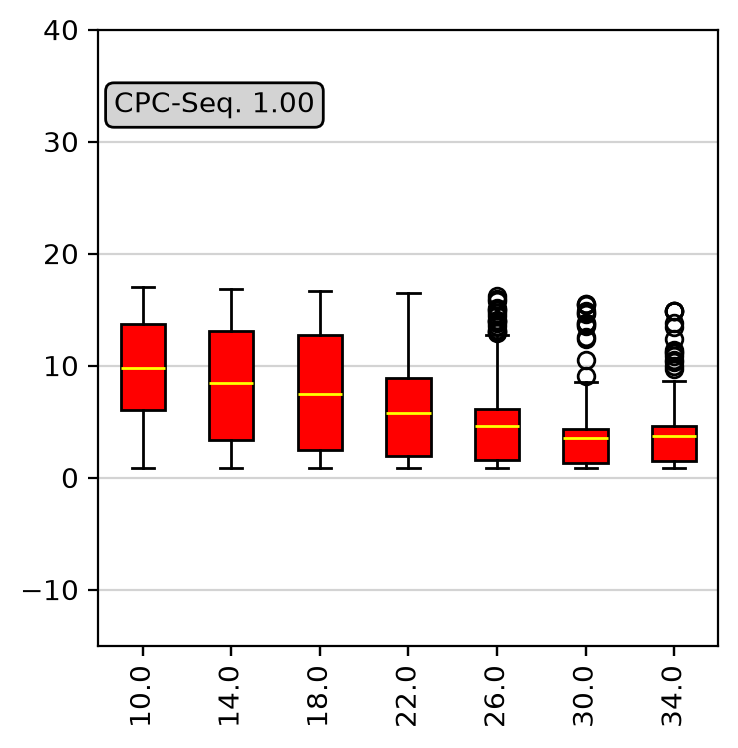}
    \end{subfigure}

    \begin{subfigure}{0.25\linewidth}
        \includegraphics[width=\linewidth, height = 4.0cm]{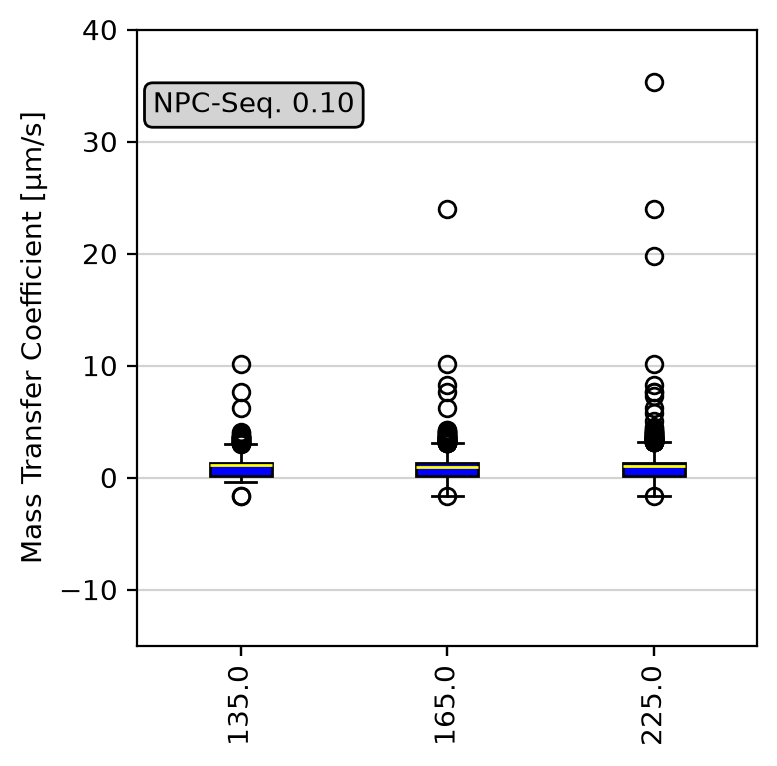}
    \end{subfigure}%
    \begin{subfigure}{0.25\linewidth}
        \includegraphics[width=\linewidth, height = 4.0cm]{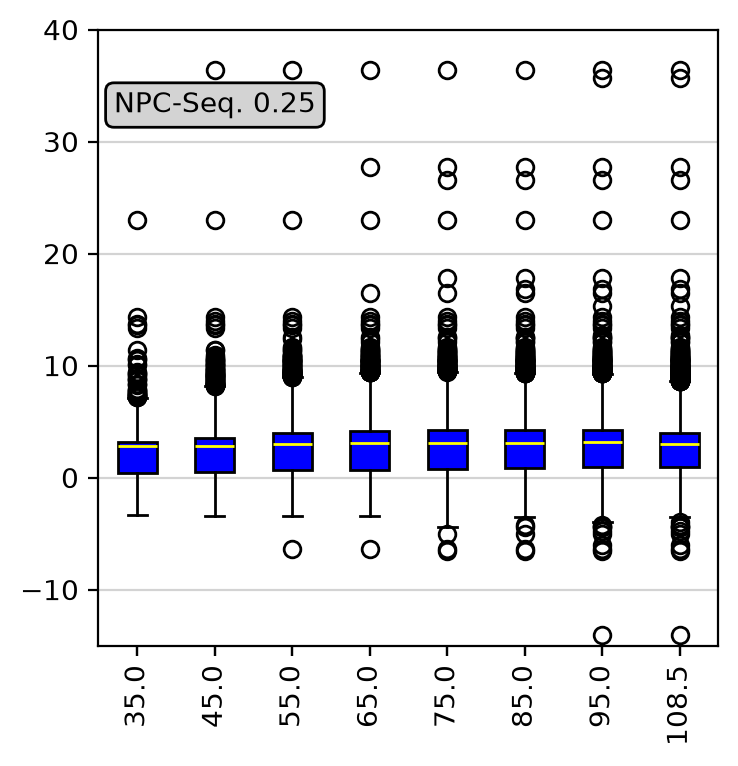}
    \end{subfigure}%
    \begin{subfigure}{.25\linewidth}
        \includegraphics[width=\linewidth, height = 4.0cm]{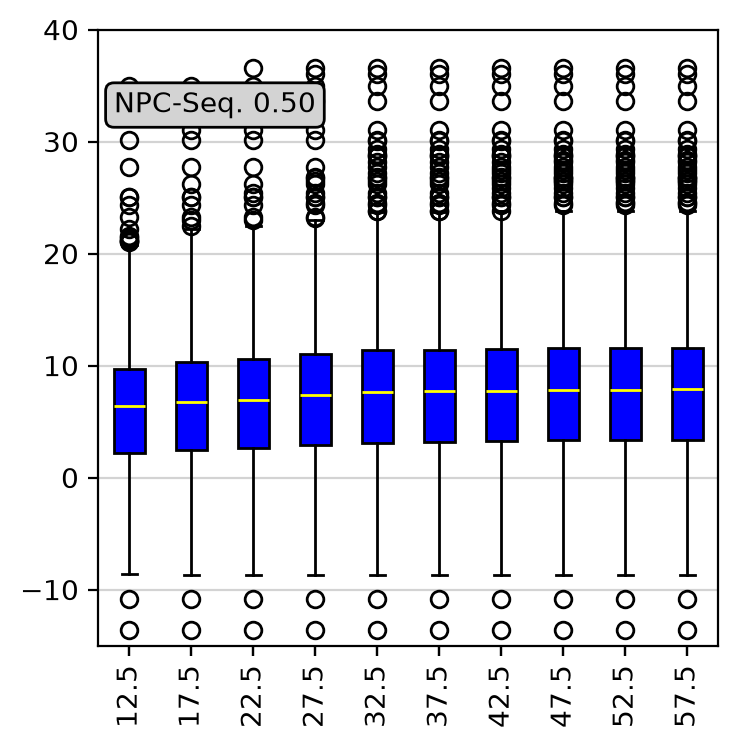}
    \end{subfigure}%
    \begin{subfigure}{0.25\linewidth}
        \includegraphics[width=\linewidth, height = 4.0cm]{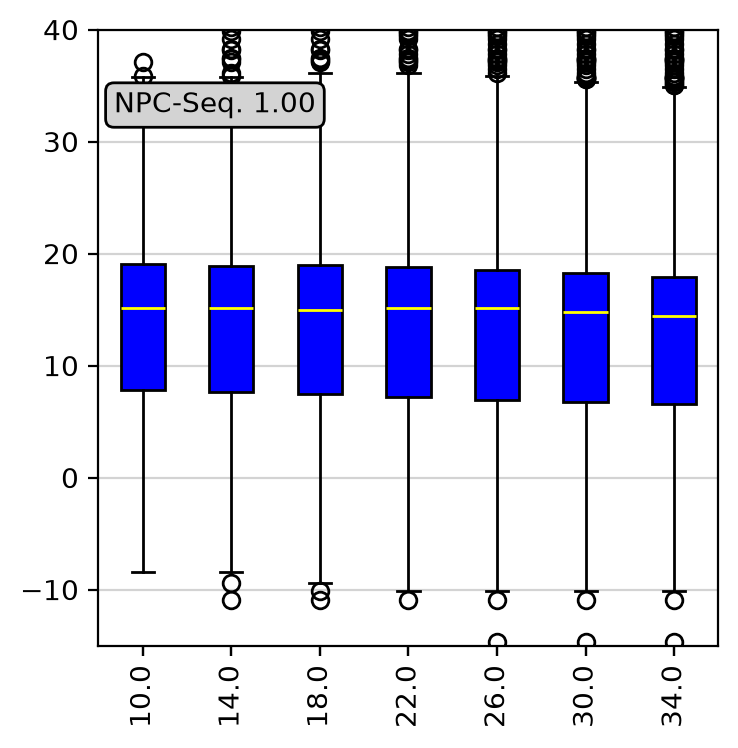}
    \end{subfigure}

    \begin{subfigure}{0.25\linewidth}
        \includegraphics[width=\linewidth, height = 4.0cm]{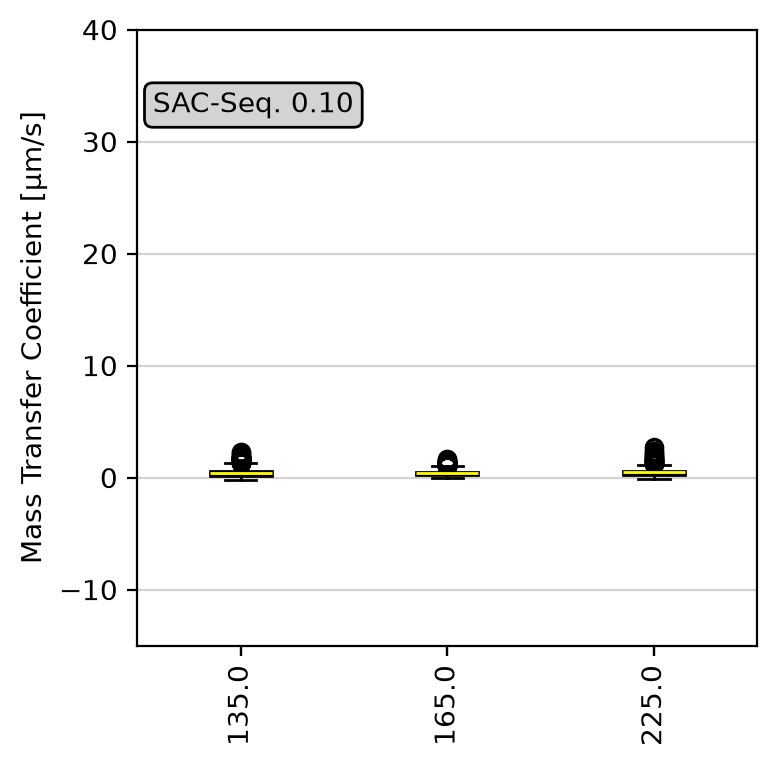}
    \end{subfigure}%
    \begin{subfigure}{0.25\linewidth}
        \includegraphics[width=\linewidth, height = 4.0cm]{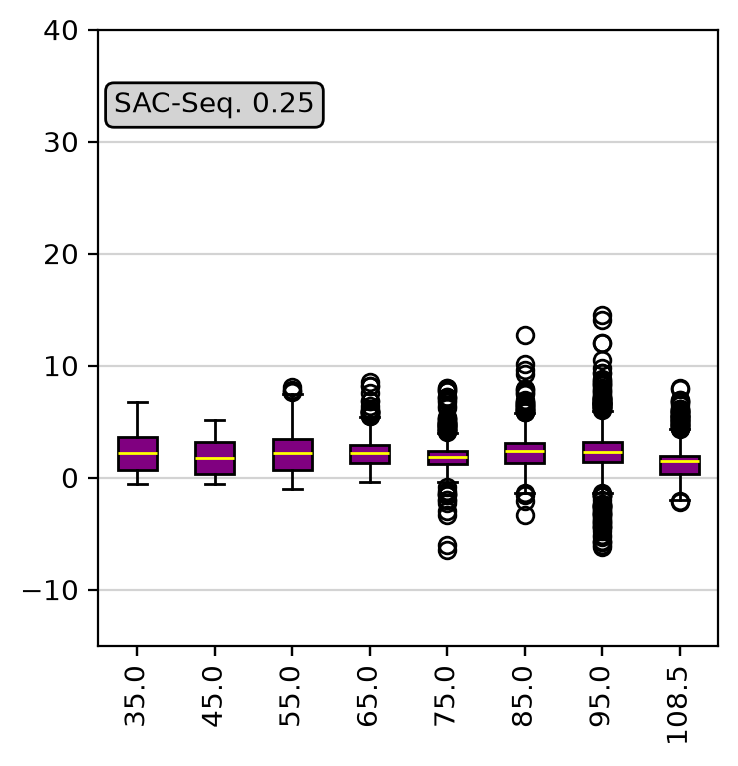}
    \end{subfigure}%
    \begin{subfigure}{.25\linewidth}
        \includegraphics[width=\linewidth, height = 4.0cm]{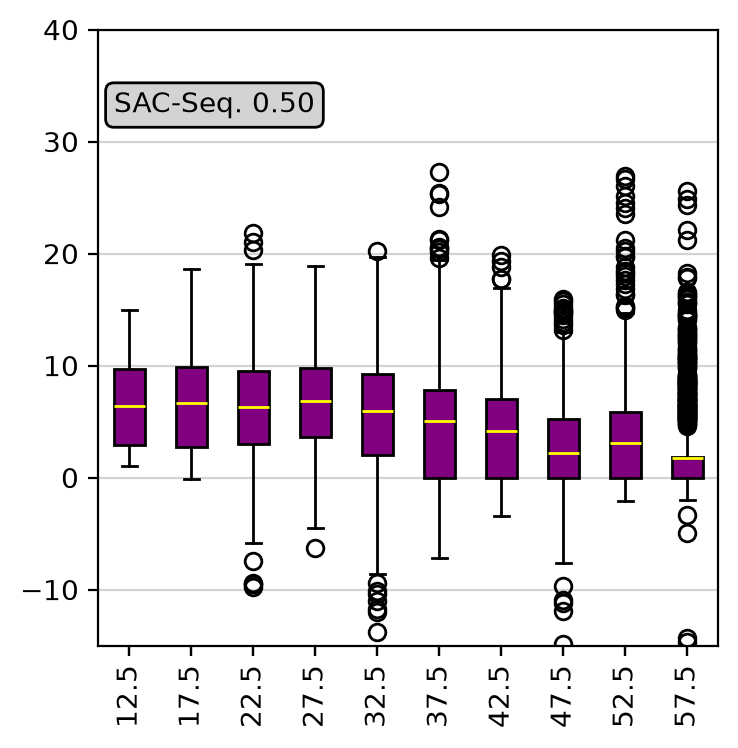}
    \end{subfigure}%
    \begin{subfigure}{0.25\linewidth}
        \includegraphics[width=\linewidth, height = 4.0cm]{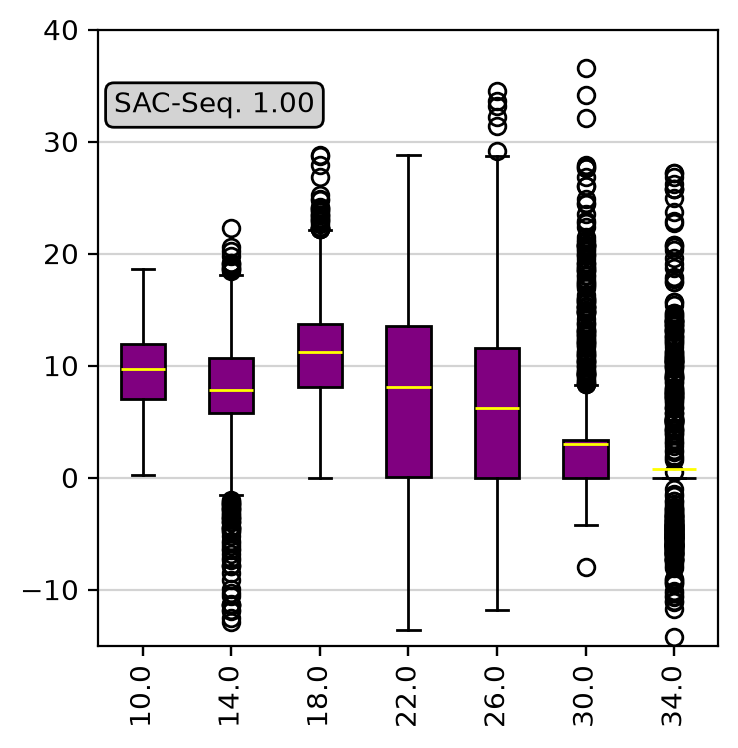}
    \end{subfigure}

    \begin{subfigure}{0.25\linewidth}
        \includegraphics[width=\linewidth, height = 4.0cm]{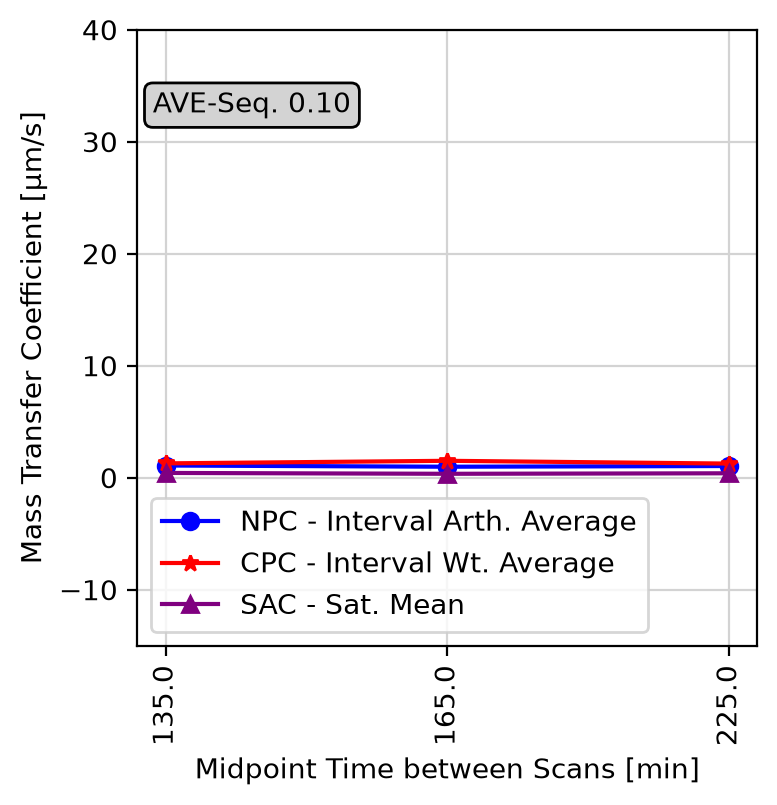}
    \end{subfigure}%
    \begin{subfigure}{0.25\linewidth}
        \includegraphics[width=\linewidth, height = 4.0cm]{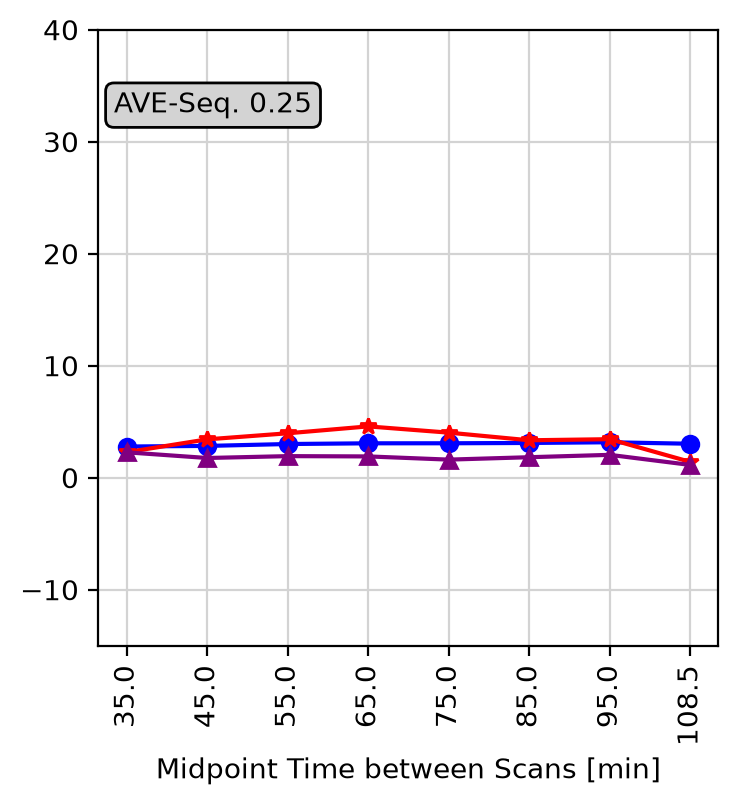}
    \end{subfigure}%
    \begin{subfigure}{.25\linewidth}
        \includegraphics[width=\linewidth, height = 4.0cm]{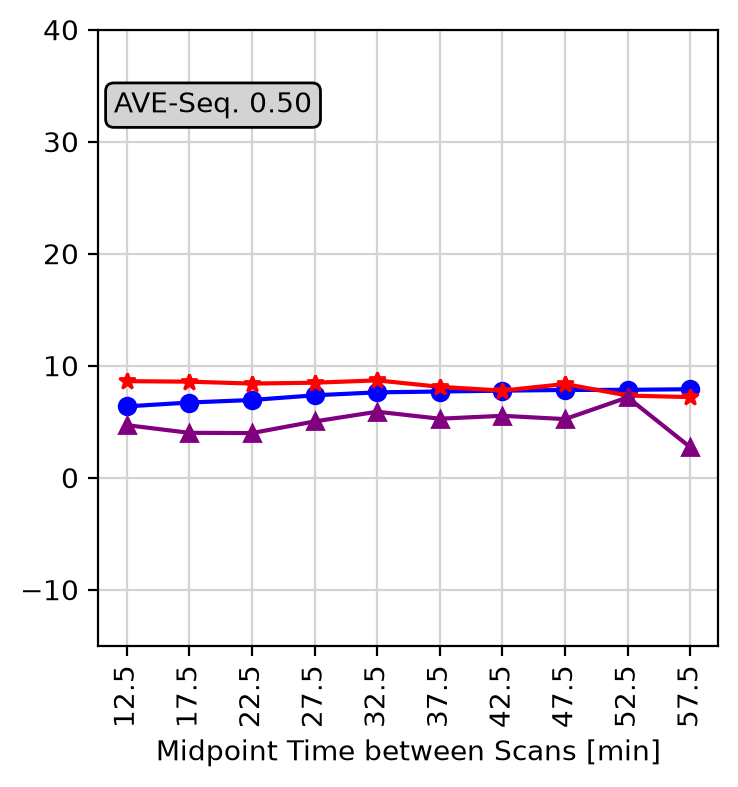}
    \end{subfigure}%
    \begin{subfigure}{0.25\linewidth}
        \includegraphics[width=\linewidth, height = 4.0cm]{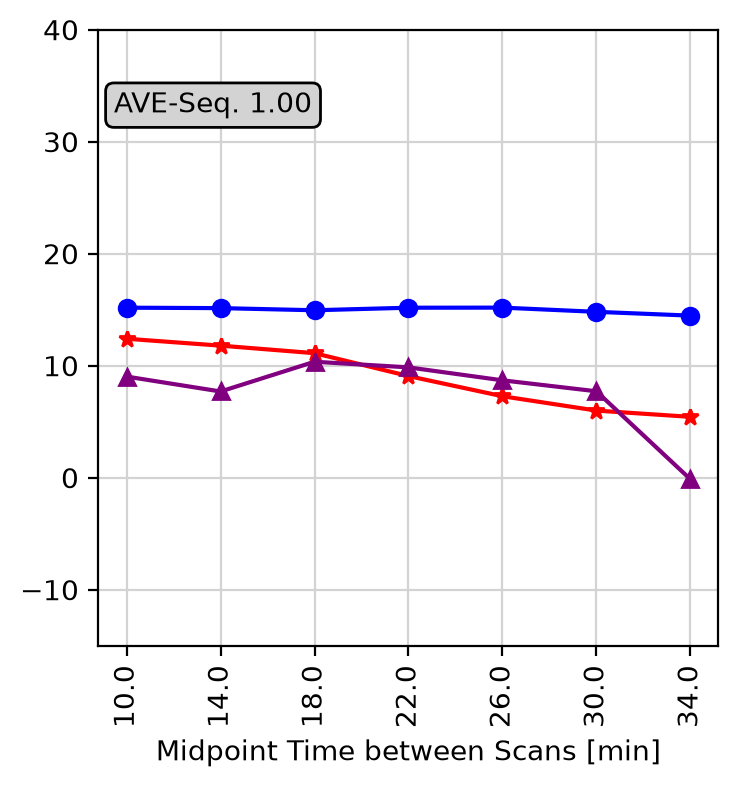}
    \end{subfigure}
    \caption{Box and whisker plots of the mass transfer coefficient ($k_{i,j}$) distributions for each mobilization filtered time-interval. Each column represents the sequence evaluated (e.g., Seqs. 0.10, 0.25, 0.50, 1.00). The first three rows show the distributions for each analytical approach used: red (CPC), blue (NPC), and purple (SAC). Row 4 plots the interval averages for each approach: the surface-area weighted average values ($K_{\text{ave}}^{\text{int}}$) for both per-Cluster approaches are shown in red (CPC) and blue (NPC), and the saturation average $K_{\text{mean}}^{\text{int}}$ values from the SAC are purple. Note: Due to outlier values ranging from -200 to +400 $\frac{\mu m}{s}$, the Y-axis is truncated to -10 and +40 $\frac{\mu m}{s}$ for all central distributions to be visible on a similar scale. Thus, truncation visually excludes outlier values in all NPC and SAC sequences. There are no such outliers in the CPC sequences.}
    \label{fig:mass_trans_dist}
\end{figure*}

Figure \ref{fig:mass_trans_dist} rows (1) to (3) display the $\mathbb{K}$ for each analytical approach as box-and-whisker plots, broken up by sequence injection rate and observation interval (after mobilization filtering). In each box plot, the yellow line represents the arithmetic mean ($K_{\text{mean}}$), the length of the box represents the central span (values within one standard deviation of the mean), the whiskers (or error bars) are the values 1.5 standard deviations from the mean, and the hollow circles represent outlier values greater than 1.5 standard deviations.\par
We have chosen to break up the distributions for each approach by time-intervals ($\mathbb{K}|_j$) in order to observe the contribution of the component sample populations to the final estimate, as an indication of the volume changes being sampled. For example, distributions with significantly different means and standard deviations across intervals would imply rapidly changing mass-transfer conditions over time, but are likely to indicate that other phenomena (e.g., remobilization) are skewing the measured distribution.\par
\subsubsection{Classified per Cluster Approach}
\begin{table*}[htb!]
    \centering
     \caption{Table of the final mass transfer estimates for each sequence, using the CPC approach, with and without mobilization filtering. The cluster and volume lost due to filtration are calculated relative to the cluster types used by each approach to estimate $k_{i,j}$ and $C_{i,j}$; for the CPC approach, these are only completely and partially dissolved clusters.}
    \resizebox{\textwidth}{!}{%
    \begin{tabular}{l c c c c c c } \hline
    Sequence & Mass Transfer & Mass Transfer   & Difference &  \multicolumn{2}{c}{Clusters } & Volume \\ 
     & Coeff. without & Coeff. with & &\multicolumn{2}{c}{Retained} &Retained \\ 
    
            & mobilization & mobilization & & & &\\ 
            & filtering $[\mu m/ s]$ & filtering $[\mu m/ s]$& [\%] & [\#] & [\%] & [\%]\\\hline \hline
        0.10  &  2.16 $\pm$ 0.60 &  1.33 $\pm$ 0.39 & - 38.2\% & 2425  & 16.6\% & 22.7\%\\
        0.25  &  3.41 $\pm$ 0.45 &  3.47 $\pm$ 0.50 & +  1.8\% & 8192  & 57.9\% & 70.7\%\\ 
        0.50  &  8.32 $\pm$ 0.48 &  8.44 $\pm$ 0.47 & +  1.4\% & 9655 & 67.7\% & 69.8\%\\
        1.00  & 11.49 $\pm$ 0.86 & 10.92 $\pm$ 1.16 & -  5.0\% & 4018  & 53.1\% & 40.2\%\\\hline \hline
    \end{tabular}}
    \label{tab:results_CPC}
\end{table*}%
The CPC distributions within each sequence (Figure \ref{fig:mass_trans_dist}) exhibit a consistent spread of overall values (the tails and outliers) among all time-intervals in a sequence, while the central quartile varies smoothly between intervals. Notably, the spread of all $\mathbb{K}_{CPC}|_j$ within a sequence consistently increases with injection rate; the movement of the central quartile from interval to interval varies depending on the sequence flow rate. The behavior of the tails indicates consistency in the range of morphology changes sampled within a sequence, while fluctuations in the central quartile reflect a shift in the prevalence of changes of magnitude ``X'', within said range. For example, in Seq. 1.00, we observe that $\mathbb{K}_{CPC}|_j$ are skewed towards larger $k_{i,j}$ values in earlier time-intervals and shift to small $k_{i,j}$ values in later time-intervals. The shift in means and the central span of the $\mathbb{K}_{CPC}|_j$ in Seq. 1.00 towards smaller $k_{i,j}$-values indicates that the results of these time-intervals contain a larger proportion of small clusters than the previous intervals. The higher prevalence of smaller clusters is likely due to the high injection rate depleting the subset of clusters that can easily interact with the preferential flow path; clusters outside of this path are not at the same mass transfer condition as the interior.\par
For both per-Cluster approaches, the magnitude of $k_{i,j}$ is strongly tied to the cluster's geometry (volume and surface area). Previous works evaluating reconstructed geometries have shown that objects on the order of the image's voxel resolution propagate far more error when approximating smooth geometries from voxelated shapes than objects several orders larger \citep{Schlüter_2014, Lin_2015, guntoro_2019}. While \citet{Ruotong_H_2023} applied the surface-area weighted average (Eq. \ref{eq:weight_ave}) to reduce the weight of errors from small clusters in the final average, the drop in the weighted interval averages in Seq. 1.00  (Figure \ref{fig:mass_trans_dist} row 4) with time indicates that even the weighting is not immune to smaller clusters if the system is being rapidly depleted of larger ones. However, this sensitivity may be a strength, as we were able to observe the depletion of a subclass of morphology changes in the system that would otherwise be lost in the full measurement pool.\par
The final $K_{\text{ave}}^{\text{seq}}$ estimates for the CPC approach are presented in Table \ref{tab:results_CPC}, with and without mobilization filtering. Each sequence estimate is accompanied by a 99\% confidence interval based on the weighted distribution, as well as the quantity of clusters and gas volume remaining after mobilization filtering. The mobilization threshold used was 0.2 ($\frac{\Delta \text{volume gained}} { |\Delta \text{volume lost}|}$). Among the four injection sequences, there is no noticeable trend in the changes pre- and post-mobilization filtering, i.e., in data loss, percent differences in estimates, or the widths of the confidence intervals. Notably, Seq. 0.10 is most sensitive to mobilization filtering, losing 15 of the 18 time-intervals (83\%) containing 78\% of the isolated gas phase volume; this removal decreased the final estimate for $K_{\text{ave}}^{\text{seq}}$ by 38\%. A sensitivity analysis of the CPC results as a function of the mobilization threshold is provided in the \textit{Mobilization Filtering Sensitivity} section.\par
\subsubsection{Non-Classified Per Cluster Approach}
\begin{table*}[htb!]
    \centering
    \caption{Table of the final mass transfer estimates for each sequence, using the NPC approach, with and without mobilization filtering. The cluster and volume lost due to filtration are calculated relative to the cluster types used by each approach to estimate $k_{i,j}$ and $C_{i,j}$. Thus, for the NPC approach, we use all clusters.}
    \resizebox{\textwidth}{!}{%
    \begin{tabular}{l c c c c c c } \hline
       Sequence & Mass Transfer & Mass Transfer   & Difference &  \multicolumn{2}{c}{Clusters } & Volume \\ 
     & Coeff. without & Coeff. with & &\multicolumn{2}{c}{Retained} &Retained \\ 
    
            & mobilization & mobilization & & & &\\ 
            & filtering $[\mu m/ s]$ & filtering $[\mu m/ s]$& [\%] & [\#] & [\%] & [\%]\\\hline \hline
        0.10  &  2.11 $\pm$ 0.66 &  1.03 $\pm$ 0.29 & -51.0\% & 2580  & 13.6\% & 20.5\%\\
        0.25  &  2.52 $\pm$ 0.20 &  3.02 $\pm$ 0.24 & + 20.1\% & 8575 & 53.3\% & 67.1\%\\ 
        0.50  &  6.38 $\pm$ 0.19&  7.89 $\pm$ 0.27 & + 23.6\% & 9903 & 65.9\% & 68.9\%\\
        1.00  & 13.5 $\pm$ 1.14 & 14.46 $\pm$ 1.99 & + 6.8\% & 4206  & 52.8\% & 39.6\% \\ \hline\hline
    \end{tabular}}
    \label{tab:results_NPC}
\end{table*}%
The behavior of the $\mathbb{K}_{NPC}|_j$ distributions in Figure \ref{fig:mass_trans_dist}, row (2), is nearly opposite to the $\mathbb{K}_{CPC}|_j$: the means and the central quartiles are fairly stable with time, while the tails and the outliers fluctuate. The increasing kurtosis of each $\mathbb{K}_{NPC}|_j$ with time indicates that previously outlying morphology changes become more prevalent over time. Despite increasing kurtosis over time, the mean and central quartiles do not appear to change significantly across sequences. Together, this consistent dual kurtosis and stable central quartiles suggest that multiple types of ``morphology changes'' occur in each measured interval, but the events that make up the central quartiles are far more numerous and dominant, preventing the distribution from flattening as rapidly as the tails grow. Additionally, note that all $\mathbb{K}_{NPC}|_j$ contain negative $k_{i,j}$-values, which result from including clusters that gained volume. Contrary to expectations, time-intervals removed by mobilization filters do not exhibit an increase in the prevalence of negative $k_{i,j}$-values; rather, the removed $\mathbb{K}_{NPC}|_j$ produce tighter distributions that rapidly expand to the more stable behavior observed in  Figure \ref{fig:mass_trans_dist}, row (2) (see also Supplementary Materials: \textit{Mass Transfer Coefficient Distributions without mobilization filtering}). While the negative $k_{i,j}$-values are not dominant, their presence may help balance the effects of the larger positive extremes and keep the means and central spans of each $\mathbb{K}_{NPC}|_j$ stable across a sequence.\par
The final $K_{\text{mean}}^{\text{seq}}$ estimates for the NPC approach are presented in Table \ref{tab:results_NPC}. We compare the NPC values with and without mobilization filtering to evaluate the effects of interval removal on a larger data pool; in theory, the NPC approach should be more resistant to threshold choice. Surprisingly, despite the stability of $\mathbb{K}_{NPC}|_j$ in each sequence, the NPC-$K_{\text{mean}}^{\text{seq}}$ is more sensitive to the mobilization filtering than CPC-$K_{\text{ave}}^{\text{seq}}$: more volume data was removed, and estimates shifted more significantly post-mobilization filtering. The NPC approach's greater sensitivity to the mobilization filtering may be due to mobilized clusters being double-counted. Since the NPC approach uses all clusters, the ``disappeared'' and ``gained'' clusters would produce both positive and negative $k_{i,j}$-values, respectively, giving this ``false'' population more weight to skew the final approximation. Regardless, the final NPC estimates for each sequence are calculated without mobilization filtering to remain closer to the original implementation.\par

\subsubsection{1-D Sliced Averaged Approach }

While we have presented the results of the SAC approach ($\mathbb{K}_{SAC}|_j$, Figure \ref{fig:mass_trans_dist} row 3) in the same manner as the CPC and NPC distributions, the position-dependent nature of the SAC approach itself prevents meaningful statistical evaluation of $\mathbb{K}_{SAC}(x)|_j$ to approximate $\mathbb{N}(k^{*}_{i,j})$. Due to the use of the 1-D advection model, SAC $k_{j}$-values all correspond to a specific axial position in the system field of view; thus, $k_{x,j}$ values (within a time-interval) are not independent of one another. The positional interdependence means that a simple average of the axial $k_{x,j}$-values will not accurately reflect overall mass transfer in the system, because clusters span multiple slices; thus, the $k_{x,j}$ of a given slice is closely related to those of neighboring slices.\par
Due to the different context of the SAC $k_{x,j}$-values, the SAC $K_{\text{mean}}^{\text{int}}$ values presented in Figure \ref{fig:mass_trans_dist} row (4) are, instead, calculated from the average change in saturation of a time-interval, following the calculations provided by \citet{Patmonoaji_2023}. This average value can be considered as a ``saturation average''. Similar to the NPC approach, the SAC approach also yielded increasingly negative $k_{x,j}$-values in slices where cluster growth is dominant. However, the inclusion of negative values affected the SAC approach differently than the NPC: the per-slice basis resulted in a smaller sample pool than the per-Cluster basis; this ultimately yielded SAC-based average $k_{x,j}$-values $\le 0$ for several time-intervals (Figure \ref{fig:mass_trans_dist}, row 3). Several negative interval averages ($K_{\text{mean}}^{\text{int}}$) remained even after mobilization filtering. While the range of values from the SAC approach is similar to both the CPC and NPC approaches, the mean of the SAC $k_{x,j}$ was found to be more variable over time for the same injection sequence.\par

\subsubsection{Comparison of Approach Estimates: Mass Transfer Coefficients}
Figure \ref{fig:mass_trans_dist} row (4) plots the post-filtered interval averages for each approach. The CPC values are plotted as surface area-weighted averages ($K_{\text{ave}}^{\text{int}}$, Eq. \ref{eq:weight_ave}), while the NPC-values are an arithmetic mean and SAC-values are calculated as the average saturation for each interval (as presented in \citet{Patmonoaji_2023}). For the same injection sequence, all three approaches yielded similar estimates of the mass transfer coefficients, ranging within one order of magnitude of each other, despite significant differences in assumptions and quantification among the approaches. The agreement between the approaches appears to vary with the solvent injection rate; the CPC approach consistently yields a higher estimate than the NPC and SAC approaches until Seq. 1.00, where the filtered interval averages for CPC and the SAC decay as the sequence progresses, and the NPC approach yields the higher estimate. Overall, each approach yields relatively stable interval averages, with Seq. 1.00 being the only sequence in which the SAC and CPC averages vary significantly over time (NPC remains stable).\par
\begin{figure}[h!]
    \centering
    \includegraphics[width=0.75\linewidth]{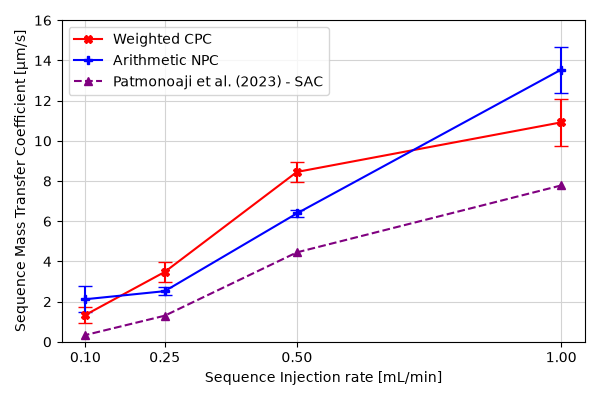}
    \caption{Comparison of the $K^{\text{seq}}$ for each of the approaches, at each injection rate. The $K_{\text{ave}}^{\text{seq}}$ for the per-Cluster approaches are shown in red (CPC) and blue (NPC). The error bars represent the 99\% confidence interval for each estimated value, with respect to the weighted-CPC and unweighted-NPC distributions. The $K^{\text{seq}}_{\text{mean}}$ SAC values obtained from \citet{Patmonoaji_2023} are in purple.}
    \label{fig:results}%
\end{figure}
\begin{figure}[ht!]
    \centering
    \includegraphics[width = 0.75\linewidth]{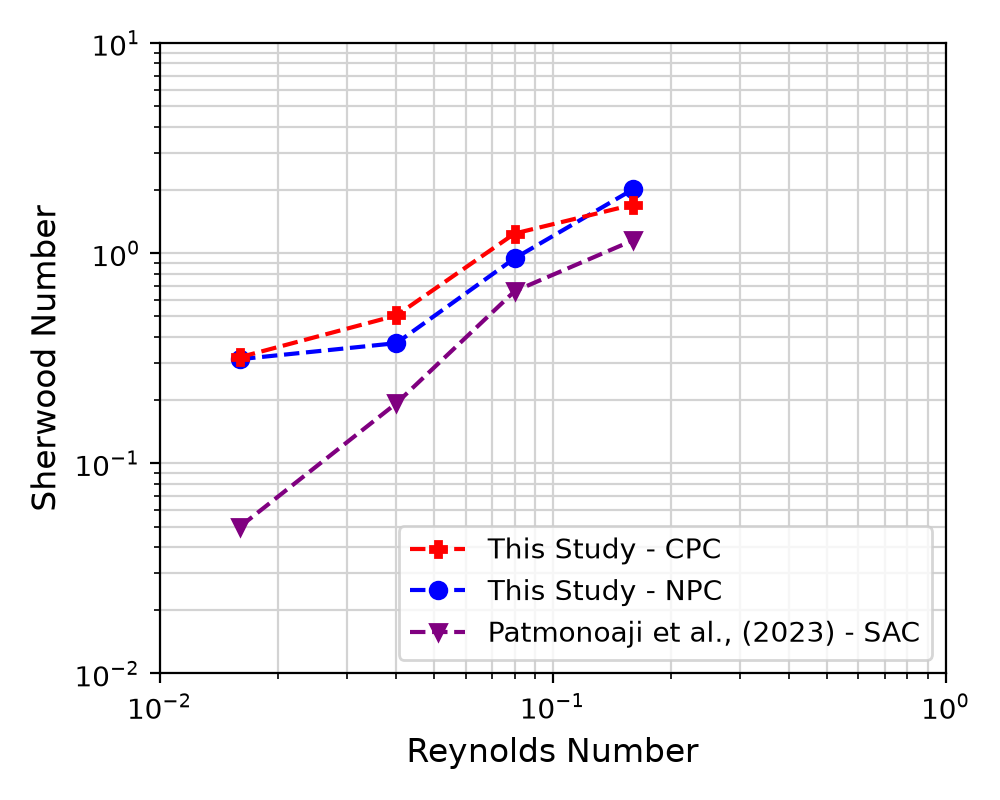}
    \caption{ Plot of the Sherwood numbers ($Sh$) versus the Reynolds numbers ($Re$) for each approach: CPC (red), NPC (blue), and the SAC (purple).}
    \label{fig:sherwood}
\end{figure}

The sequence average mass transfer coefficients ($K_{\text{ave}}^{\text{seq}}$ for the CPC (post-mobilization filtering) and the $K_{\text{mean}}^{\text{seq}}$ for the NPC, and SAC (no mobilization filtering) for each approach as a function of the solvent injection rate are plotted in dimensional form in Figure \ref{fig:results}, and dimensionless form in Figure \ref{fig:sherwood}.\par
To evaluate the estimates from each approach in dimensionless form, the Sherwood number (Sh) reflects $K^{\text{seq}}$, while the Reynolds number (Re) (Figure \ref{fig:sherwood}) represents the solvent injection rate. From Figures \ref{fig:results} and \ref{fig:sherwood}, we observe similar behavior in the sequence averages as in interval-averaged data, showing the following consistent trends:
\begin{itemize}
    \item All three approaches yielded similar estimates (within an order of magnitude) for the same injection rate.
    \item All estimated $K^{\text{seq}}$ exhibit a positive relationship with injection rate.
    \item The CPC and NPC per-Cluster approaches consistently yielded higher.
\end{itemize}
Visually, CPC $K_{\text{ave}}^{\text{seq}}$ \& $K_{\text{mean}}^{\text{seq}}$ > SAC $K_{\text{mean}}^{\text{seq}}$ for all injection rates; and whether the CPC or the NPC produces a higher estimate alternates. The similarity between the $K^{\text{seq}}$ estimates across approaches depends on whether a linear or a proportional basis is used.  With a linear basis, the distance (difference) between the CPC and SAC estimates increases with injection rate, whereas with a proportional basis, the difference between estimates is small at higher injection rates. Even in dimensionless form, the estimates for the different approaches remain within an order of magnitude for a given Reynolds number; the similarity between the approaches' Sherwood numbers increases with the injection rate (on a proportional basis).\par
\subsubsection{Mobilization Filtering Sensitivity Analysis}

\begin{figure*}[hbt]
    \begin{subfigure}{0.5\linewidth}
        \includegraphics[width=\linewidth]{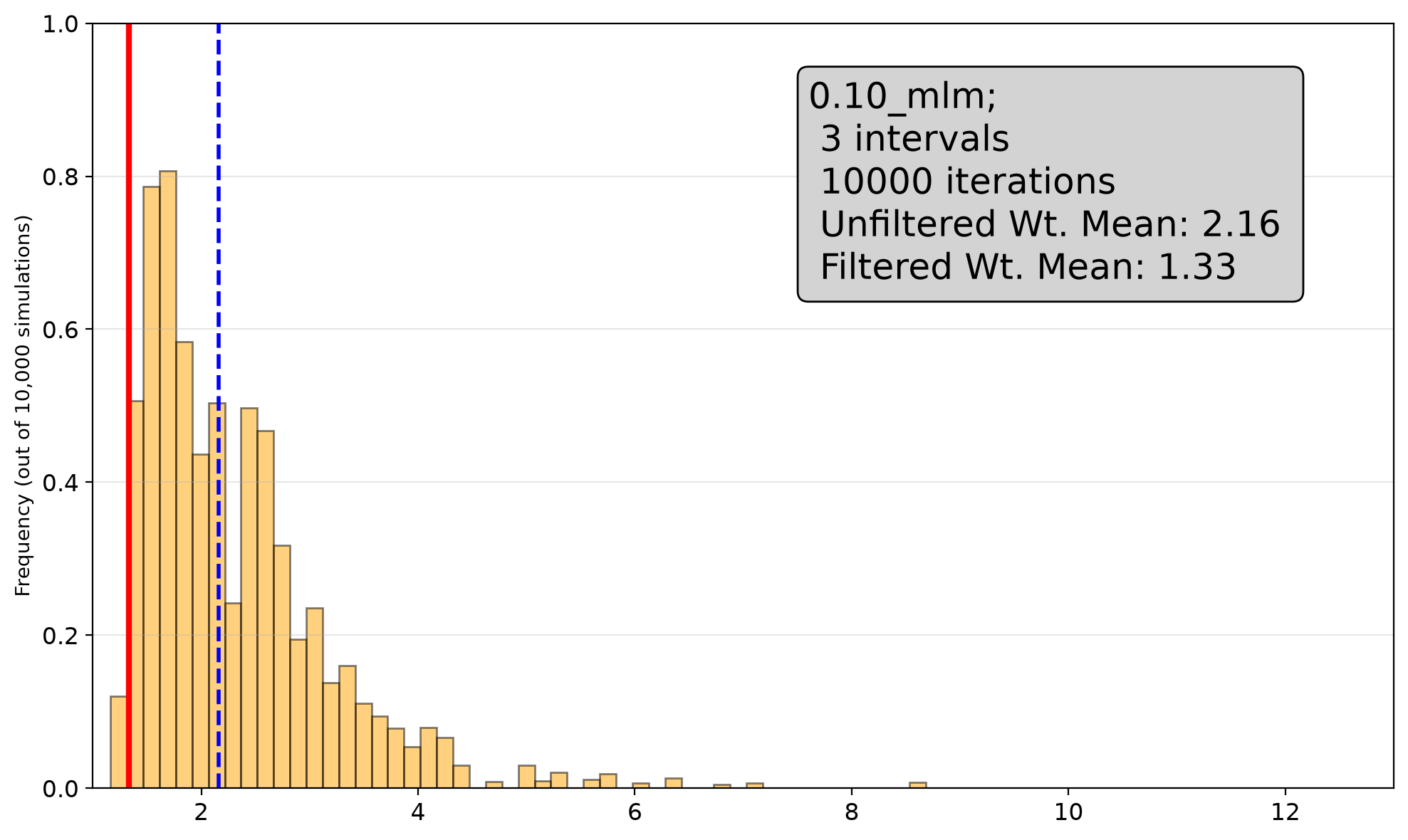}
    \end{subfigure}%
    \begin{subfigure}{0.5\linewidth}
       \includegraphics[width=\linewidth]{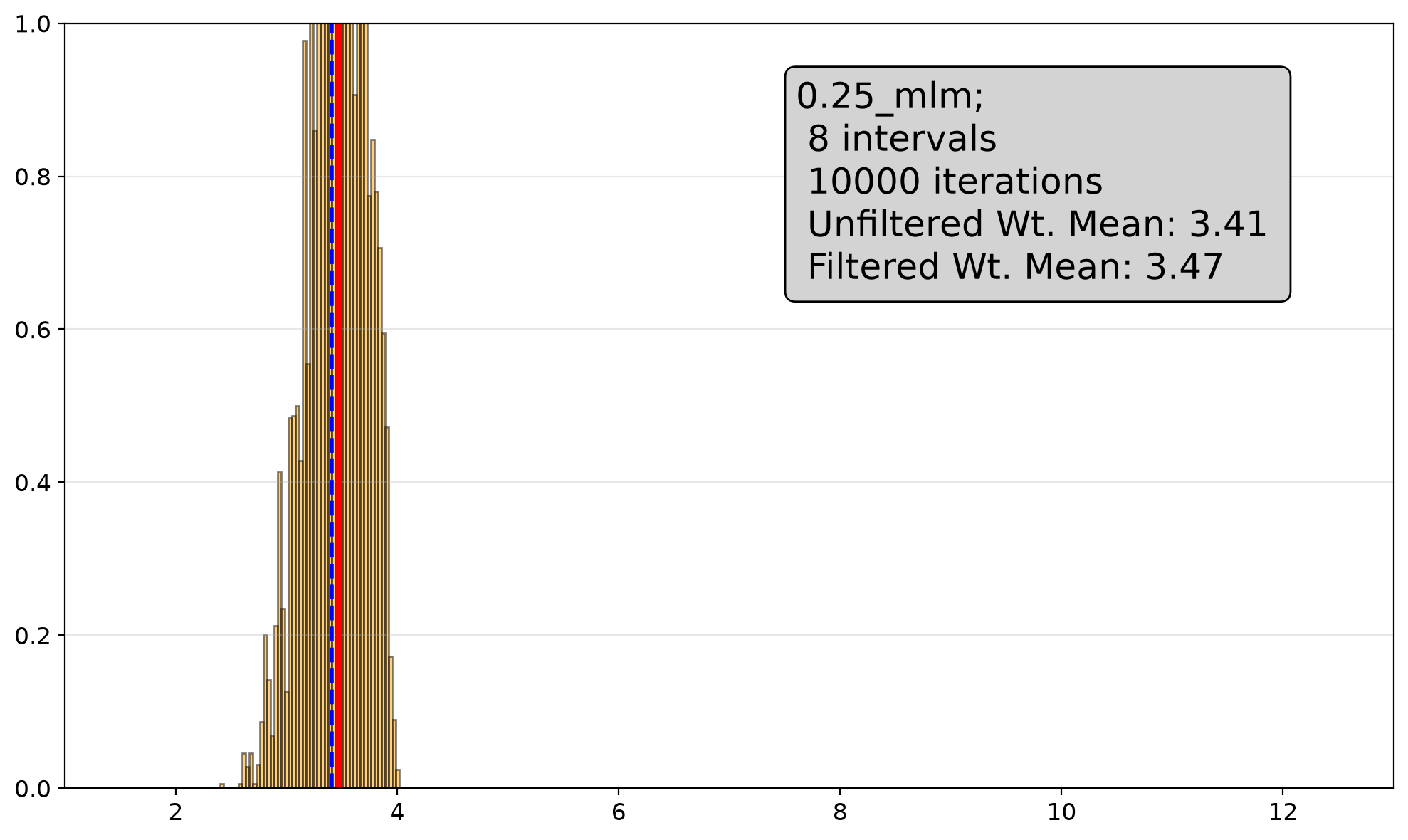}
    \end{subfigure}
    \begin{subfigure}{0.5\linewidth}
        \includegraphics[width=\linewidth]{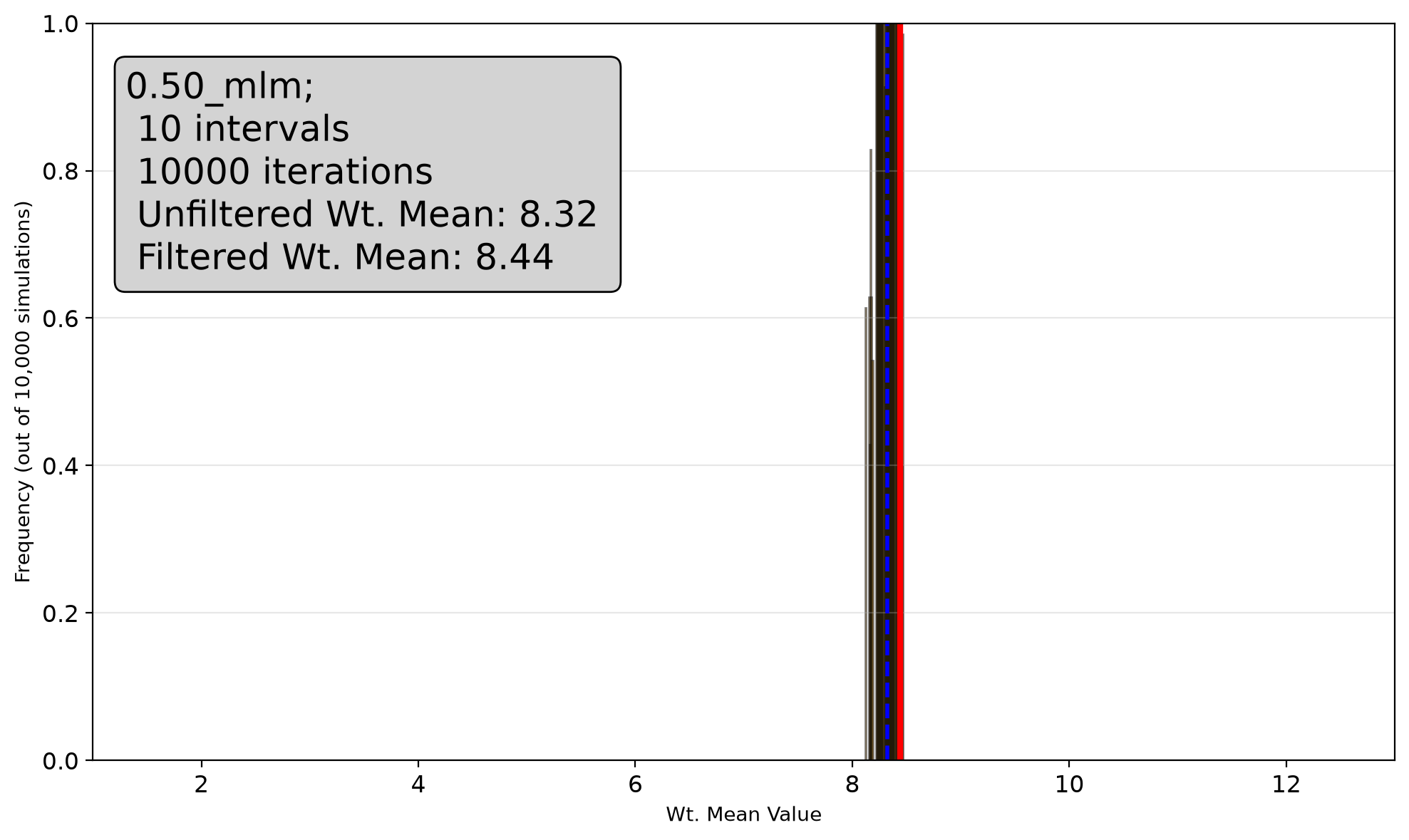}
    \end{subfigure}%
    \begin{subfigure}{0.5\linewidth}
        \includegraphics[width=\linewidth]{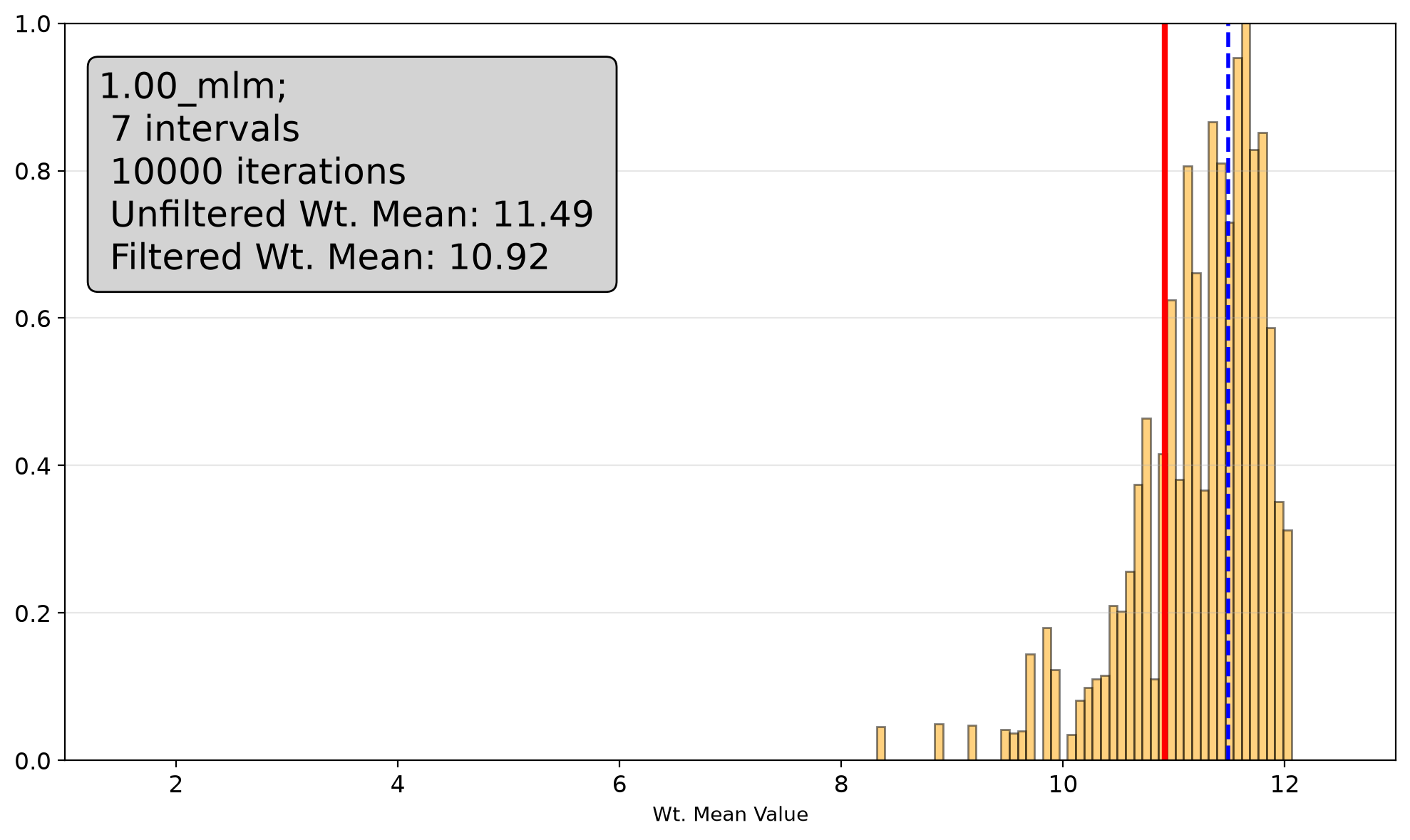}
    \end{subfigure}
    \caption{Plots of the weighted sequence average CPC mass transfer coefficients calculated from randomly resampled time-intervals (without replacement and without mobilization filtering) for 10000 iterations. The selection size for each sequence is based on the remaining number of time-intervals post-filtration, in Figure \ref{fig:per_chng_bars}}
    \label{fig:mass_trans_boot}
\end{figure*}
\begin{table*}[hbt!]
    \centering
    \caption{Random Resampling of CPC Estimates - Results}
    \resizebox{\textwidth}{!}{%
    \begin{tabular}{l c c | c c c c}\hline
         & \multicolumn{2}{c|}{K - Estimates} & \multicolumn{4}{c}{Resampling Statistics}\\
    Seq. &  Unfiltered $K_{\text{ave}}^{\text{seq}}$ & Filtered $K_{\text{ave}}^{\text{seq}}$ & Mean & Left 95\% C.I. & Right 95\%C.I. & Interquartile Range\\\hline\hline
    0.10 & 2.16  & 1.33  & 2.12  & 1.33 & 4.38  & 1.03\\
    0.25 & 3.41  & 3.47  & 3.48  & 2.91 & 3.88  & 0.36\\
    0.50 & 8.32  & 8.44  & 8.34  & 8.22 & 8.44  & 0.07\\
    1.00 & 11.49 & 10.92 & 11.36 & 9.74 & 11.94 & 0.72\\\hline 
    \end{tabular}}
    Note: all values are in µm/s.
    \label{tab:CPC_boot_stats}
\end{table*}
Whether CPC estimates remain representative after filtering requires evaluation, given that mobilization filtering removes intervals and thus data. Without an actual ground-truth measurement or another way to physically verify remobilization in an optically obscure system, we must numerically evaluate the estimates by randomly resampling each image sequence to understand how estimates change as a function of the remaining data window. Figure \ref{fig:mass_trans_boot} plots the resampled $K_{\text{ave}}^{\text{seq}}$-CPC values for the same number of remaining time-intervals indicated in Figure \ref{fig:per_chng_bars}, the statistics of which are summarized in Table \ref{tab:CPC_boot_stats}. Given that each Seq. was resampled relative to the remaining time-interval after mobilization filtering, the 95\% confidence intervals for the resampled $K_{\text{ave}}^{\text{seq}}$-CPC values have different implications (plots for the same resampling of each Seq. are given in the \textit{Supplementary Materials: Additional Mobilization Resampling}).\par
\begin{figure*}[hbt!]
    \begin{subfigure}{\quarter}
        \includegraphics[width = \linewidth]{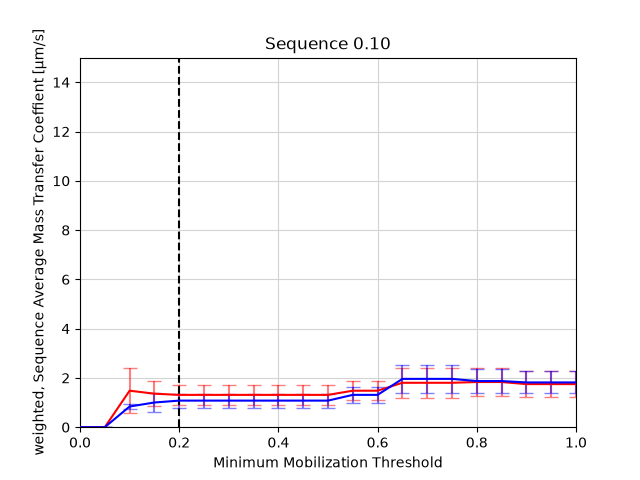}
    \end{subfigure}
    \begin{subfigure}{\quarter}
        \includegraphics[width = \linewidth]{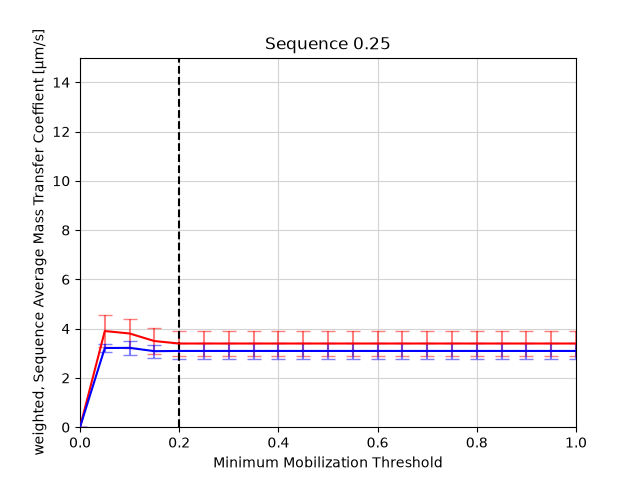}
    \end{subfigure}

    \begin{subfigure}{\quarter}
        \includegraphics[width = \linewidth]{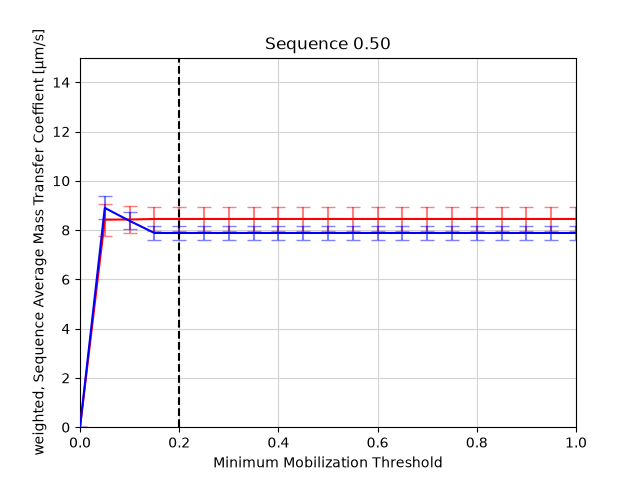}
    \end{subfigure}
    \begin{subfigure}{\quarter}
        \includegraphics[width = \linewidth]{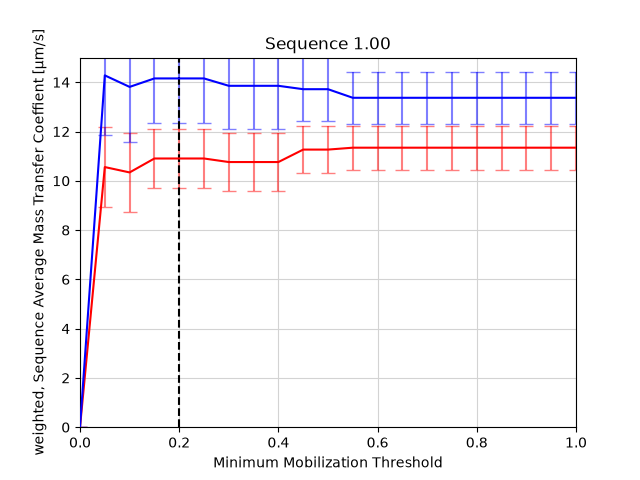}
    \end{subfigure}
    \caption{Plots of the weighted sequence average mass transfer coefficient, with 99\% confidence intervals, for multiple mobilization threshold values. The CPC approach values are in red, while the NPC values are in blue; the dashed black line indicates the mobilization threshold value for the current study (0.2).}
    \label{fig:mass_trans_sense}
\end{figure*}

 
Seqs. 0.25 and 1.00 both fall within their respective confidence intervals, indicating that the filtering is almost indistinguishable from resample selection of intervals. However, the Seq. 0.25 distribution is clearly much tighter (the 0.36 µm/s interquartile span), despite the resampling using a lower overall percentage (57\%) of the 14 time intervals, as opposed to the 72\% of intervals participating in Seq. 1.00's resampling. This implies that mobilization filtering removed different information depending on the injection rate. The clearly right-skewed distribution in Seq. 1.00 indicates that the removed intervals contained far more completely dissolved clusters, which may have counteracted the increased prevalence of smaller clusters in the later intervals that lessened the estimated $K_{\text{ave}}^{1.00}$. Observing clusters growing in size, is a strong indication that mobilization is occurring in the system.  However, without a means of validating the what of the ``dissolved clusters'' actually dissolved and what mobilized, it is unclear if our thresholding is too conservative. Recalculating $K_{\text{ave}}^{\text{1.00}}$-CPC as a function of the mobilization threshold (Figure \ref{fig:mass_trans_sense}) does suggest that 0.2 may be too conservative of the threshold for Seq. 1.00, though the difference is not statistically significant (indicated by the confidence intervals in Figures \ref{fig:mass_trans_boot} \& \ref{fig:mass_trans_sense} and Table \ref{tab:CPC_boot_stats}).\par
Meanwhile, Table \ref{tab:CPC_boot_stats} indicates that Seq. 0.10 and 0.50 fall just outside of their respective 95\% confidence intervals, in which the statistical significance of $K_{\text{ave}}^{\text{seq}}$-CPC is up to the selection of confidence interval. Again, the placement of the filtered $K_{\text{ave}}^{\text{seq}}$-CPC relative to the interquartile span similarly implies that the mobilization filtering removes different information, correlating with injection rate. Seq. 0.50 is the most tightly bound distribution (0.07 µm/s), and Seq. 0.10 is the least tightly bound (1.03 µm/s). The clear left-bound skew and apparent multi-modality of Seq. 0.10 in Figure \ref{fig:mass_trans_boot} indicates that there are separate populations of events within the sampled intervals. From Figure \ref{fig:mass_trans_sense}, we see that Seq. 0.10 also benefits from a relaxation of the mobilization threshold; however, for the purposes of this study, this difference does not significantly change how the approach estimates relate to one another. For the purposes of estimating the mass transfer coefficient with the CPC approach, we recommend a more relaxed mobilization threshold ( < 1.0).\par

\subsection{Concentration Estimates}
\label{subsec:conc_grad}
\begin{figure*} [h!]
    \begin{subfigure}{0.250\linewidth}
        \includegraphics[width=\linewidth, height = 4.0cm]{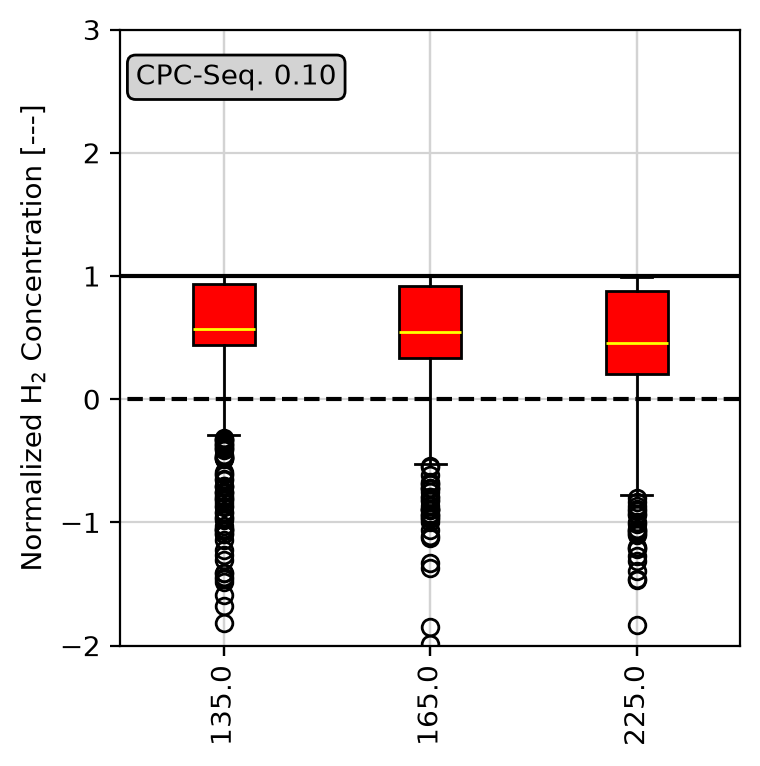 }
    \end{subfigure}%
    \begin{subfigure}{0.25\linewidth}
        \includegraphics[width=\linewidth, height = 4.0cm]{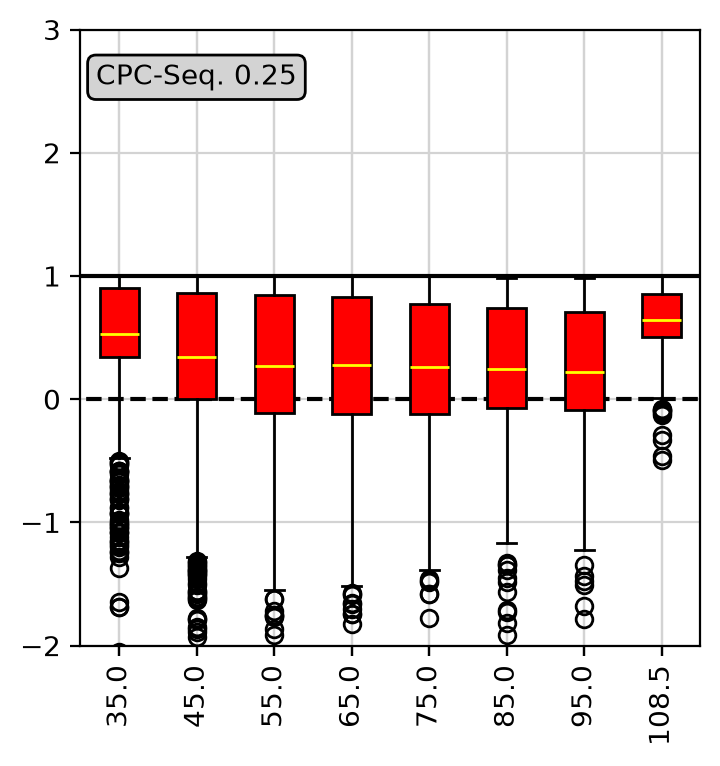}
    \end{subfigure}%
    \begin{subfigure}{0.25\linewidth}
        \includegraphics[width=\linewidth, height = 4.0cm]{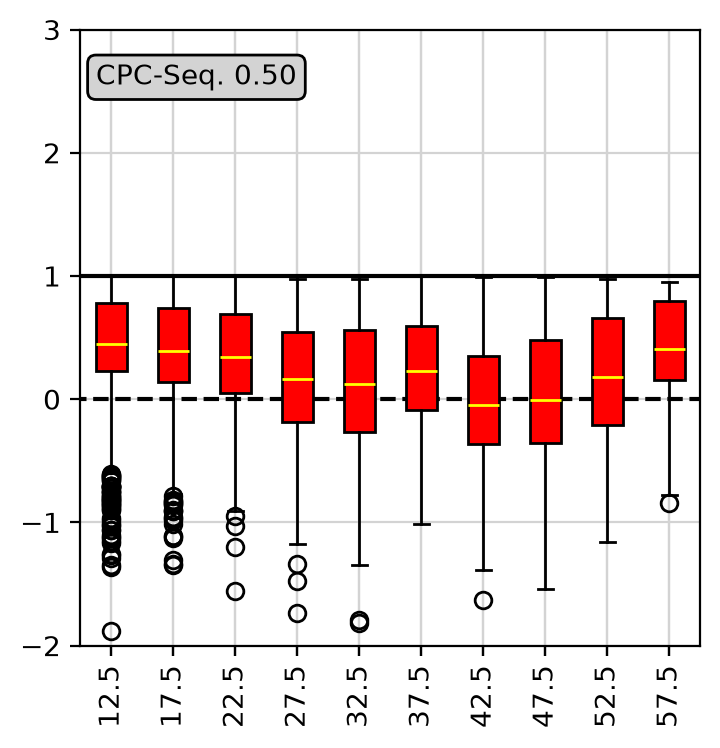}
    \end{subfigure}%
    \begin{subfigure}{0.25\linewidth}
        \includegraphics[width=\linewidth, height = 4.0cm]{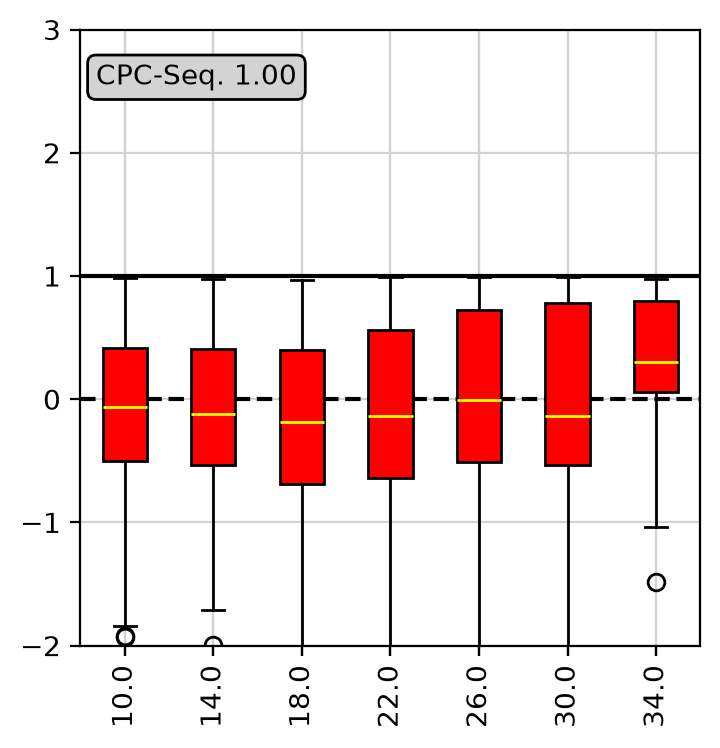}
    \end{subfigure}

    \begin{subfigure}{0.25\linewidth}
        \includegraphics[width=\linewidth, height = 4.0cm]{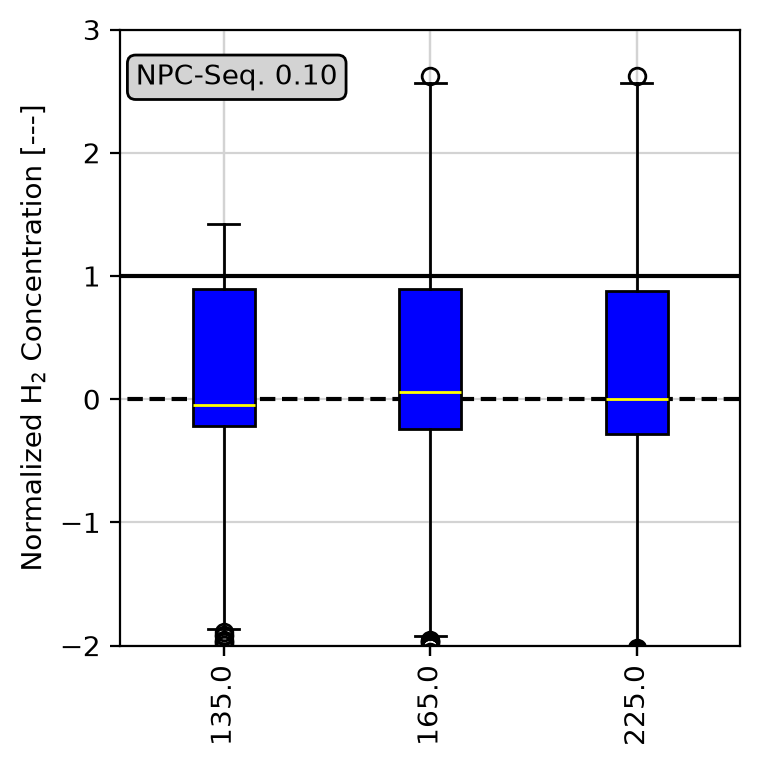}
    \end{subfigure}%
    \begin{subfigure}{0.25\linewidth}
        \includegraphics[width=\linewidth, height = 4.0cm]{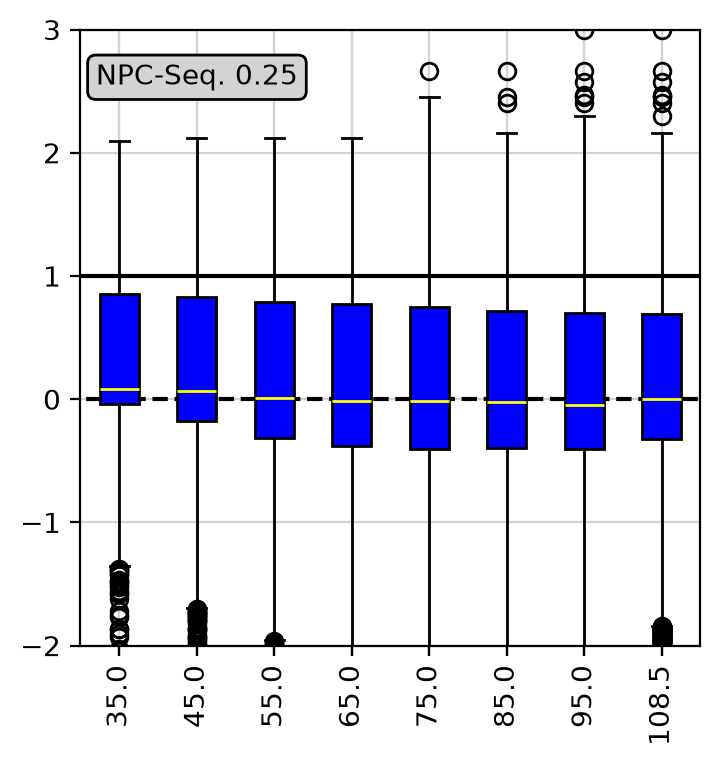}
    \end{subfigure}%
    \begin{subfigure}{.25\linewidth}
        \includegraphics[width=\linewidth, height = 4.0cm]{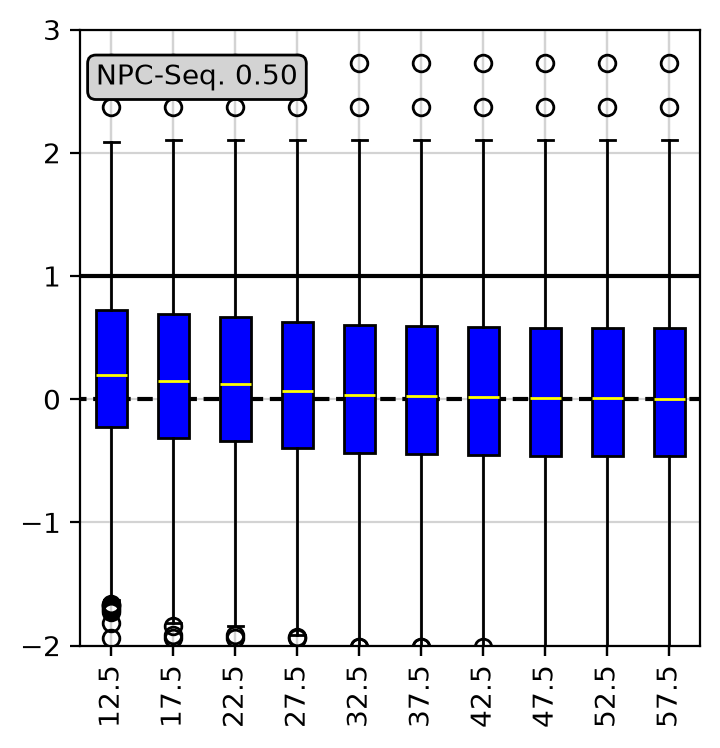}
    \end{subfigure}%
    \begin{subfigure}{0.25\linewidth}
        \includegraphics[width=\linewidth, height = 4.0cm]{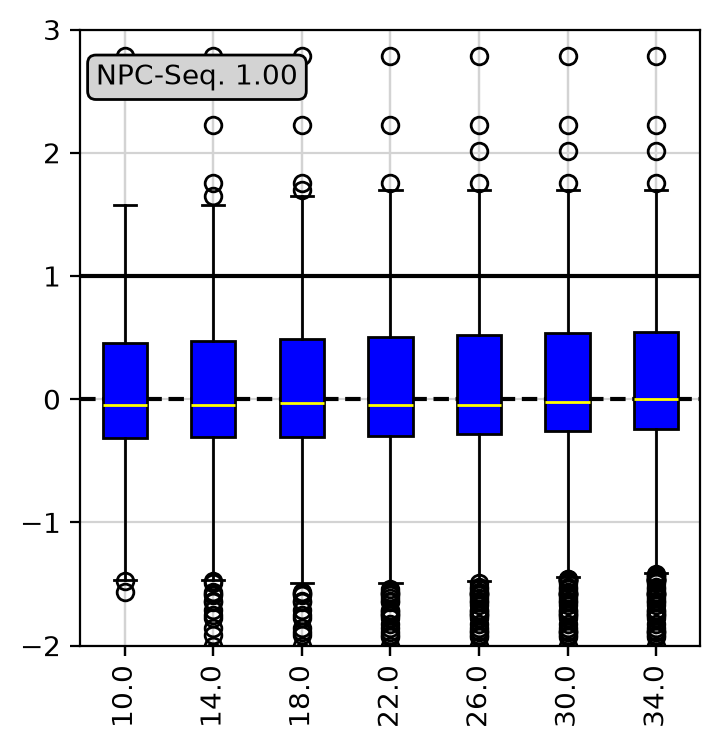}
    \end{subfigure}

    \begin{subfigure}{0.25\linewidth}
        \includegraphics[width=\linewidth, height = 4.0cm]{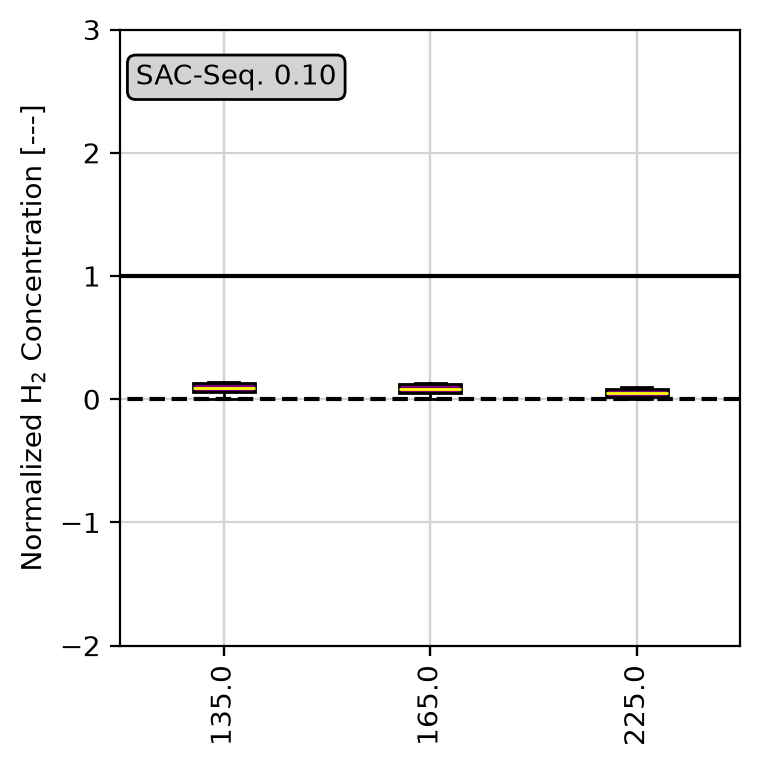}
    \end{subfigure}%
    \begin{subfigure}{0.25\linewidth}
        \includegraphics[width=\linewidth, height = 4.0cm]{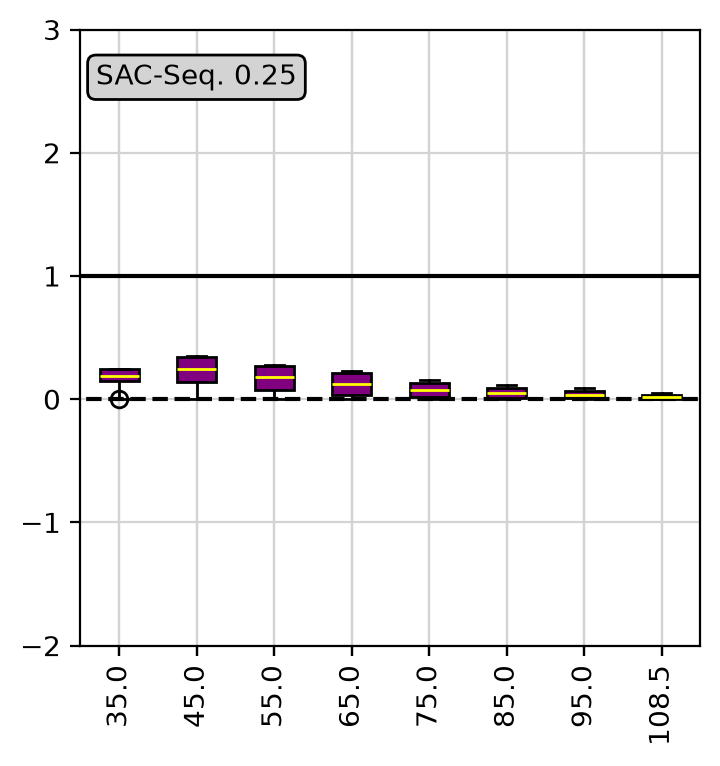}
    \end{subfigure}%
    \begin{subfigure}{.25\linewidth}
        \includegraphics[width=\linewidth, height = 4.0cm]{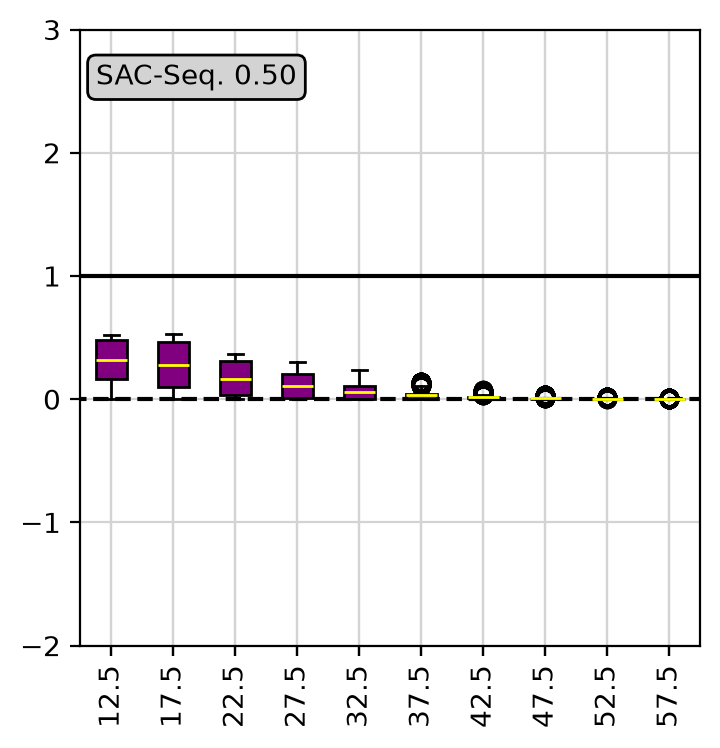}
    \end{subfigure}%
    \begin{subfigure}{0.25\linewidth}
        \includegraphics[width=\linewidth, height = 4.0cm]{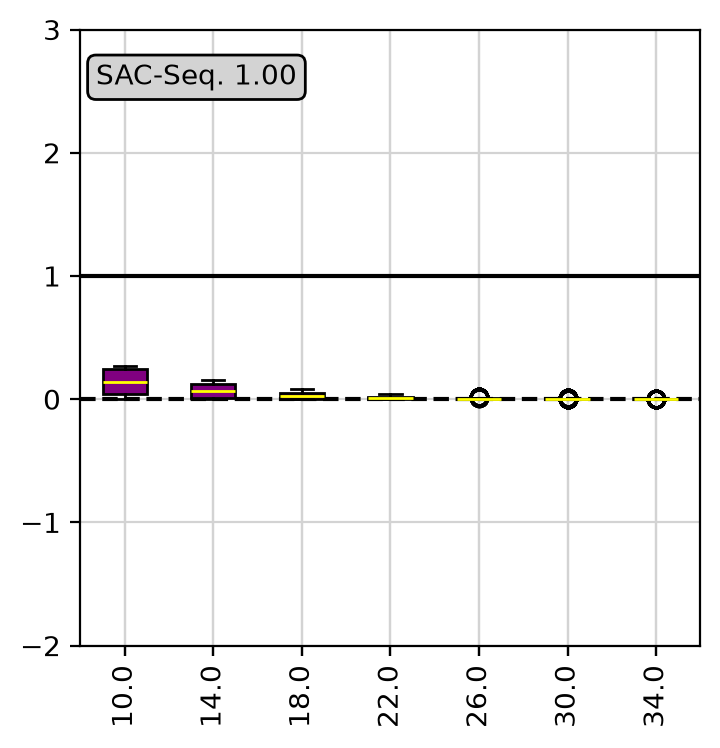}
    \end{subfigure}
    \caption{The ranges of estimated concentrations from each approach, normalized by the solubility limit. The back-calculated concentrations around each cluster for the CPC approach are shown in red, and for the NPC approach in blue. The pre-calculated concentrations used in the SAC approach are in purple. The dashed black line indicates where $\frac{C_{i,j}}{C_{\text{sol}}} = 0$, the solid black line indicates where $\frac{C_{i,j}}{C_{\text{sol}}} = 1$. Plots are truncated and exclude outliers from the CPC and NPC approaches.}
    \label{fig:conc_grads}
\end{figure*}

A motivation behind the development of the three approaches discussed in the current study is the inability to resolve the aqueous concentration of the analyte gas with the same spatial resolution as we can for the gas clusters and their respective interfaces. The research to cross-verify these concentration estimates in 3D does not currently exist, and thus it is unknown how the assumptions to estimate the mass transfer coefficients affect the accuracy of the estimated concentrations. The normalized aqueous concentrations ($\frac{C_{i,j}}{C_{\text{sol}}}$) for all approaches are given in Figure \ref{fig:conc_grads}. Given a constant gas solubility, it is expected that all real concentrations would fall between $0 \leq \frac{C_{i,j}}{C_{\text{sol}}} \leq 1$, where one indicates 100\% saturation of solute, and zero indicates no dissolved solute is present. Unlike the mass transfer coefficient, the aqueous concentration is a localized value in space and time (for an unsteady state and unmixed system), and thus there is little value in describing the aqueous concentration profile as a bulk system-level value. As such, while we have displayed the $\frac{C_{i,j}}{C_{\text{sol}}}$ values as box-and-whisker plots, the value of each plot will come from the range (the values of the tails and the 1st and 3rd quartiles) of aqueous concentrations within each.\par
\subsubsection{Classified Per Cluster Approach}
The CPC approach back-calculates the local concentration using the previously estimated $K_{\text{ave}}^{\text{seq}}$,  Eq. (\ref{eq:indi_con_grad}), and the volume change of clusters that partially dissolved in the observed time-interval. In actuality, each partially dissolved cluster has its own mass transfer coefficient ($k^{C*}_{i,j} \text{, where }  k^{C*}_{i,j} \in \mathbb{N}(k^{*}_{i,j})$) associated with the observed volume change. $k^{C*}_{i,j}$ cannot be calculated from the partially dissolved clusters, as the CPC approach assumes that $C^{gas}_{i,j} \approx 0$ is only true for the concentration around the completely dissolved clusters. However, the the CPC approach posits that $K_{\text{ave}}^{\text{seq}}$-CPC may approximate $k^{C*}_{i,j}$, if $K_{\text{ave}}^{\text{seq}}$-CPC accurately approximates the expected value of $\mathbb{N}(k^{*}_{i,j})$.\par
The negative concentration values ($\frac{C_{i,j}}{C_{\text{sol}}} < 0$) in Figure \ref{fig:conc_grads} rows (1) indicate clusters for which $K_{\text{ave}}^{\text{seq}}\text{-CPC} < k^{C*}_{i,j}$, i.e., the sequence average is under-representative of some individual clusters. The increasing amount of negative $\frac{C_{i,j}}{C_{\text{sol}}}$-values with increasing injection rate, in Figure \ref{fig:conc_grads} rows (1), indicates that small, completely dissolved clusters in the mass transfer coefficient estimation are likely responsible for $K_{\text{ave}}^{\text{seq}}$-CPC under-representing some partially dissolved clusters.\par
\begin{figure*}
    \centering
    \begin{subfigure}{0.48\linewidth}
      \includegraphics[width=\linewidth]{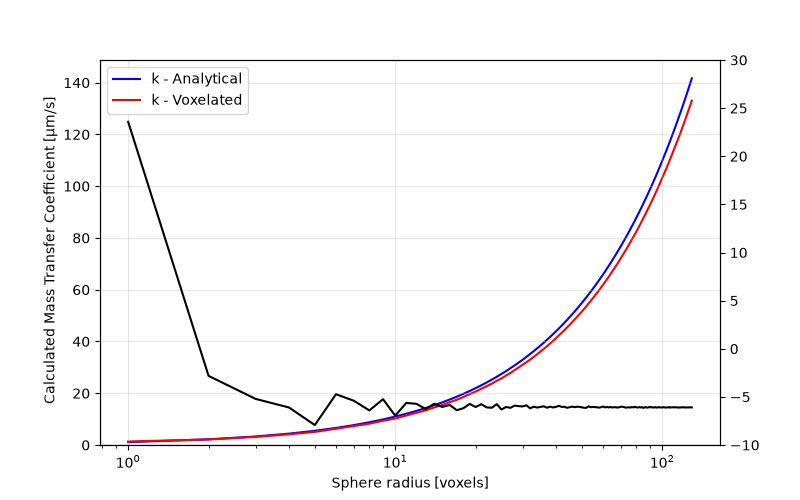} 
    \end{subfigure}
    \begin{subfigure}{0.50\linewidth}
        \includegraphics[width=\linewidth]{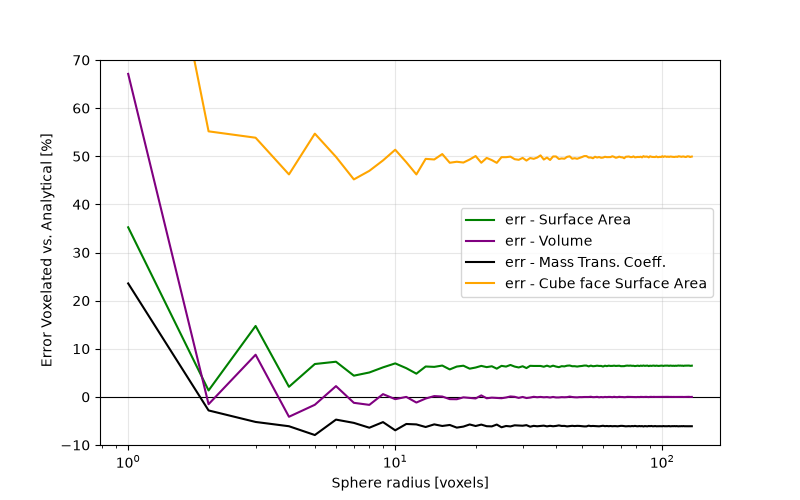}   
    \end{subfigure}
    \caption{(left) Mass transfer coefficients estimated from spheres of radius ($r$), using analytical geometries (red) and geometric approximations (blue) of the same voxelated spheres, and the absolute error (black) between the resulting mass transfer coefficients. (right) Zoomed out, comparison of the separate errors from geometric voxelization for surface area (Crofton Formulation: green, Exposed Cube Surface: orange), volume (purple), and mass transfer coefficient (black), only the Crofton values were used in the mass transfer estimates.}
    \label{fig:pixel_err}
\end{figure*}

As discussed in the previous section, measuring continuous geometries from a discrete domain introduces an inherent error, where values may be under- or overestimated due to pixelation of the object, and the stringency of the segmentation algorithm. In Figure \ref{fig:pixel_err}, we generated spheres with different radii and voxelized them to compare exact analytical geometries (i.e., surface area and volume) with geometric approximations. For comparison, we implement both naive estimation techniques (surface area: exposed cubic face counts) and common, higher-accuracy estimation techniques (volume: voxel counts; surface area: Crofton formulation \citet{crofton_form}) to compare with the exact analytical values. The error trend shows that all estimation techniques converge to a stable asymptotic value as the spherical radii increase, indicating the inherent error in any geometric approximation. For the same geometric approximation, we observe higher error variability at smaller radii, fluctuating above and below the stable value by several percent to fractions of a percent. The oscillation comes from the binary categorization of voxels (segmentation) as either sphere or void, and from truncating any voxels that would otherwise be included in the analytical volume. While smaller clusters may inherit more geometric error, the errors in the mass-transfer estimates from those smaller geometries are not substantially larger than the stable error of the larger geometries, to result in the underestimation of $K_{\text{ave}}^{\text{seq}}$-CPC on their own. The main reason behind the underestimation is the time interval; all the spherical volume changes in Figure \ref{fig:pixel_err} are over the same interval, implying that the smaller clusters took just as long to dissolve as the largest clusters, which yields a lower $k_i$-value $\approx$ 1 to 2 orders of magnitude smaller than the larger volumes, should both exist in the observation window. \citet{Ruotong_H_2023} implemented a surface-area weighted average to reduce the influence of small clusters on the final estimate; however, this does not prevent a large population of smaller clusters from shifting the final average $K_{\text{avg}}^{\text{seq}}$-CPC. For example, in Seq 1.00 $K_{\text{ave}}^{\text{int}}$-CPC decreases with time (Figure \ref{fig:mass_trans_dist}, row 4). The observed decline in the $K_{\text{ave}}^{\text{int}}$-CPC is a strong indication that smaller, completely dissolved clusters skew the weighted average of $\mathbb{K}_{CPC}$to a lower $K_{\text{avg}}^{\text{seq}}$-CPC values and result in $K_{\text{avg}}^{\text{seq}}$-CPC under representing $K^{C*}_{i,j}$. This is why the $\frac{C_{i,j}}{C_{\text{sol}}}\text{ range}< 0$ are more prevalent in CPC-Seq 1.00 that the other sequences.\par
\subsubsection{Non-Classified Per Cluster Approach}
The dilute concentration assumption ($C_{i,j} = 0$) is the primary difference between the NPC and CPC approaches, and applying the resulting NPC estimates to calculate concentration profiles is thus an internal inconsistency. However, the NPC-derived concentration field provides a means to evaluate how $K_{\text{mean}}^{\text{seq}}$-NPC approximates $\mathbb{N}(k^{*}_{i,j})$. From Figure \ref{fig:conc_grads}, row (2), we observe that the NPC approach yields both $\frac{C_{i,j}}{C_{\text{sol}}} < 0$ and $\frac{C_{i,j}}{C_{\text{sol}}} > 0$. We have already discussed that $\frac{C_{i,j}}{C_{\text{sol}}} < 0$ indicates that $K_{\text{mean}}^{\text{seq}}$-NPC under-approximates $k^{C*}_{i,j}$; thus, evaluating Eq. (\ref{eq:indi_con_grad}), we find that $\frac{C_{i,j}}{C_{\text{sol}}} > 1$ will only occur when the change in cluster volume is positive (i.e., the cluster grew in size), given that all other parameters are positive. Explicitly, the CPC approach uses only decreasing-size clusters to calculate concentration and thus cannot yield an oversaturated concentration value; in contrast, the NPC approach uses all volume changes. From the previous sections, we also found that clusters that grew in size yielded negative mass transfer coefficients, further complicating any meaningful interpretation of the NPC concentrations. Ultimately, as expected, the means of all intervals approach or equal zero, indicating, despite the non-physical concentration estimates, that the calculation is at least self-consistent.
\subsubsection{1-D Sliced Averaged Approach }
The concentration values from the SAC are forward calculated, based on the change in saturation within a slice, added to the concentration of the slice before it, using the 1-D advection equation (Eq. \ref{eq:pat_conc_x}). The different ranges of normalized concentrations produced by the SAC approach and the per-Cluster approaches stem from the different spatial domains of the underlying equations. The per-Cluster approaches assess the concentration of the bulk advecting fluid across the stagnant film surrounding each cluster; no supposition is made about the volume or region of fluid over which these concentration values exist. In contrast, the dimensional reduction of the 1-D advection model in the SAC approach has the effect of diluting the solute concentration over the entirety of each cross-section. The dimensional reduction (assumption of radial uniformity) dilutes the concentration estimates, as it assumes that all solvent voxels in the cross-section contribute equally to dissolving the lost mass, rather than that the dilution occurs in diffusion-limited layers, effectively treating the system as well-mixed in the cross-section. From the forward propagation of the axial concentrations, starting from the initial boundary condition $C(x=0, t=0) = 0$, the dilute concentrations additively propagate to the next slice. Further, the SAC model does \textbf{not} account for the saturated solvent ($\frac{C_{i,j}}{C_{\text{sol}}} = 1$) that initially occupies the space around the trapped clusters; rather, the accompanying small amounts of mass dissolving in these saturated regions are also diluted over each slice.\par
\subsection{Comparison of Approaches}
The SAC approach only requires segmenting the wet-scans into 2 phases before iteratively processing the data in time as 2-D slices (Figure \ref{fig:img_pro_pipes}). The iteration supplements much of the image processing required in the other approaches through the 1-D advection model (Eq. \ref{eq:pat_conc_x}), which allows the SAC approach to circumvent unknown system variables by assuming knowledge of the system (advection dominance). A consequence of this simplification is that values from the SAC approach are limited to a macroscale understanding of the system, which is useful when estimating more uniform/generalizable system parameters, such as the mass transfer coefficient. Less generalizable finer-scale parameters, such as the concentration profile or cluster remobilization, are inadvertently averaged out or lost during the simplification to larger-scale geometries. The 1-D advection model additionally requires that the observation-window boundaries align with boundary conditions (i.e., that $C(x^*,t) = 0$), so that portion must be usable data and not be cropped.\par
The per-Cluster approaches forgo knowledge of the macroscale process in favor of focusing on microscale process information contained within the image data, though they require statistics to obtain more uniform/generalizable parameters, such as the mass transfer coefficient. As published by \citet{Lv_2024}, the NPC approach required image registration, 2-phase wet-scan segmentation, and cluster labeling and matching. As implemented in the current study, the NPC approach required image registration, dry-scan 2-phase segmentation, wet-scan 2-phase segmentation, and 3-phase segmentation, as well as cluster labeling and matching. As such, the lower computational cost of the NPC approach as originally published is not emphasized in the current study, however, given that the approach estimated mass transfer coefficients similar to those of the SAC and CPC. The trade-off of the somewhat simpler NPC workflow is that it is similarly unable to resolve fine-scale phenomena, such as the aqueous concentrations or cluster remobilization, though for different reasons. The NPC does not average out the information contributed by each cluster, but by not distinguishing volume changes by event class, much of the microscale information is lost in the large number of clusters and morphology changes.\par
The CPC approach (which requires additional cluster classification, mobilization filtering, and iterative refinement) has the highest computational cost among the approaches outlined, at the benefit of preserving finer transient phenomena, such as non-uniform dissolution fronts arising from dissolution fingering \citep{Patmonoaji_2021, Ruotong_H_2023}. Figure \ref{fig:con_front} is a simple visualization of some of these phenomena that were preserved, reconstructed from data across different change-cluster groups, much of which is otherwise not captured by our subsequent statistical descriptions of the system or is lost in simplifications of the other approaches. However, due to the numerous and dense nature of trapped clusters in porous networks, the 3-D visuals presented do not readily lend themselves to anything beyond qualitative interpretation.\par
Figure \ref{fig:con_front}(a) visualizes cluster remobilization in column 1, where we observe distinct, concurrent stratification between the two cluster groups. The gained volume (red, scaled volume gained) and the ``lost'' volume (blue, scaled by the volume lost) are on distinct sides of the field of view, but loosely scattered within their respective regions. In the neighboring ( 2nd to 4th columns) visualizations not flagged for mobilization, substantially less stratification is present, and the ``lost'' clusters densely populate their respective regions.\par
Figure \ref{fig:con_front}(b)  shows only the completely dissolved clusters, but now each cluster is scaled by the mass transfer coefficient value obtained from that cluster. Note the apparently uniform propagation of the completely dissolved clusters in the direction of flow, suggesting that these clusters exist in a specific condition, the dissolution front.\par
Lastly, Figure \ref{fig:con_front}(c) shows the normalized aqueous concentration values, plotted at the centers of the geometry of the partially dissolved clusters. While $\frac{C_{i,j}}{C_{\text{sol}}} \text{-values } < 0$ are fictitious, Figure \ref{fig:con_front}(c) illustrates that the CPC classified clusters with  $C_{i,j} <0$ (blue) are localized to a moving region, rather than distributed throughout the field of view. The moving region of blue clusters aligns with the leading edge of the reconstructed cloud of completely dissolved clusters (Figure \ref{fig:con_front}.b), suggesting that the leading edge of the dissolution front is increasingly underestimated as the injection rate increases. Note the spatial heterogeneity in concentrations that propagates in the direction of flow, where the partially dissolved clusters are downstream of the injection inlet, while the completely dissolved clusters remain upstream, but only in the time-interval that passes the mobilization threshold (i.e., not column 1).\par 
\newcommand{\scale}{.75\linewidth}
\begin{figure*}[h!]
    \centering
        \rule{\scale}{0.25pt}\\ 
    a)\includegraphics[width=\scale]{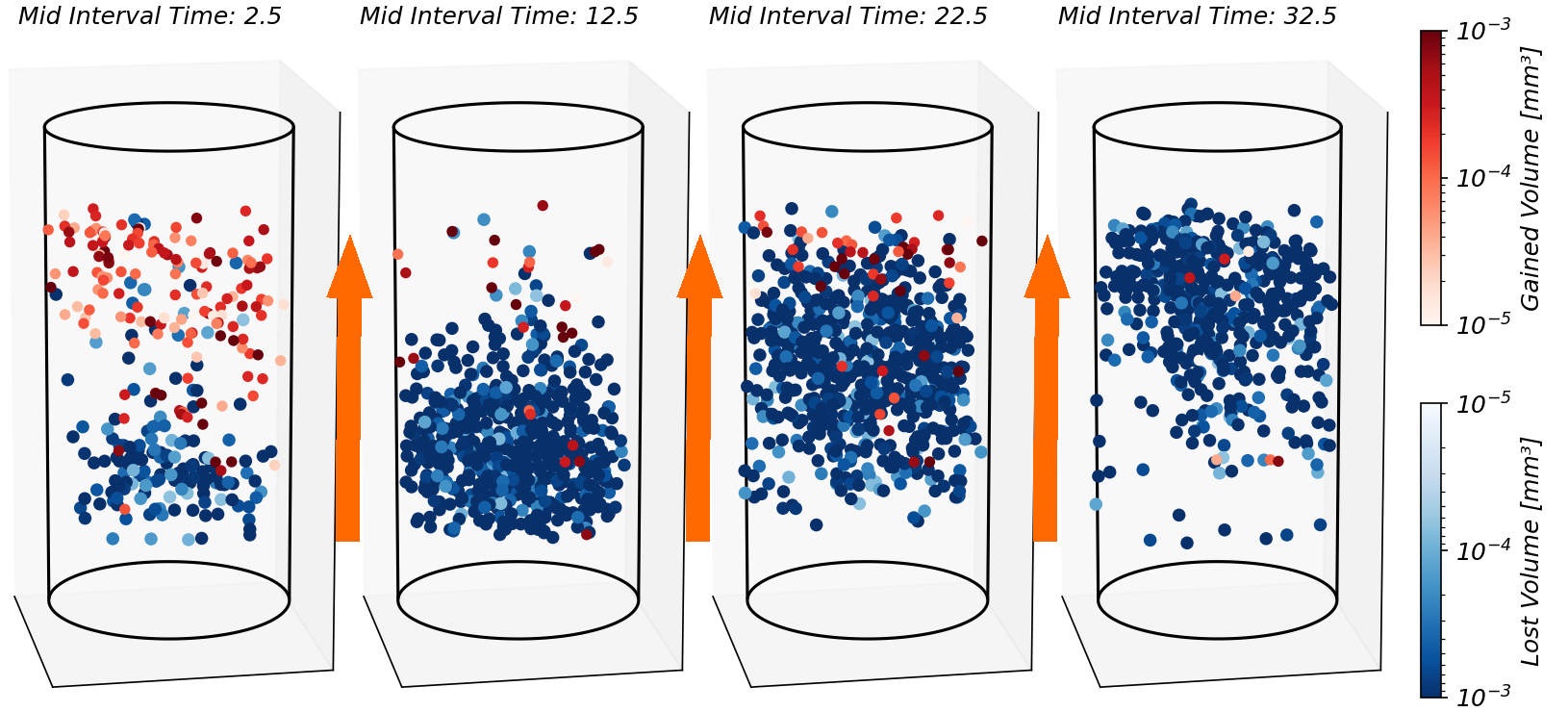}\\
        \rule{\scale}{0.25pt}\\ 
    b)\includegraphics[width=\scale]{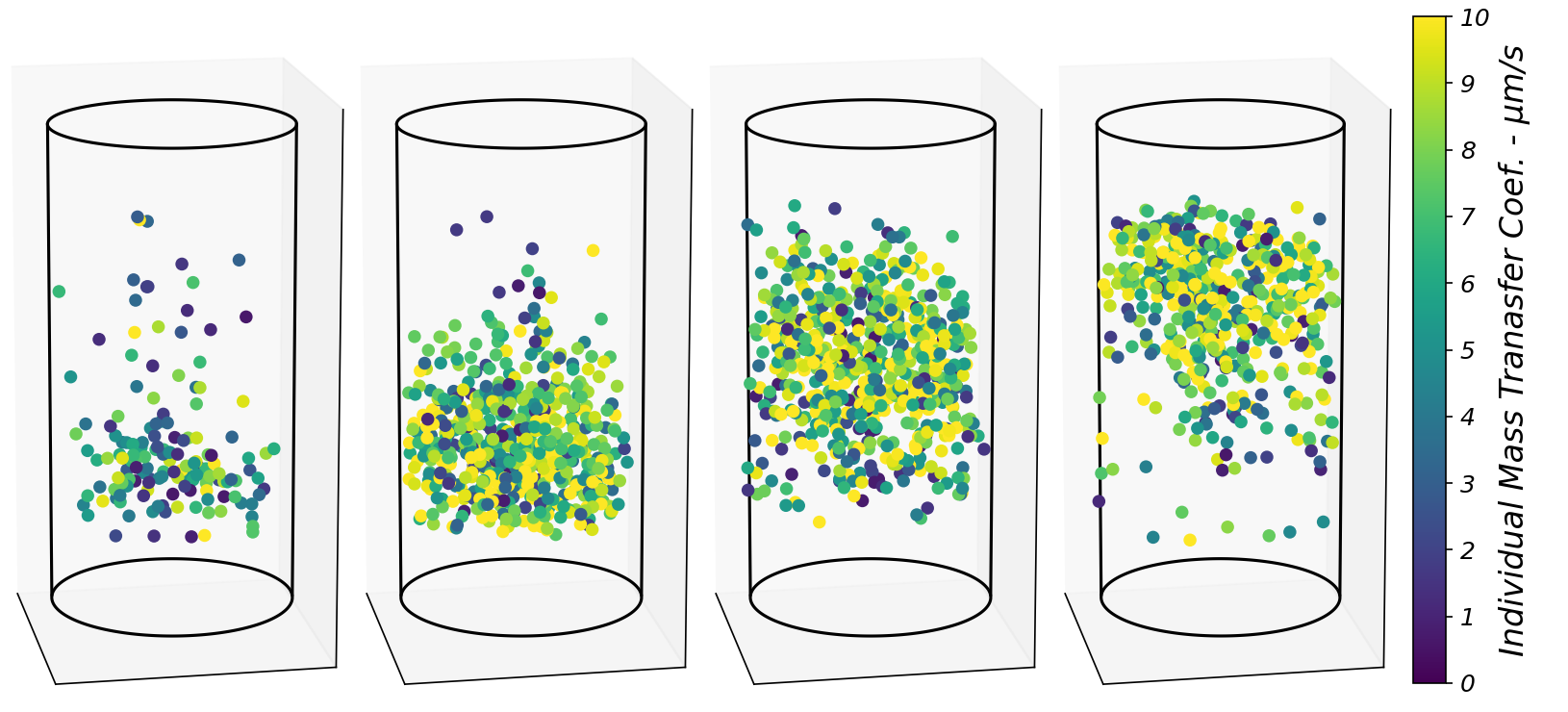}\\
        \rule{\scale}{0.25pt}\\
    c)\includegraphics[width=\scale]{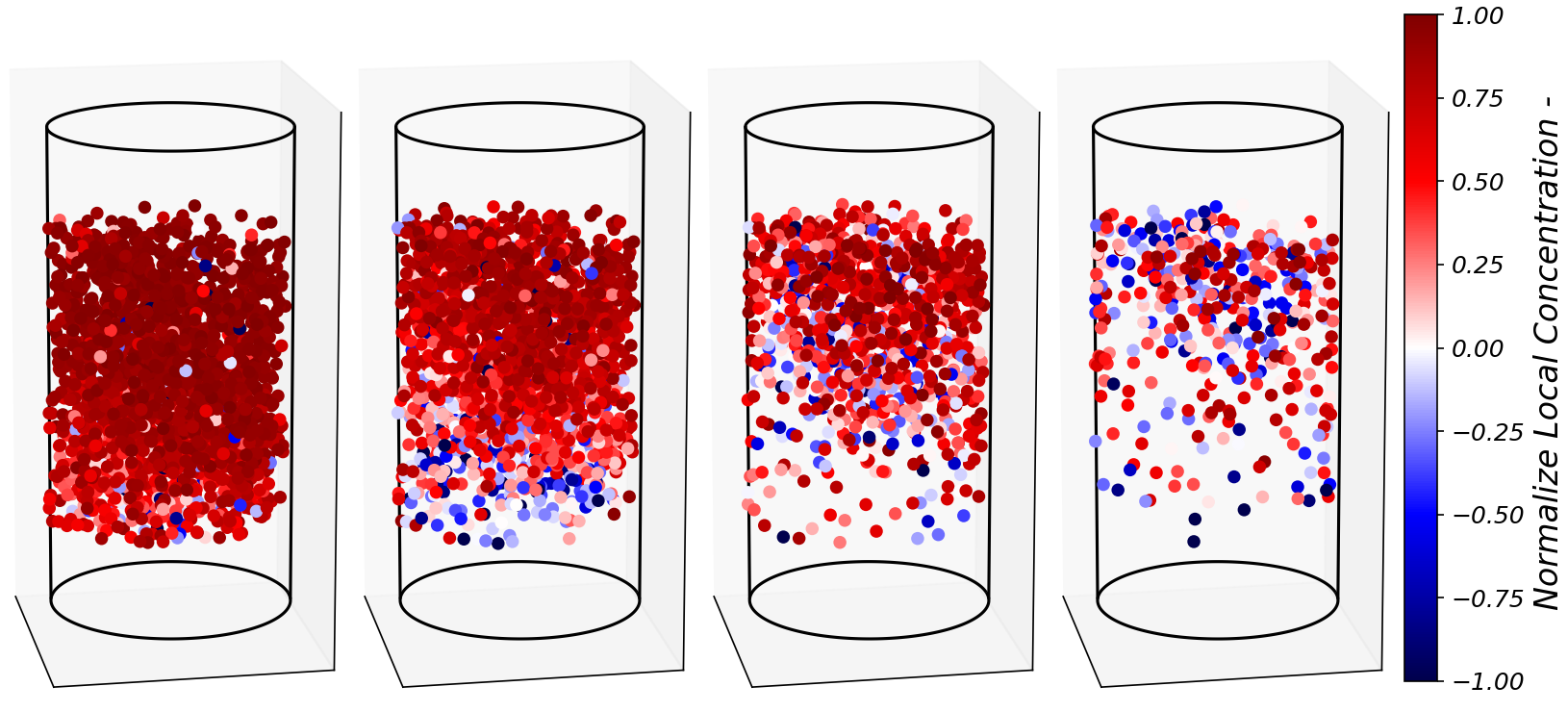}\\
        \rule{\scale}{0.25pt} 
    \caption{Time-lapsed reconstructions of cluster positions, from sequence 0.50 mL/min, using the classified point cloud approach. Note: the Z-axis has been stretched (x2) to assist in visualizing the upward movement of the dissolution front. Fresh solvent is injected from the bottom and flows in the direction of the orange arrows in (a). The mid-interval time is at the top in  (a), varying from left to right, but is constant for each column. Each row shows different cluster properties, as indicated by the color bars on the right. Each mid time-interval corresponds to a set of time stamps given in Tables \ref{tab:scans_times} and \ref{tab:interval_times}.}
    \label{fig:con_front}
\end{figure*}

However, the CPC's greater reliance on cluster-scale information increases the approach's sensitivity to the quality of the image data and the fidelity of the image-processing pipeline used to obtain it. This increased quality sensitivity is a side effect of the CPC approach's cluster classification significantly reducing the sampled measurement pool, and poorly resolved clusters potentially having greater weight in the final estimates, as we saw with Seq. 1.00. Though the SAC and NPC approaches may be more resilient to the same image-quality issues, both depend on mass-transfer-related events being the dominant measurement and similar measurement errors being outweighed by the law of large numbers. The SAC's requirement that the boundary condition align with a known point in the image places its own requirements during image acquisition, and dictates that the SAC approach cannot be applied to a ``region of interest'' away from the injection inlet. The NPC approach's reliance on large numbers of events may prove problematic in ``region of interest'' scans where there may be too few clusters in the field of view to balance out the sampling of non-mass transfer related volume changes. Region of interest scans could be used to acquire high resolution subsamples for greater clarity of events that are difficult to resolve at the scale of the entire system. Nonetheless, image quality and sample size will affect all approaches.\par  
Despite differences in assumptions and calculation frameworks, the three approaches estimated mass transfer coefficients within an order of magnitude of each other. These results indicate the choice of analytical approach used to estimate the interphase mass transfer coefficient (a system-scale value) from X-ray $\mu$CT data does not significantly impact (to within an order of magnitude) the $K_{\text{ave}}$-value yielded. The variation increases as the injection rate decreases, and the trend observed in Figure \ref{fig:sherwood} suggests the difference between Sherwood number values (dimensionless $K_{\text{ave}} $-values) may begin to exceed one order of magnitude for Reynolds numbers < 0.016 (Seq. 0.10). We find that the most apparent distinction between approaches arose when calculating pore-scale properties, such as the aqueous concentration field, because the effect that the assumption frameworks have on the individual data points (cluster changes) is no longer averaged out, as it was with the mass transfer coefficient. Thus, for the injection rates and system conditions covered in the H$_2$ data set, the selection of approach should be based on the additional information desired from the system (i.e., the aqueous concentration field, the cluster mobilization, or dissolution path) and the computational resources available to process the image data. The SAC approach required the least image processing and estimated a mass transfer coefficient similar to the CPC and NPC approaches, but could not resolve the other pore-scale phenomena. The NPC approach preserved more pore-scale phenomena and estimated similar mass transfer coefficients, with more image processing than the SAC approach, but considerably less than the CPC. The CPC approach preserved the most pore-scale phenomena of the three approaches, enabling quantification of important localized physicochemical processes such as dissolution fingering and cluster remobilization (Figure \ref{fig:con_front}), but required the most image processing to do so.\par  
\newcommand{\quack}{\dimexpr0.5\linewidth\relax}
\begin{figure*}[htb]
    \begin{subfigure}{\quack}
        \includegraphics[width=\linewidth]{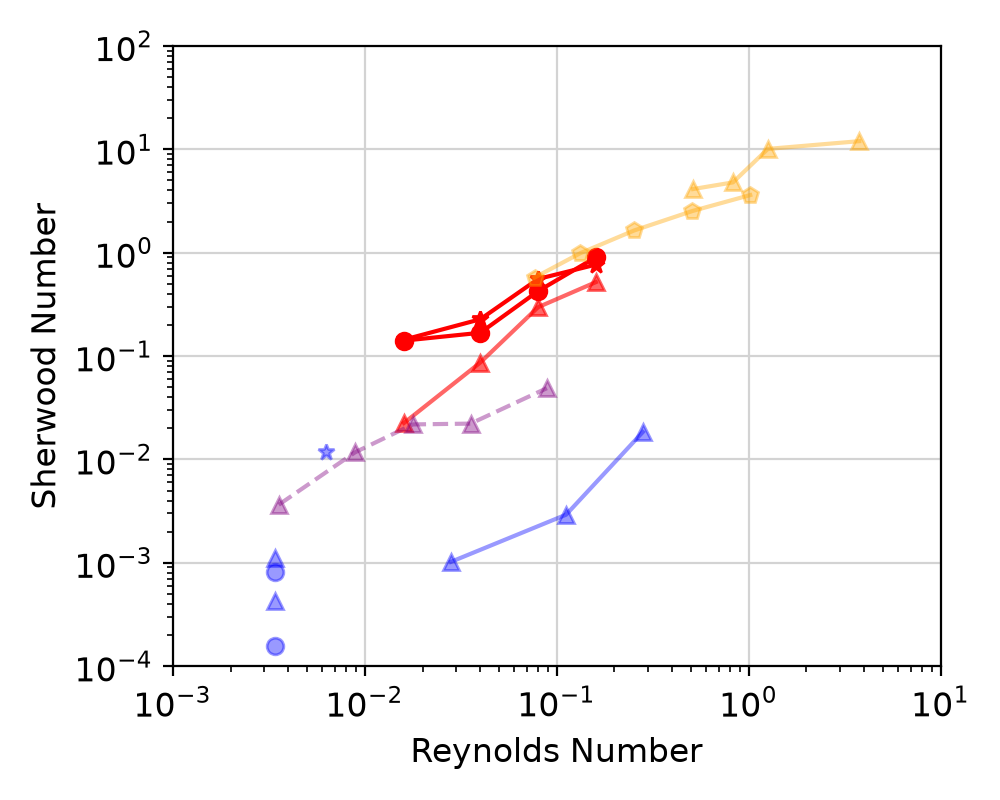}
    \end{subfigure}%
    \begin{subfigure}{\quack}
        \includegraphics[width=\linewidth]{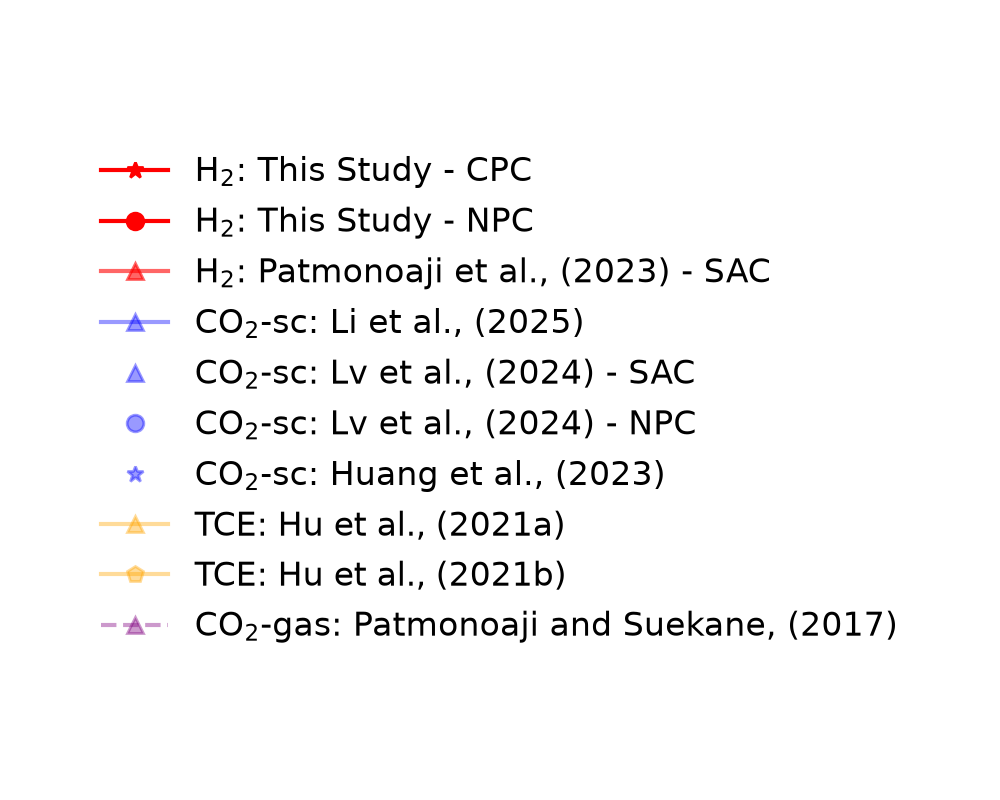}
    \end{subfigure}%
    \caption{Figure \ref{fig:cross_study_plots} (b) updated with the results from the CPC and NPC approaches on the $H_2$ data set.}
    \label{fig:plus_cross_study_plots}
\end{figure*}
 Figure \ref{fig:plus_cross_study_plots} is the plot of other studies evaluating mass transfer in porous media with $\mu$CT, now updated with the data from this study. While the pool of existing studies are still too disparate to interpret whether the spread in values is due to the methods used, or the conditions evaluated in each experiment; however, there are discernible patterns in the dimensionless trends that are more apparent with our comparison data included. First, the H$_2$ (the current study and \citet{Patmonoaji_2023}) and Trichloroethylene (TCE, \citet{Hu_2021a, Hu_2021b}) data exhibit a nearly linear trend. Notably, these studies are conducted at atmospheric conditions and in a grain pack. The gaseous CO$_2$ data by \citet{Pat_Suek_2017}, also conducted at atmospheric conditions and in a grain pack, loosely follow the same log-log linear trend as the H$_2$ and TCE data at the lower Reynolds number injections, but deviate slightly at the higher Reynolds numbers. The studies using supercritical CO$_2$ exhibit significantly more spread than other studies with a common analyte; notably, the Sherwood numbers from sand pack data by \citet{Li_2025} are an order of magnitude lower than the gaseous CO$_2$ data from \citet{Pat_Suek_2017}. Despite operating at similar injection rates and in granular packing, the increased discrepancy between the estimates from \citet{Pat_Suek_2017} and \citet{Li_2025} indicates that more study will be needed to delineate the effect of the estimation approaches on systems at different thermodynamic conditions, to understand if these differences are method artifacts or real variation.\par
\section{Conclusion}
\label{sec:conclusion}
\textit{In situ} measurement of interphase mass transfer properties in porous media is challenging due to dependencies on gas-solvent geometries and properties, the potentially complex distribution of the aqueous analyte concentration, and the tortuous, opaque nature of porous media. Measuring mass transfer via X-ray $\mu$CT enables measurement of distinct phase geometries in porous media, but with accompanying limitations in what can be resolved in space and time, such as the aqueous concentrations of many solutes of interest, a key factor in mass transfer calculations. Multiple analytical approaches have been developed to leverage information from time-lapsed sequences of X-ray $\mu$CT scans to estimate interphase mass-transfer properties, employing various models and assumption frameworks to address unknown variables.\par
At the time of the current study, no other studies have evaluated the physical implications of specific conditions on the mass transfer estimated by existing assumption frameworks. \citet{Lv_2024} is the only other study to use multiple analytical approaches on the same time-lapsed µCT data set. Having used the common Slice-Averaged Concentration (SAC) approach \citep{Pat_Suek_2017, jiang_2017, Hu_2021a, Hu_2021b, Patmonoaji_2021, Patmonoaji_2023, Lv_2024, Li_2025} and their own Non-Classified per-Cluster (NPC) approach \citet{Lv_2024} found that both approaches yield estimates within an order of magnitude, and showed that the difference between approach estimates differed with the porous media, even under the same experimental conditions a phenomenon similarly observed by \citet{Patmonoaji_2021}. Given that the existing body of studies using X-ray $\mu$CT to measure interphase mass transfer properties in porous media (Table \ref{tab:cross_study}) spans highly dissimilar combinations of porous media and experimental conditions, the three analytical approaches must first be evaluated on the same data set.\par
The current study compares the results of three analytical approaches on the same set of experimental image data observing injection-driven gas dissolution across multiple solvent injection rates. The data used were originally obtained by \citet{Patmonoaji_2023}, who utilized time-lapsed $\mu$CT to evaluate dissolution-driven interphase mass transfer of several gases trapped in a water-filled porous medium using the SAC approach. The current study reprocessed the four time-lapse sequences in the hydrogen gas (H$_2$) data set using the NPC approach of \citet{Lv_2024} and the Classified Per Cluster (CPC) approach of \citet{Ruotong_H_2023}. The current study focused on the different assumptions and frameworks underlying each approach: from simplifying the spatial domain to a one-dimensional representation to assuming that every observed morphology change is related to mass transfer. To evaluate these assumptions, we look at key mass-transfer properties, such as the system-scale interphase mass-transfer coefficient and the locally distributed aqueous solute concentration.\par
In estimating the mass transfer coefficient, the current study found that for the injection rates and experimental conditions evaluated all three approaches yielded estimates within one order of magnitude of each other. While the estimates remain within an order of magnitude, we observed a clear distinction between the SAC and per-Cluster approaches as a function of injection rate. Linearly, the difference between the SAC and the per-Cluster mass transfer estimates increased with the injection rate; while proportionally, the differences decrease as the injection rates increase.\par
The differences between the approaches became most apparent when comparing the aqueous concentration profiles, a pore-scale parameter. The SAC approach yielded concentration estimates far more dilute than expected, which is attributed to the cross-sectional averaging required to simplify the flow model. The NPC approach produced highly variable distributions that exceeded the expected bounds and were difficult to interpret, and doing so contradicts the initial assumption key to the NPC approach, that $C^{gas}_i \approx 0$ throughout the system. The CPC approach produced a range of concentrations closer to the expected bounds, but also yielded nonphysical, negative aqueous concentrations due to application of an average value ($K_{\text{ave}}$) to all clusters, including those in regions of higher individual values ($k_i$). From the spatial plots of the aqueous concentrations, we see that the negative values were localized to a region upstream to the leading edge of the injected solvent front --i.e., where quickest dissolving clusters are.\par 
From these analyses of different scale mass transfer properties, the choice of analytical approach was found to be a balance between the amount of image processing required and the level of detail desired. The SAC uses a 1-D advection model to compute a linear concentration profile for each time-interval, resulting in mass transfer values that are intrinsically tied to the macroscopic scale of the process model. When the full field of view can be used, the SAC approach is less computationally intensive and yields mass transfer coefficient estimates of the same order of magnitude as the other approaches. However, the dimensional simplifications of the SAC limit the resolution of the results to more macroscale details, as other microscopic details are inadvertently averaged over the geometry; the model's boundary condition requires the field of view to align with the injection inlet, preventing cropping of region-of-interest evaluations without prior knowledge of the inbound concentrations.\par
The NPC and CPC approaches both assess changes at the level of individual solute clusters to approximate the conditions of the bulk advecting fluid surrounding each cluster. The NPC approach is unique in that it can be applied as published by \citet{Lv_2024}, which is less computationally intensive, or as applied in the current study as a sister process to the CPC \citep{Ruotong_H_2023} approach, which is more computationally intensive. The NPC approach applies a system-wide dilute aqueous concentration assumption ($C^{gas}_i \approx 0$), which is the pivotal distinction between the NPC and the CPC approaches and changes how each per-Cluster approach addresses morphology changes observed in a sequence to approximate system conditions. The NPC approach, similar to the SAC approach, treats all morphology changes as potential instances of mass transfer and assumes that enough of the observed volume changes are due to mass transfer for the sample distribution to be representative of the population distribution via the law of large numbers. The bulk use of all morphology changes makes NPC approach estimates robust to outlier values within the observed time-intervals, erroneous cluster measurements, and shifts in system dynamics, at the cost of high variance, negative mass-transfer values, and reduced resolution of distinct phenomena within the system dynamics. Despite the relaxation of data sampling (cluster classification) and the greater variability in the sample distributions, the mean NPC mass transfer coefficients are nearly indistinguishable from the weighted average CPC estimates for the evaluated conditions.\par 
The CPC approach significantly limits the sampled morphology changes via classification, and time-interval filtration, to events that potentially approximate specific mass transfer conditions. The classification of events by observed volume change requires assurance that mass transfer is the dominant phenomenon, rather than cluster remobilization, and flagged time-intervals must be excluded from the final calculations. The CPC approach significantly reduces the measurement pool of observed events to reduce the uncertainties in the system, given the limitations of $\mu$CT resolution in time and space, at the cost of being less robust to outlier events and erroneous measurements. The benefit of all of the image processing and data filtration required by the CPC approach is that the classified morphology changes enable visualization of complex small-scale phenomena, such as the non-uniform propagation of the injected solvent or the cluster remobilization used to filter the data, imaged information not preserved or readily apparent with other approaches.\par
Our results provide guidance for selecting an analytical approach capable of evaluating interphase mass transfer phenomena in porous media at various scales, given the constraints inherent in $\mu$CT-based imaging and limited computational resources. For the conditions evaluated in this study, estimates of system-scale properties such as the mass transfer coefficient are similar within an order of magnitude. However, resolving pore-scale properties such as the aqueous concentration field or cluster remobilization requires a more granular approach, such as the CPC.\par


\section{Credit authorship statement}
\textbf{Christopher A. Allison}: Conceptualization, Methodology, Investigation, Visualization, Writing - Original Draft, Review \& Editing.
\textbf{Anindityo Patmonoaji}: Conceptualization, Data Curation, Writing - Review \& Editing.
\textbf{Ruotong Huang}: Methodology, Writing - Review \& Editing.
\textbf{Lydia Knuefing}: Software, Writing - Review \& Editing.
\textbf{Anna L. Herring}: Conceptualization, Resources, Supervision, Project administration, Writing - Review \& Editing, Funding acquisition. 

\section{Declaration of Competing Interests}
The authors declare that they have no known competing financial interests or personal relationships that could have influenced the work reported in this paper.

\section{Acknowledgments}
\begin{itemize}
    \item C. Allison and A. Herring acknowledge support by the U.S. Department of Energy, Office of Science, Office of Basic Energy Sciences, Geosciences program under Award Number: DE-SC0025400.
    
    \item We gratefully acknowledge the use of WebMango, a web-based image processing platform hosted and supported by the Department of Materials Physics at the Australian National University, with assistance of resources from the National Computational Infrastructure (NCI Australia), an NCRIS-enabled capability supported by the Australian Government.

    \item We gratefully acknowledge the help of Undergraduate Research Assistant Katie Bush in processing the tomogram data and refining our image processing workflow.
\end{itemize}


\end{document}


\section{Supplementary Material}
\subsection{Segmentation Criteria}
\label{subsec:seg_critera_meth}
\begin{figure}[!ht]
    \centering
    \includegraphics[width=0.75\columnwidth]{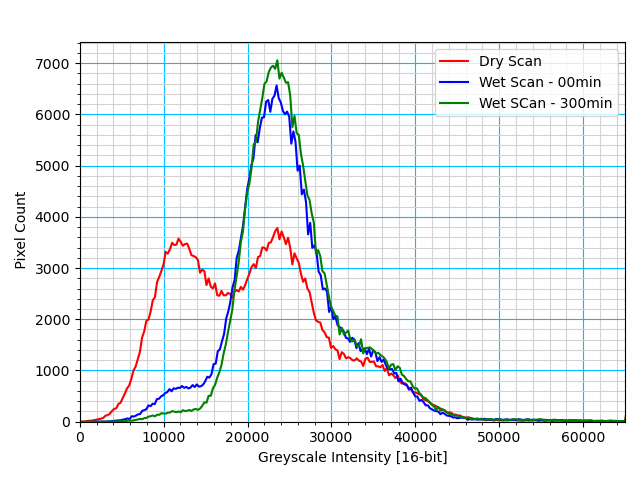}
        \caption{Tomographic 16-bit grayscale intensity histogram from the dry scan (red), 00 min wet scan (blue), 300 min wet scan (green).}
    \label{fig:inten_histo}
\end{figure}
The initial 2-phase segmentation is completed using a Converging Active Contours algorithm \citep{SHEPPARD_2004} threshold. The Converging Active Contours algorithm uses upper and lower intensity thresholds to seed voxels that definitively belong to phase 0 or phase 1 intensity. Intensities less than the lower thresholds are labeled phase 0, intensities above the upper threshold are labeled phase 1, and intensities between the thresholds are left undefined. The algorithm then spatially expands the seeded voxel regions into the undefined regions, as determined by local intensity gradients. This expansion rate is inversely proportional to the local intensity gradient, with regions of larger gradient showing higher contrast and indicating phase boundaries. The initial phase thresholds for the dry scan are selected from the grayscale intensity histogram (Figure \ref{fig:inten_histo}, red histogram). The thresholds were chosen as linearly equidistant values centered on the local minimum, minimizing the output sensitivity as measured by the Euler characteristic of the resultant gas phase. The Euler characteristic ($\chi$) is a topological invariant, measuring the structure and shape of an object, regardless of its size. Ultimately, the optimal thresholds were determined to be the two intensity maxima flanking our initial, local minimum. Erroneous segmentation objects, such as clusters entirely enclosed within the grain phase, were removed from the measurement pool to avoid propagation of error. Additionally, clusters at the edges of the field of view were excluded from the calculations, as they yielded fictitious surface areas relative to the measured volume (i.e., a surface-area-to-volume ratio lower than that of a sphere of equivalent volume).

\subsubsection{3-phase Segmentation}
Here, the gas phase = 0 and the grain phase = 1. The wet scans were segmented into two phases (0 = gas, 1 = else) using criteria similar to those of the dry scan method. Each wet scan is segmented into three phases using a logical separation informed by the segmented dry scan, as shown in Table \ref{tab:3phase_table} (0 = gas, 1 = water, 2 = grain). This method was selected over an additional Converging Active Contours algorithm because it produced significantly fewer false gas clusters and interfaces, due to packing movement and the algorithm's difficulty with overlapping intensities. The ``false'' clusters resulting from the merger method are more easily identified because they lack a wetted surface area (i.e., no contact with the water phase). Investigating the original tomograms, these clusters are real and have wetted surface areas in the wet scans, but exist in a space previously occupied by the grain phase in the dry scan. The choice of logic, as shown in Table \ref{tab:3phase_table}, prioritizes preserving the gas phase in the wet scan, resulting in clusters encased in the grain phase. 

\begin{table}[h]
    \centering
    \caption{Phase arithmetic used in 3-phase segmentation }
    \begin{tabular}{c c|c}\hline
        Dry & Wet & Result \\ \hline\hline
        0 & 0 & 0 \\ \hline
        0 & 1 & 2 \\ \hline
        1 & 0 & 0 \\ \hline
        1 & 1 & 1 \\ \hline
    \end{tabular}
    
    \label{tab:3phase_table}
\end{table}

\section{CPC Surface Area Calculations}
\begin{figure}[htb!]
    \centering
    \includegraphics[width = 0.75\linewidth]{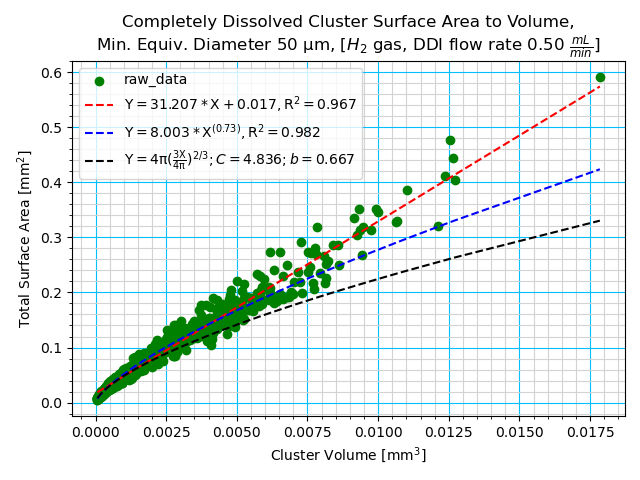}
    \caption{ Sequence 0.50 $\frac{mL}{min}$ }
    \label{fig:vol_TSA_050}
\end{figure}

\begin{figure}[htb!]
    \centering
    \includegraphics[width = 0.75\linewidth]{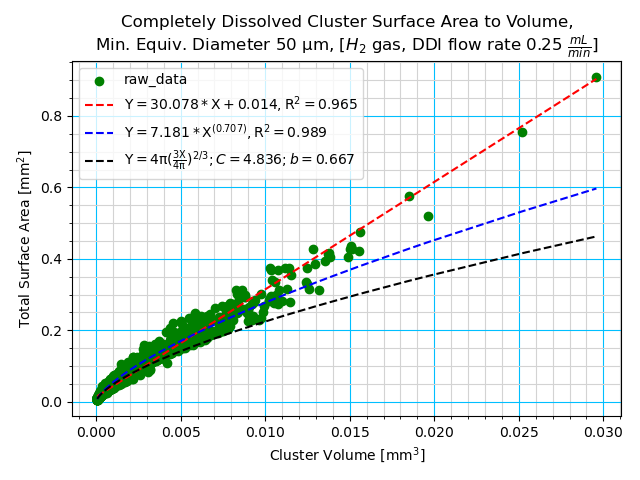}
    \caption{ Sequence 0.25 $\frac{mL}{min}$ }
    \label{fig:vol_TSA_025}
\end{figure}

\begin{figure}[htb!]
    \centering
    \includegraphics[width = 0.75\linewidth]{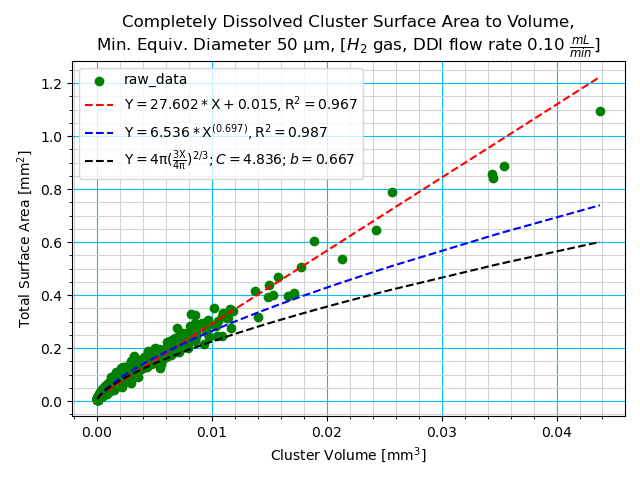}
    \caption{ Sequence 0.10 $\frac{mL}{min}$ }
    \label{fig:vol_TSA_010}
\end{figure}

Next, we do not know how the cluster's fluid-fluid interfacial area and total surface area evolved with respect to each other and time. To solve this, \citet{Ruotong_H_2023} performed the calculation assuming the entire cluster surface is in contact with the solvent. This is assumed to occur when the cluster's volume has decreased by half, in the middle of the time interval. Since we do not know the surface area at this mid-point volume, we develop a relationship between the measured volumes and total surface areas of all completely dissolved clusters in the data sequence; see Figures \ref{fig:vol_TSA_050} to \ref{fig:vol_TSA_010} . Note$^1$: \citet{Ruotong_H_2023} found a best-fit relationship using a power law; this study used a linear fit because it included more data points from large clusters. Note$^2$: We recalculate this relationship for each data sequence.\\ 

The fitted Relationships between gas cluster volume and total surface area. Raw completely dissolved cluster data (green), Linear fit (red), Power-law fit (blue), equation of a sphere's surface area to its volume (black) for reference
\section{Cluster Matching and Classification}
\begin{table*}[htb!]
    \centering
    \caption{Table of Scan time within a sequence}
    \resizebox{\textwidth}{!}{%
    \begin{tabular}{l|p{\wide} p{\wide} p{\wide} p{\wide} p{\wide} p{\wide} p{\wide} p{\wide} p{\wide} p{\wide} p{\wide} p{\wide} p{\wide} p{\wide} p{\wide} p{\wide} p{\wide} p{\wide} p{\wide} p{\wide} p{\wide} } \hline
         scan \# & 1 & 2 & 3 & 4 & 5 & 6 & 7 & 8 & 9 & 10 & 11 & 12 & 13 & 14 & 15 & 16 & 17 & 18 & 19  \\ \hline \hline
        Sequence & \multicolumn{19}{c}{Time of scan in the sequence }\\
        \text{[ml/min]} & \multicolumn{19}{c}{\text{[minutes]}}\\\hline 
        0.10& 00 & 05 & 10 & 15 & 20 & 25& 30 & 40 & 50 & 60 & 80 & 100 & 120 & 150 & 180 & 210 & 240 & 270 & 300\\
         0.25& 00 & 05 & 10 & 15 & 20 & 25& 30 & 40 & 50 & 60 & 70 & 80 & 90 & 100 & 117 & - & - & - & - \\
         0.50& 00 & 05 & 10 & 15 & 20 & 25& 30 & 35 & 40 & 45 & 50 & 55 & 60 & - & - & - & - & - & - \\
         1.00& 00 & 04 & 08 & 12 & 16 & 20& 24 & 28 & 32 & 36 & 40 & - & - & - & - & - & - & -  & - \\ \hline \hline
    \end{tabular}}
     Note: the scan duration and image settings of each scan are constant. This table is also in the main manuscript, but added here for convenience.
     \label{tab:sup_scan_time}
\end{table*}
\begin{table*}
    \centering
    \caption{Table of time-intervals for each Sequence}
    \resizebox{\textwidth}{!}{%
    \begin{tabular}{l|p{\wide} p{\wide} p{\wide} p{\wide} p{\wide} p{\wide} p{\wide} p{\wide} p{\wide} p{\wide} p{\wide} p{\wide} p{\wide} p{\wide} p{\wide} p{\wide} p{\wide} p{\wide} } \hline
         Interval Time & 1 & 2 & 3 & 4 & 5 & 6 & 7 & 8 & 9 & 10 & 11 & 12 & 13 & 14 & 15 & 16 & 17 & 18  \\ \hline \hline
        Sequence & \multicolumn{18}{c}{Mid-interval time}\\ 
        \text{[ml/min]} & \multicolumn{18}{c}{[minutes]}\\\hline 
        0.10 & 2.5 & 7.5 & 12.5 & 17.5 & 22.5 & 27.5 & 35 & 45 & 55 & 70 & 90 & 110 & 135 & 165 & 195 & 225 & 255 & 285\\
        0.25& 2.5 & 7.5 & 12.5 & 17.5 & 22.5 & 27.5 & 35 & 45 & 55 & 65 & 75 & 85 & 95 & 108.5 & - & - & - & - \\
        0.50 & 2.5 & 7.5 & 12.5 & 17.5 & 22.5 & 27.5 & 32.5 & 37.5 & 42.5 & 47.5 & 52.5 & 57.5 & - & - & - & - & -  & -\\
        1.00& 2.0 & 6.0 & 10.0 & 14.0 & 18.0 & 22.0 & 26.0 & 30.0 & 34.0 & 38.0 & - & - & - & -  & - & - & - & - \\ \hline \hline     
    \end{tabular}}
     This table is also in the main manuscript, but added here for convenience.
     \label{tab:sup_int_times}
\end{table*}

Cluster categorization began by matching clusters across scan pairs ``time-intervals'', Tables \ref{tab:sup_scan_time} and \ref{tab:sup_int_times} based on their centers of geometry and volume. The centers of geometry are allowed to vary by a maximum of four voxels, and the cluster volumes are allowed to change by 99\% (both increasing and decreasing in volume). Clusters that merge or fragment into multiple clusters between scans were identified using additional cluster-matching steps, following the process described in \citet{Ruotong_H_2023}. Matched clusters were categorized into five sub-populations based on their volume change. Clusters that decreased in volume by more than 10\% are labeled ``partially dissolved clusters''. Clusters that increased in volume by more than 10\% are labeled ``grown clusters''. Lastly, clusters whose volume changed less than 10\%, regardless of direction, were labeled as non-changed clusters.\par
This matching routine finds multiple clusters that ``converge'' or ``diverge''. Converging matches are when multiple clusters merge in a time-interval, while diverging matches are the reverse: a cluster splits. Converging clusters were found to make up a small percentage of observed events ($\approx 1\%$), while no diverging matches were observed. When categorizing converging and diverging clusters, it is important to account for the total volume merged/separated, as the net volume may increase, decrease, or remain constant. Although converging clusters represented a small fraction of observed events, they accounted for a significant volume change.\par
Non-matched clusters fall into two categories based on the timestamp at which they are observed. Non-matched clusters that appear in the first timestamp, but not the second, are assumed to have completely dissolved in the time between scans, and are labeled ``completely dissolved clusters''. Non-matched clusters that appear only in the later scan are labeled as ``snapped-off clusters''. Snapped-off clusters differ from diverging clusters because they \textbf{do not match} with a cluster in the previous scan, though they are likely the result of diverging clusters and mobilization.\par

\section{Mass Transfer Coefficient Distributions without mobilization filtering}
\begin{figure*} [htb!]
    \begin{subfigure}{0.25\linewidth}
        \includegraphics[width=\linewidth, height = 3.5cm]{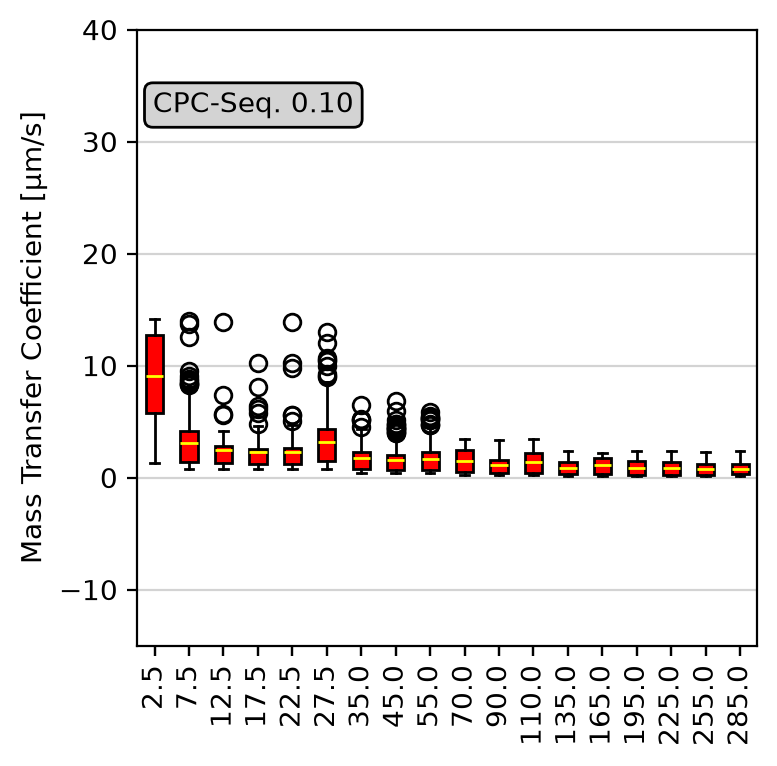}
    \end{subfigure}%
    \begin{subfigure}{0.25\linewidth}
        \includegraphics[width=0.96\linewidth, height = 3.5cm]{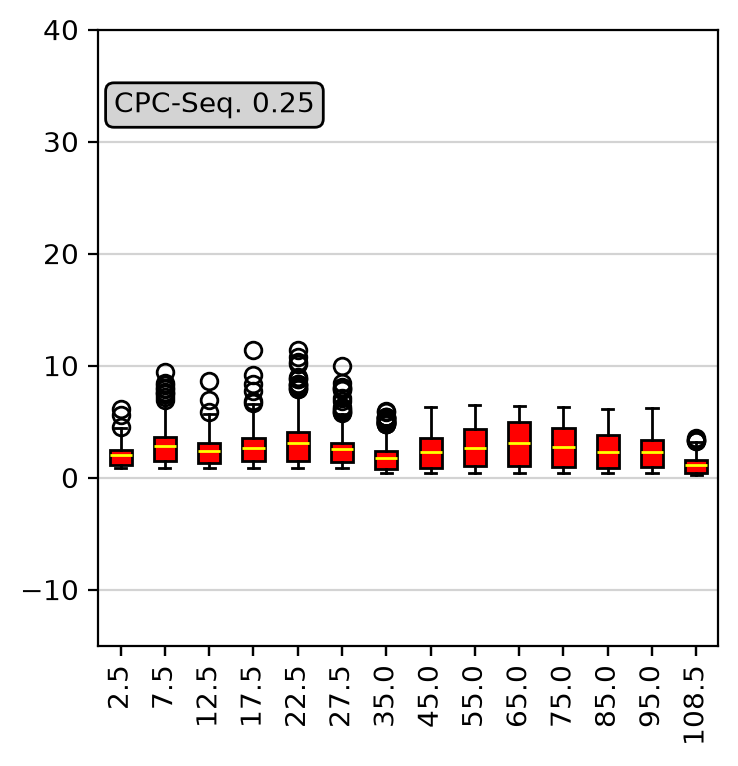}
    \end{subfigure}%
    \begin{subfigure}{0.25\linewidth}
        \includegraphics[width=\linewidth, height = 3.5cm]{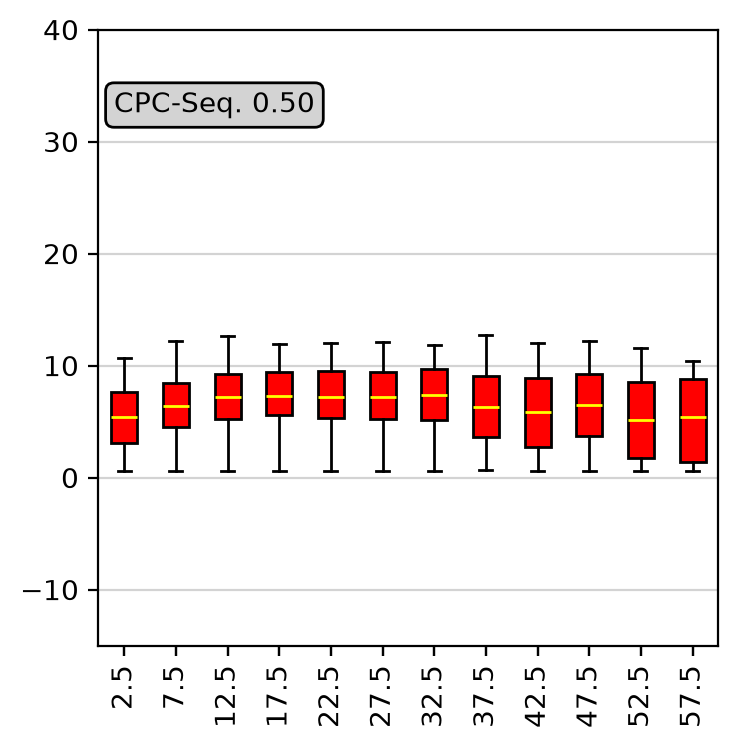}
    \end{subfigure}%
    \begin{subfigure}{0.25\linewidth}
        \includegraphics[width=\linewidth, height = 3.5cm]{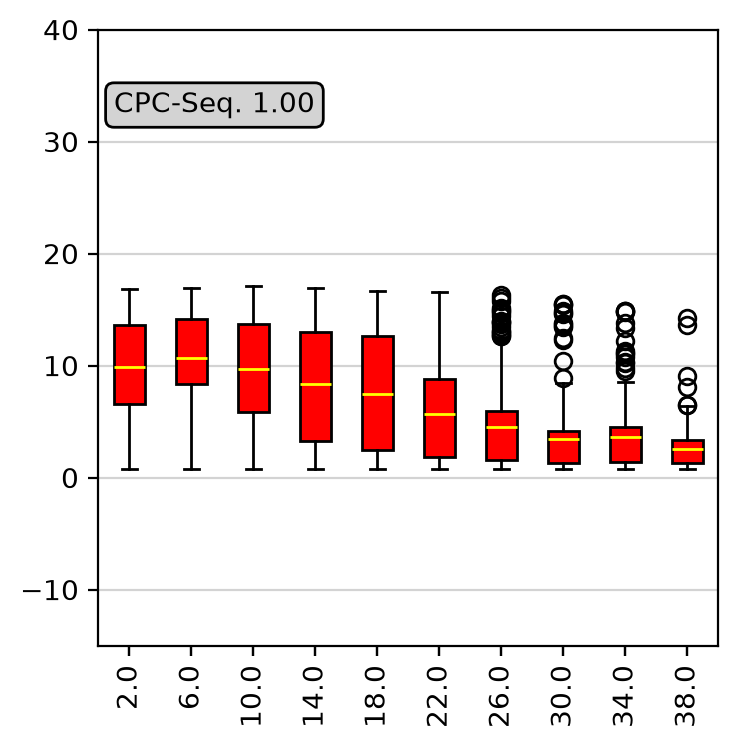}
    \end{subfigure}

    \begin{subfigure}{0.25\linewidth}
        \includegraphics[width=\linewidth, height = 3.5cm]{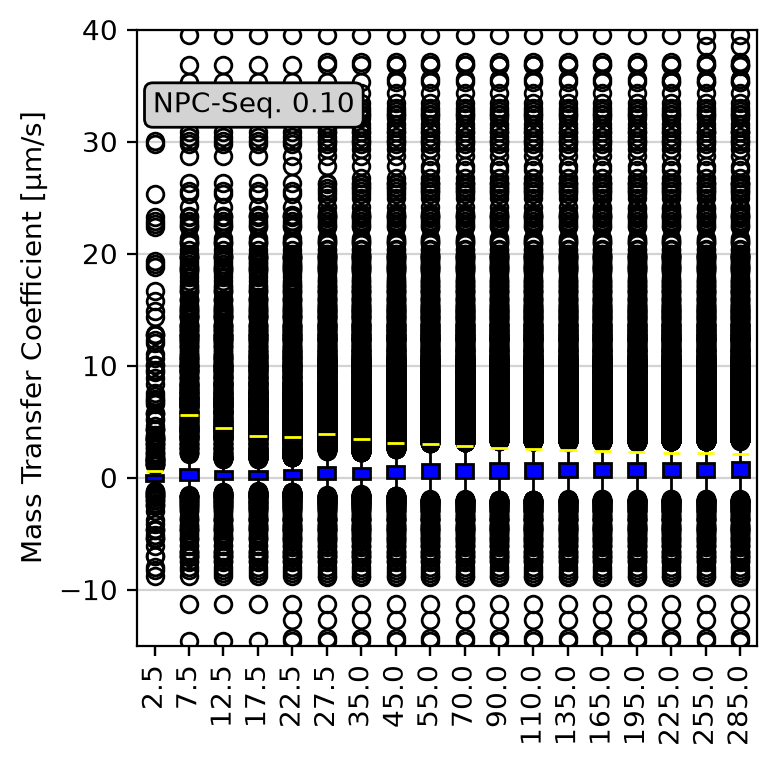}
    \end{subfigure}%
    \begin{subfigure}{0.25\linewidth}
        \includegraphics[width=\linewidth, height = 3.5cm]{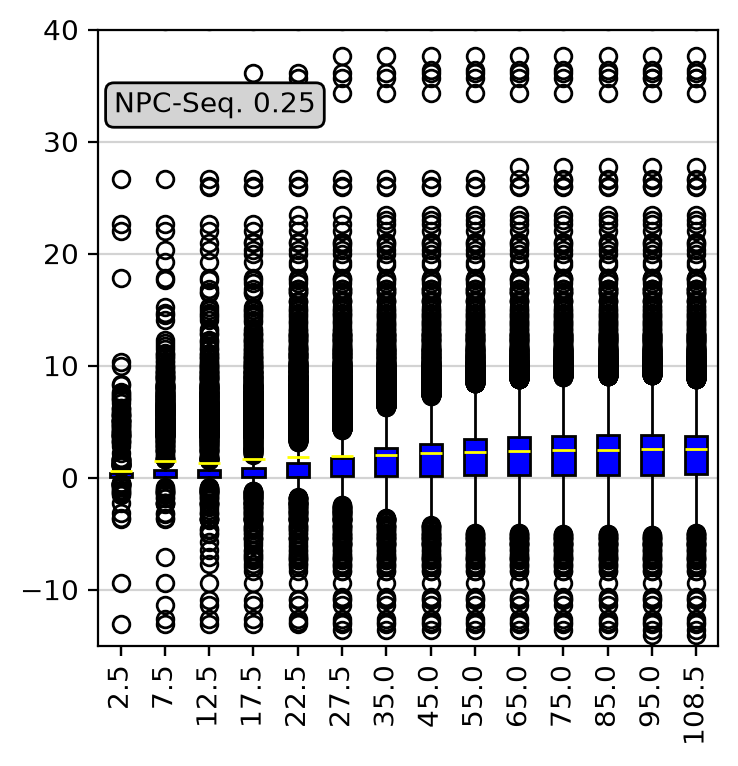}
    \end{subfigure}%
    \begin{subfigure}{0.25\linewidth}
        \includegraphics[width=\linewidth, height = 3.5cm]{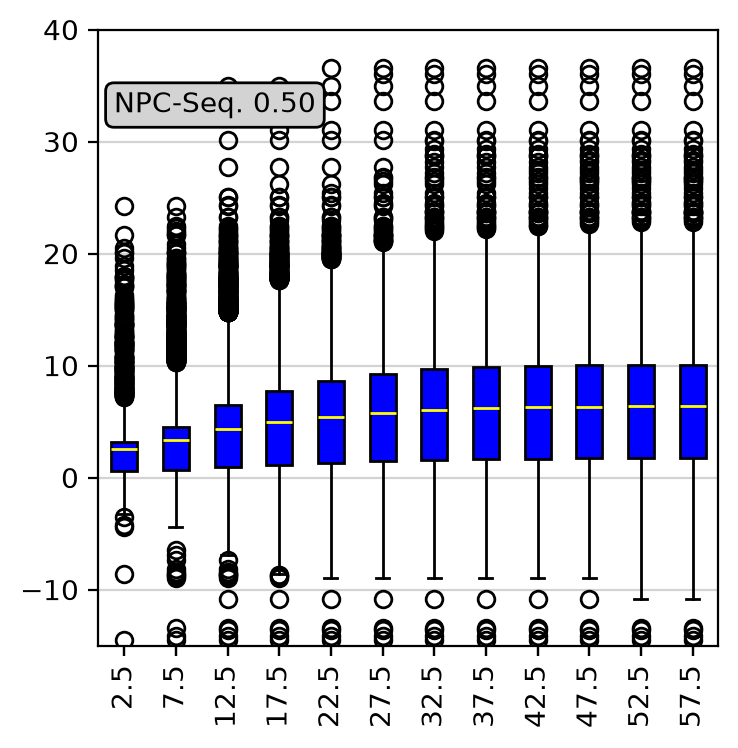}
    \end{subfigure}%
    \begin{subfigure}{0.25\linewidth}
        \includegraphics[width=\linewidth, height = 3.5cm]{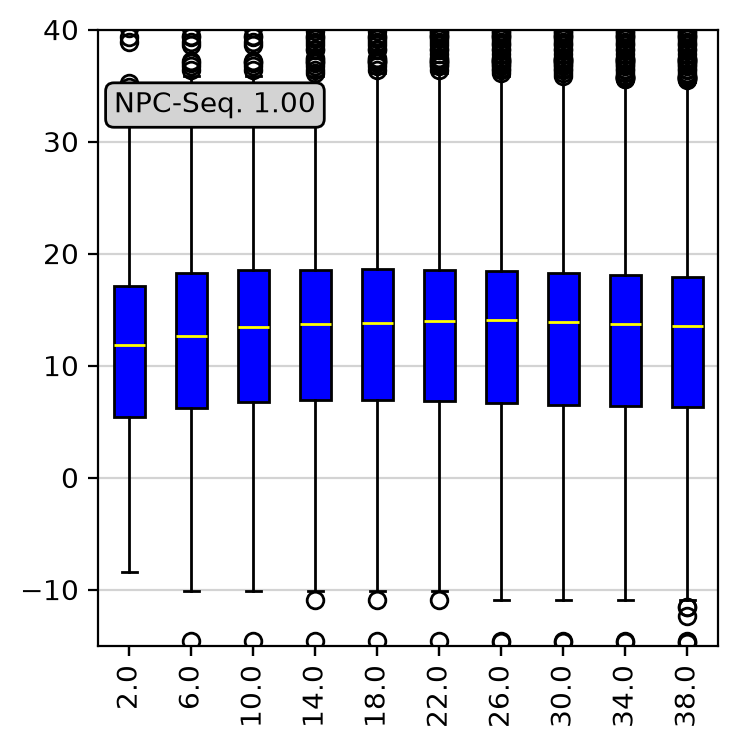}
    \end{subfigure}

    \begin{subfigure}{0.25\linewidth}
        \includegraphics[width=\linewidth, height = 3.5cm]{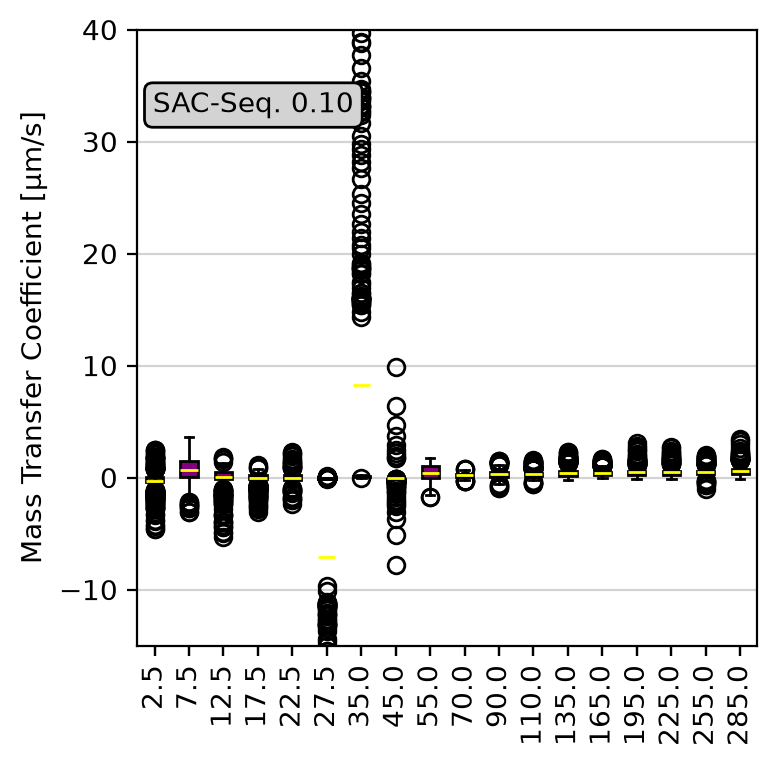}
    \end{subfigure}%
    \begin{subfigure}{0.25\linewidth}
        \includegraphics[width=\linewidth, height = 3.5cm]{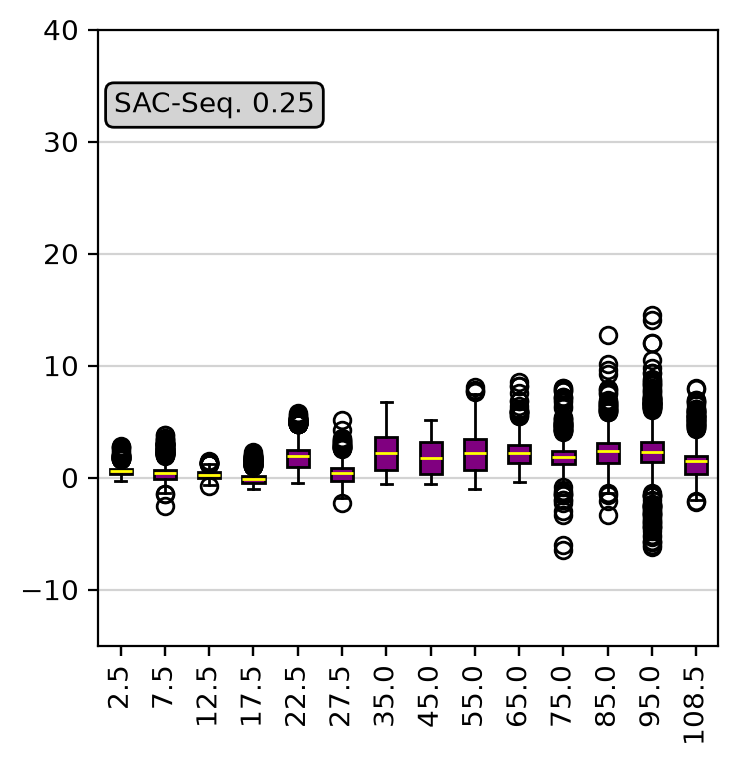}
    \end{subfigure}%
    \begin{subfigure}{.24\linewidth}
        \includegraphics[width=\linewidth, height = 3.5cm]{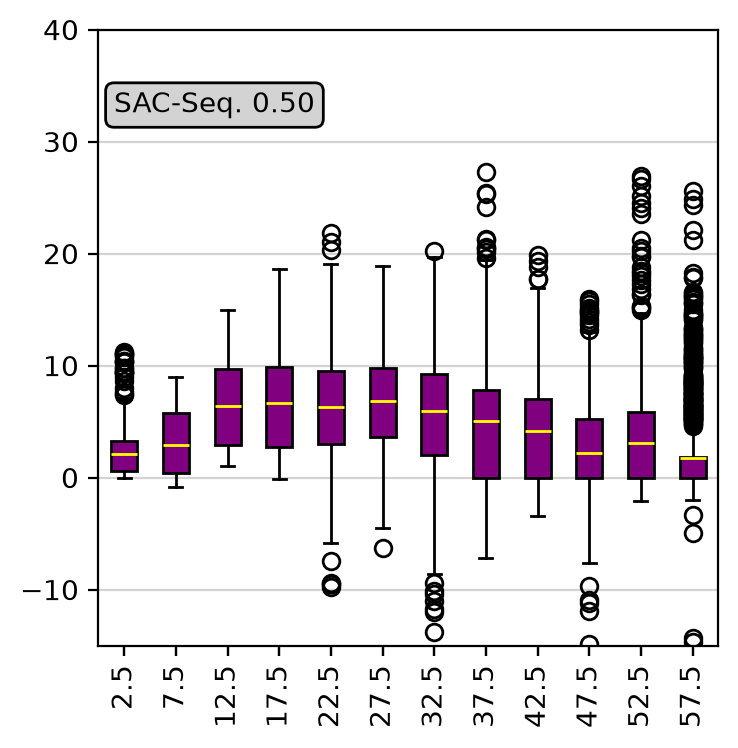}
    \end{subfigure}%
    \begin{subfigure}{0.25\linewidth}
        \includegraphics[width=\linewidth, height = 3.5cm]{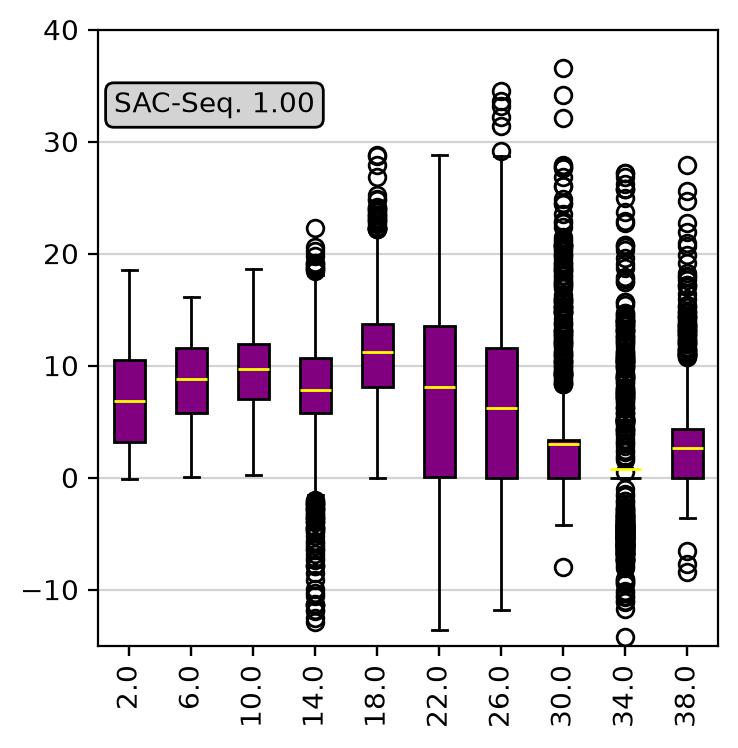}
    \end{subfigure}

    \begin{subfigure}{0.25\linewidth}
        \includegraphics[width=\linewidth, height = 3.5cm]{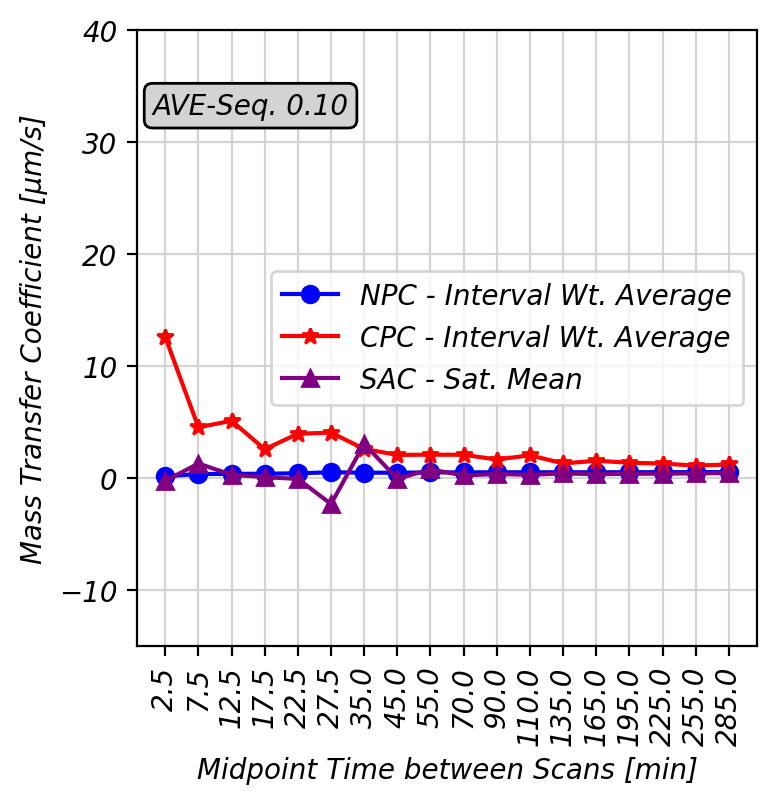}
    \end{subfigure}%
    \begin{subfigure}{0.25\linewidth}
        \includegraphics[width=\linewidth, height = 3.5cm]{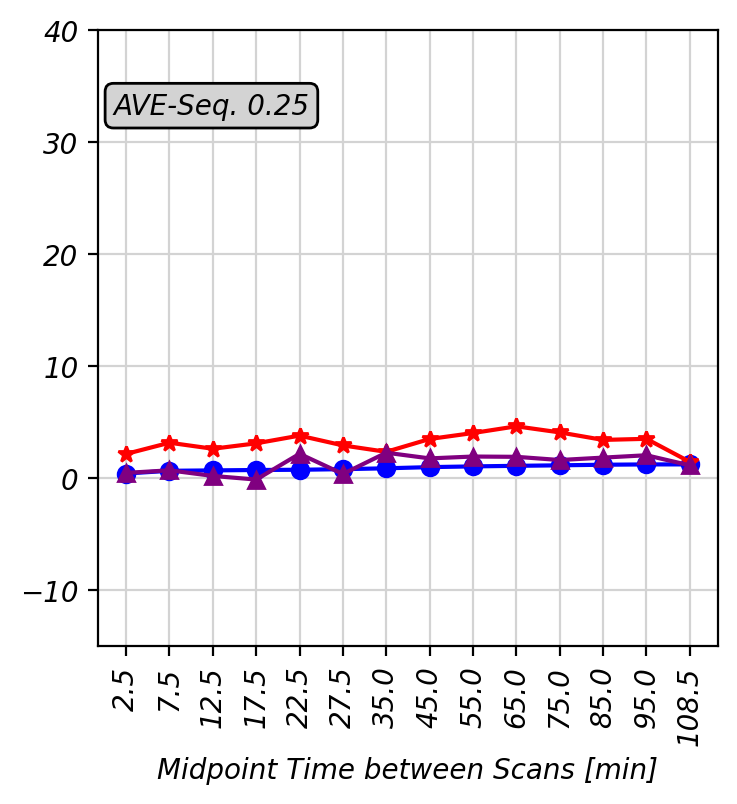}
    \end{subfigure}%
    \begin{subfigure}{.24\linewidth}
        \includegraphics[width=\linewidth, height = 3.5cm]{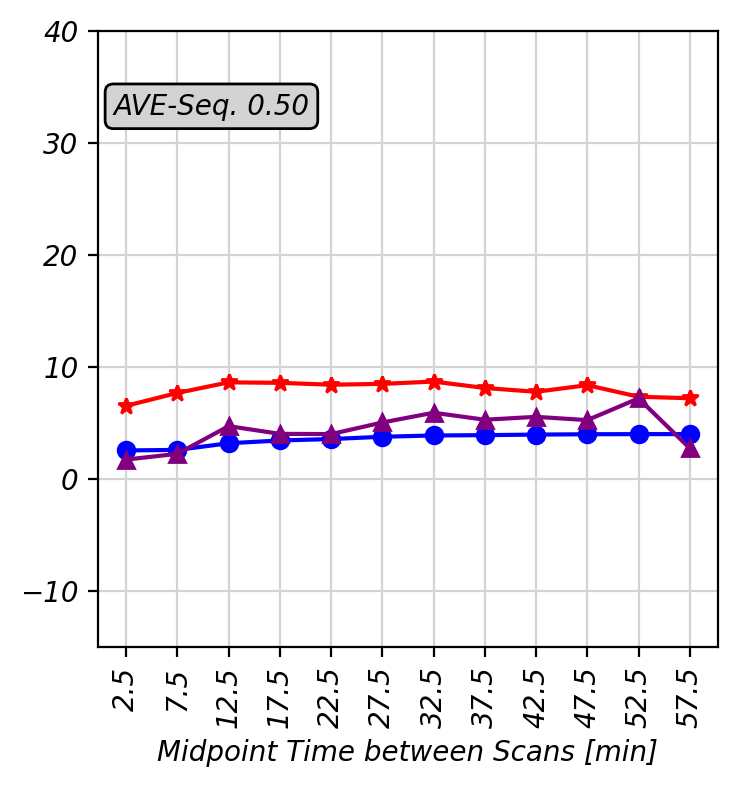}
    \end{subfigure}%
    \begin{subfigure}{0.25\linewidth}
        \includegraphics[width=\linewidth, height = 3.5cm]{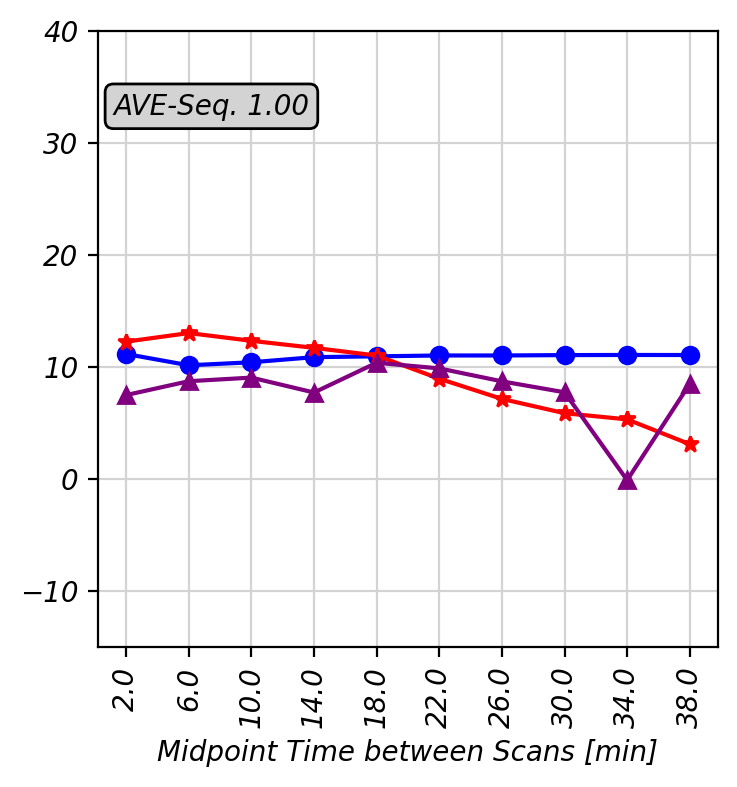}
    \end{subfigure}
    \caption {Box-and-whisker plots of the mass transfer coefficient ($k_{i,j}$) distributions for each time-interval (no mobilization filtering).}
    \label{fig:no_filt_mass_trans_dist}
\end{figure*}

\section{Additional Mobilization Resampling Results}
\begin{figure*}[hbt]
    \begin{subfigure}{0.5\linewidth}
        \includegraphics[width=\linewidth]{figures/CPC_0_10_mlm_wt_mean_bootstrap_sel3_it10000.png}
    \end{subfigure}%
    \begin{subfigure}{0.5\linewidth}
       \includegraphics[width=\linewidth]{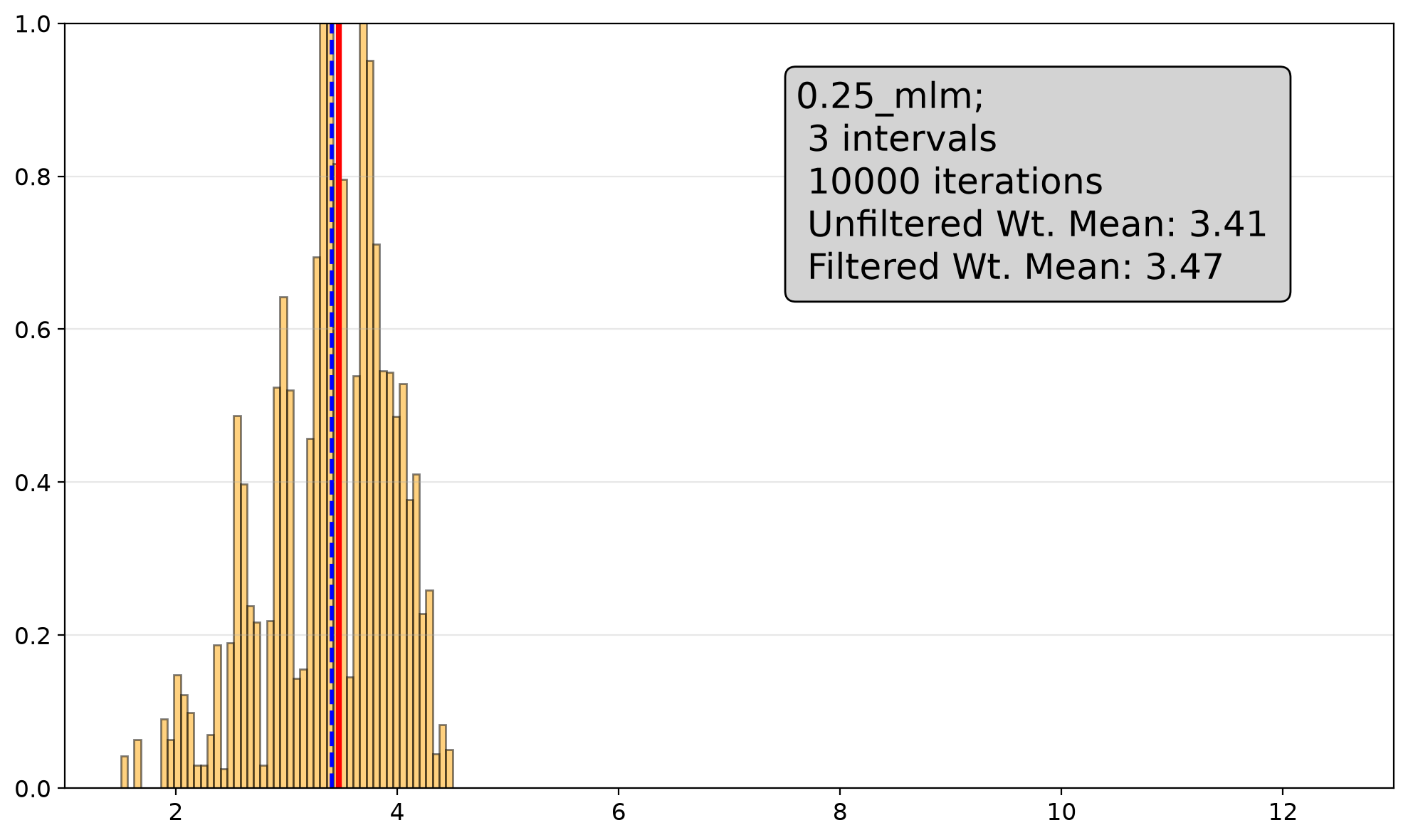}
    \end{subfigure}
    \begin{subfigure}{0.5\linewidth}
        \includegraphics[width=\linewidth]{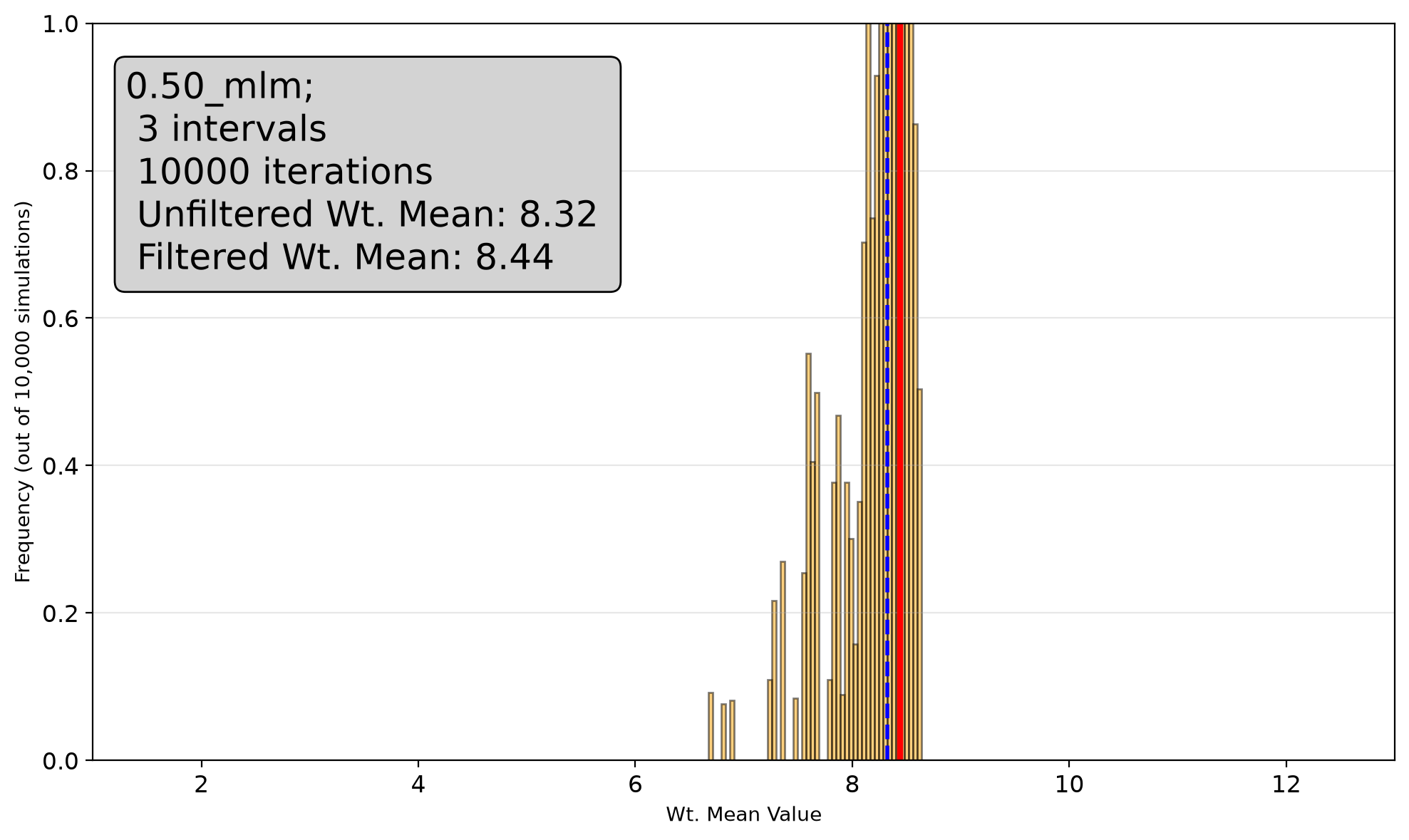}
    \end{subfigure}%
    \begin{subfigure}{0.5\linewidth}
        \includegraphics[width=\linewidth]{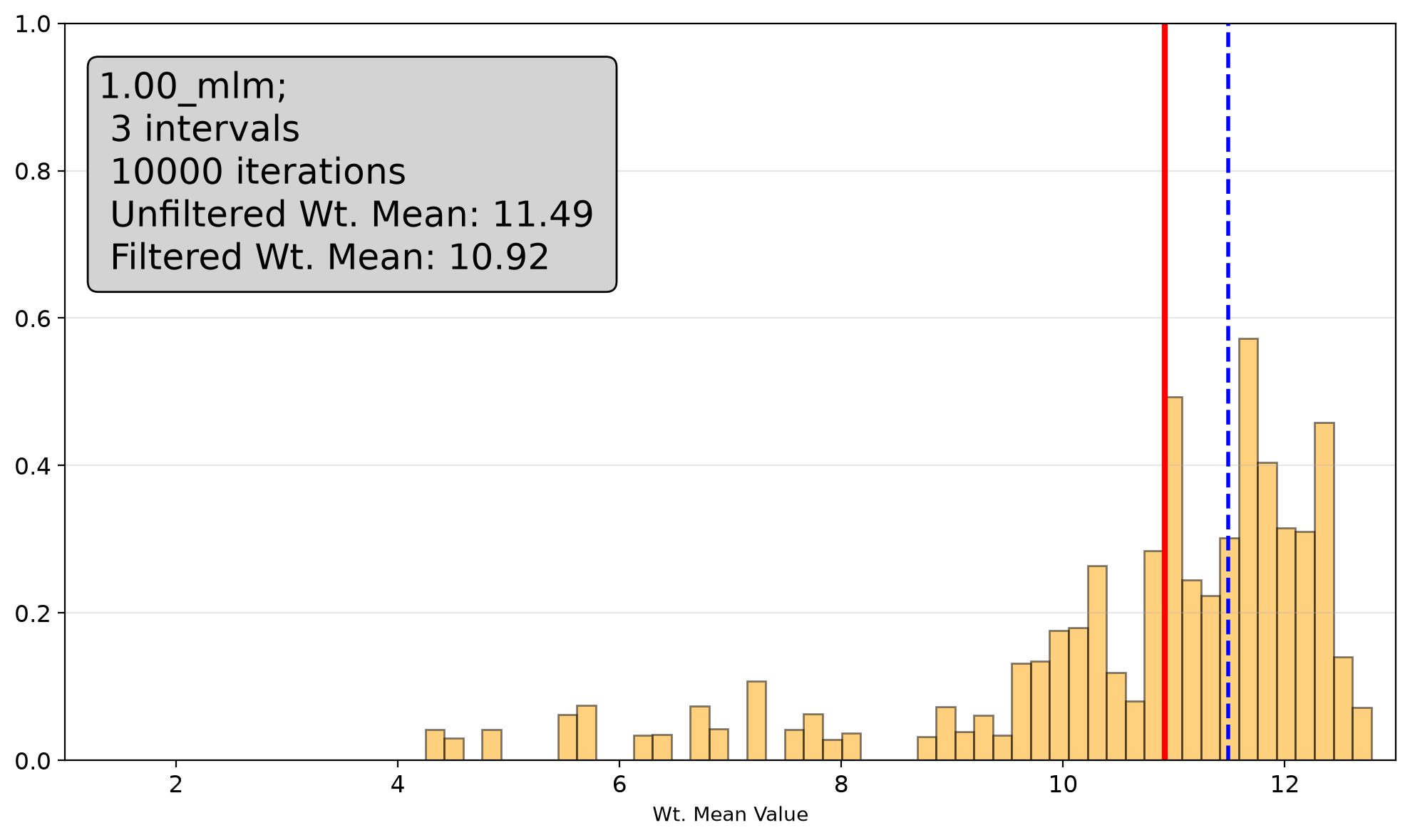}
    \end{subfigure}
    \caption{Plots of the weighted sequence average CPC mass transfer coefficients calculated from randomly resampled time-intervals (without replacement, and without mobilization filtering) for 10000 iterations. The selection size for each sequence is limited to 3 intervals to observe the effect of Seq. 0.10 smaller sampling size on the other Seq. estimates.}
    \label{fig:mass_trans_sense_pick3}
\end{figure*}